\documentclass[12pt]{article}

\usepackage{amsmath}
\usepackage{epsfig}
\usepackage{epsf}
\usepackage{psfrag}
\usepackage{amssymb}

\usepackage{graphicx}
\usepackage{array}
\usepackage{subcaption}
\usepackage{natbib}
\usepackage{algorithm}
\usepackage{algorithmic}
\usepackage[dvipsnames]{xcolor}
\usepackage{amsfonts}
\usepackage{multirow}
\usepackage{chngcntr}
\usepackage{apptools}
\usepackage{adjustbox} 

\usepackage[margin=1.0in]{geometry}

\newcommand{\xx}{\mbox{\boldmath $x$}}
\newcommand{\XX}{\mbox{\boldmath $X$}}

\newcommand{\bzero}{\mbox{\boldmath $0$}}

\newcommand{\btheta}{\mbox{\boldmath $\theta$}}
\newcommand{\bbeta}{\mbox{\boldmath $\beta$}}

\newcommand{\bmu}{\mbox{\boldmath $\mu$}}

\newcommand{\bSigma}{\mbox{\boldmath $\Sigma$}}

\def\XX{\textbf{X}}
\def\xx{\textbf{x}}

\def\calD{\mathcal{D} }

\def\DD{\textbf{D}}
\def\ww{\textbf{w}}

\def\calL{\mathcal{L} }
\def\calI{\mathcal{I} }
\def\calJ{\mathcal{J} }
\def\calS{\mathcal{S} }

\def\calA{\mathcal{A} }
\def\calB{\mathcal{B} }

\def\tbSigma{\tilde{\pmb\Sigma}}

\def\hsigma{\hat{\sigma}}

\def\tbmu{\tilde{\pmb\mu}}
\def\hbmu{\hat{\pmb\mu}}
\def\hmu{\hat{\mu}}

\def\btheta{\pmb\theta}
\def\hbtheta{\hat{\pmb\theta}}

\def\bbeps{\bar{\pmb{\epsilon}}}

\def\Dp2{\Delta_p^2}

\usepackage[figuresright]{rotating}

\renewcommand{\hat}{\widehat}

\makeatletter
\renewcommand{\section}{\@startsection{section}{1}{0mm}
   {0.5\baselineskip}{0.4\baselineskip}{\bf\normalsize\uppercase}}%
\renewcommand{\subsection}{\@startsection{subsection}{1}{0mm}
   {0.25\baselineskip}{0.25\baselineskip}{\bf\normalsize}}%
\makeatother

\usepackage{changes}

\usepackage{scalerel,stackengine}
\stackMath
\newcommand\reallywidehat[1]{%
\savestack{\tmpbox}{\stretchto{%
  \scaleto{%
    \scalerel*[\widthof{\ensuremath{#1}}]{\kern-.6pt\bigwedge\kern-.6pt}%
    {\rule[-\textheight/2]{1ex}{\textheight}}
  }{\textheight}%
}{0.5ex}}%
\stackon[1pt]{#1}{\tmpbox}%
}

\numberwithin{equation}{section}
\newtheorem{thm}{Theorem}[section]
\newtheorem{dfn}{Definition}
\newtheorem{prn}{Proposition}[section]
\newtheorem{rmk}{Remark}[section]
\newtheorem{lem}{Lemma}[section]
\AtAppendix{\counterwithin{lem}{section}}

\begin{document}

\title{
New Hard-thresholding Rules based on Data Splitting in High-dimensional 
Imbalanced Classification
\vspace{-.1in}
}
\author{Arezou Mojiri\footnote{arezu.mojiri@math.iut.ac.ir},
\\  Department of Mathematical Sciences, Isfahan University of Technology, 
\and
Abbas Khalili\footnote{abbas.khalili@mcgill.ca}, 
\\ Department of Statistics and Mathematics, McGill University, 
\and
Ali Zeinal Hamadani\footnote{hamadani@cc.iut.ac.ir},\\
 Department of Industrial and Systems Engineering, 
Isfahan University of Technology}

\maketitle
\noindent{\centerline{\bf Abstract}

In binary classification, imbalance refers to 
situations in which one class is heavily under-represented. 
This issue is due to either a data collection process or because one  
class is indeed rare in a population. Imbalanced classification 
frequently arises in applications 
such as biology, medicine, engineering, and  
social sciences. In this paper, for the first time, 
we theoretically study the impact of imbalance class sizes on the 
linear discriminant analysis (LDA) in high dimensions.  
We show that due to data scarcity in one class, referred to as the minority class, 
and high-dimensionality of the feature space, 
the LDA ignores the minority class yielding a 
maximum misclassification rate. 
We then propose a new construction of hard-thresholding rules   
based on a data splitting technique that 
reduces the large difference between the misclassification rates. We 
show that the proposed method is asymptotically optimal. 
We further study two well-known sparse versions of the LDA in imbalanced cases. 
We evaluate the finite-sample performance of 
different methods using simulations and by analyzing two real data sets. 
The results show that our method either outperforms 
its competitors or has comparable performance  
based on a much smaller subset of selected features, 
while being computationally more efficient.


\section{Introduction}
\label{sec:Introduction}

The rise of high-dimensional data has affected many areas of research in 
statistics and machine learning, including classification.  
Linear Discriminant Analysis (LDA) 
has been extensively studied in high-dimensional classification.   
\cite{bickel2004some}, \cite{fan2008high},  
and \cite{shao2011sparse} showed that 
when the number of features is 
larger than the sample size, the LDA can perform as badly 
as a random guess. To deal with the curse of dimensionality,  
several developments have been made over the last decade or so. 
For example, among others, new developments include 
the nearest shrunken centroids \cite{tibshirani2002diagnosis}, 
shrunken centroids regularized discriminant analysis 
\cite{guo2006regularized}, features annealed independence rule 
({\sc fair}) \cite{fan2008high}, 
sparse and penalized LDA \cite{shao2011sparse, witten2011penalized},  
regularized optimal affine discriminant 
({\sc road}) \cite{fan2012road}, 
multi-group sparse discriminant analysis \cite{gaynanova2015optimal}, 
pairwise sure independent screening \cite{pan2016ultrahigh}, and the 
ultra high-dimensional multiclass LDA \cite{li2019multiclass}. 
The general idea of these methods is to incorporate a feature selection strategy in 
a classifier in order to obtain certain 
optimality properties in the sense of misclassification rates. 

To the best of our knowledge, most of the existing developments 
in high dimensions focus on problems with comparable class sizes in the training data.   
However, in applications such as 
clinical diagnosis \citep{bach2017study}, 
fraud detection \citep{bolton2002statistical}, 
drug discovery \citep{zhu2006lago},  
or equipment malfunction detection \citep{park2013design}, 
classification  
often suffers from imbalanced class sizes 
where, for example, in a binary problem one class 
(referred to as the minority class) is heavily under-represented.
This 
is due to either a data collection process or because 
one class is indeed rare in a population. 
In such situations, the minority class is of primary interest as 
it carries substantial information, and often has higher 
misclassification costs compared to the larger class, referred to as the majority class. 
For example, in a study of a certain rare disease, the cost of misclassifying 
a positive case is often higher than the cost of misclassifying a negative one
\citep{ramaswamy2002molecular}. 
In banking or telecommunication studies, few customers 
are voluntarily willing to terminate their contracts and leave their provider.
In these applications, misclassification of a potential churner is more expensive 
than that of a non-churner for a provider \citep{verbeke2012new}. 
Due to data scarcity in the minority class, 
conventional discriminant methods are often biased toward the majority class 
resulting in much higher misclassification rate for the minority class. 
This error dramatically increases in high-dimensional cases, 
as empirically shown by \cite{Blagus2010}. 
In this paper, we study 
imbalanced 
binary classification with the class sizes $n_2 << n_1$, when 
the number of features, $p_n$, grows to infinity 
as the total sample size $n = (n_1 + n_2)$ grows to infinity. 
We refer to Class 1 with size $n_1$ 
as the majority class, and Class 2 with size $n_2$ as the minority class. A specific limiting 
relationship between $n_1$ and $n_2$ is given in Section \ref{sec:impact2}.

Imbalanced classification under various settings
have attracted attention in recent years. 
A common approach to deal with the imbalanced issue 
is to make virtual class sizes comparable by using resampling methods, for example, 
the synthetic minority over-sampling technique 
({\sc smote}) of \citep{chawla2002smote}. 
The recent work of \cite{feng2020imbalanced} provides a  
review of the common re-sampling techniques 
for fixed dimensional imbalanced problems.
In other methods, such as the 
weighted extreme learning machine \cite{zong2013weighted} and the
cost-sensitive support vector machine ({\sc svm}) 
\cite{iranmehr2019cost}, the idea is 
to strengthen the relative impact of the minority class 
by either assigning different weights to sample units or 
different costs to misclassification instances in each class. 
\cite{bak2016high}  
studied distributional properties of the correct classification probabilities 
of the minority and majority classes of 
a hard-thresholding independence rule.  
Using a non-asymptotic approach, they adjusted the bias of 
correct classification probabilities which is 
rooted on the imbalanced class sizes.  
\cite{huang2010bias} and \cite{pang2013block} 
proposed bias-corrected discriminant functions. 
\cite{owen2007infinitely} studied limiting form of the 
logistic regression under a so-called infinitely imbalanced case in which 
the size of one class is fixed and the other grows to infinity. 
\cite{qiao2009adaptive} proposed new evaluation 
criteria and weighted learning procedures that increase the impact of 
a minority class.   
\cite{qiao2010weighted} developed a 
distance weighted discrimination method 
({\sc dwd}), originally proposed to overcome 
the well-known data-piling issue \cite{ahn2010maximal}
in high-dimensions, 
by an adaptive  
weighting scheme to reduce sensitivity to unequal class sizes. 
\cite{qiao2013distance} proposed a linear classifier 
that is a hybrid of {\sc dwd} and {\sc svm}, thus haivng advantages of both techniques.
\cite{qiao2015flexible} introduced a new family of 
classifiers including {\sc svm} and {\sc dwd} that 
provides a trade-off between imbalanced and high-dimensionality. 
\cite{hall2005geometric,nakayama2017support} theoretically
showed that under certain conditions, {\sc svm} 
suffers from data-piling in high-dimensions, meaning that all 
data points become the support vectors
which may result in ignorance of one of the classes. 
\cite{nakayama2017support} proposed a biased-corrected 
{\sc svm} that improves its performance even when the class 
sizes are imbalanced. 
\cite{nakayama2020support}  proposed a robust {\sc svm}  	
which is less sensitive to class sizes and choice of a regularization parameter. 	 
\cite{xie2020fused} 
used a repeated case-control sampling technique coupled with 
a fused feature screening procedure to deal with imbalanced  and high-dimensionality.

The behaviour of LDA in high-dimensional imbalanced classification 
has often been studied empirically. In this paper, we first theoretically 
show that in such cases this classifier 
ignores the minority class, 
yielding a maximum misclassification rate for this class. On the other hand, 
a common approach to deal with high-dimensionality 
is to use a hard-thresholding operator for feature selection. 
However, our simulations 
show large differences between the misclassification rates  
of the hard-thresholding rule ({\sc hr}) in imbalanced settings.
Thus, we face both high-dimensionality and an inflated bias 
in the difference between the two misclassification rates.
To address the issues, we propose a new construction of the 
{\sc hr} based a multiple data splitting (Msplit)  
technique as described below, and thus called Msplit-{\sc hr}. 
We randomly split the training data in each class 
into two parts of sizes $\lfloor n_k/2 \rfloor, k=1, 2$, 
and use one part 
only for feature selection and the other part is then used 
to construct a bias-corrected classifier based on the selected features.  
As shown in Section \ref{sec:DAC}, the splitting 
facilitates the correction of 
the inflated bias 
in the difference between the two misclassification rates.  
To reduce the effect of randomness in single-split, 
we repeat the process several ($\calL$) times which  
maximizes the usage of training data in finite-sample 
situations. In general, 
as pointed out by \cite{meinshausen2009p}, 
multiple splitting also helps 
reproducibility of finite sample results. 
As shown numerically in 
Figures \ref{fig:effect_T_1} and \ref{fig:effect_T_2}, 
respectively 
discussed in Sections \ref{subsec:indep} and 
\ref{subsec:general_sigma}, 
the classification results of Msplit-{\sc hr} 
corresponding to $\calL \approx 30$ 
are unsurprisingly 
more powerful than a single-split ($\calL=1$). 	 
We show that our method 
is asymptotically optimal. 
We also study asymptotic properties of two well-known 
linear classifiers, 
namely the sparse LDA 
\cite{shao2011sparse}, and the 
regularized optimal affine discriminant analysis  
\cite{fan2012road}, 
under the imbalanced setting. Our simulations show that Msplit-{\sc hr} 
either outperforms its competitors or has comparable performance
based on a much smaller subset of selected features, while 
being computationally more efficient as discussed in Section \ref{sec:sim}. 

The rest of the paper is organized as follows. 
Section \ref{sec:Impact} gives the problem setup and 
investigates the behaviour of the LDA in high-dimensional 
imbalanced binary classification. 
Section \ref{sec:DAC} introduces our proposed method, Msplit-{\sc hr}.   
Large-sample properties of the method are also discussed in this section. 
Two well-known high-dimensional variants of the LDA, under the imbalanced setting,  
are studied in Section \ref{sec:variants}. 
The finite-sample performance of several 
binary classifiers is examined 
using simulations in Section \ref{sec:sim}. Analysis of two real data sets 
are given in Section \ref{sec:real}.
A summary and discussion are given in Section \ref{sec:discuss}. 
Technical Lemmas and proofs of our main 
results are given in Appendices A-C. 

\noindent
{\bf Notation:}
All vectors and matrices are shown in bold letters.
For any vector $\textbf{a}\in \mathbb{R} ^p$,  
$\Vert \textbf{a} \Vert_0=\# \lbrace j : \ a_j\neq 0 \rbrace$,
$\Vert \textbf{a} \Vert_1=\sum_{j=1}^p\vert a_j \vert$, 
$\Vert \textbf{a} \Vert_2=(\sum_{j=1}^p a_j ^2)^{1/2}$, 
$\Vert \textbf{a} \Vert_{\infty}=\max _{j=1,...,p}\vert a_j \vert$. 
For any symmetric matrix $\textbf{A}\in \mathbb{R} ^{p\times p}~$, 
$\Vert\textbf{A} \Vert_1=\max_{i=1,...,p}\sum_{j=1}^p \vert a_{ij} \vert$, 
$\Vert\textbf{A} \Vert 
=\max_{j=1,...,p} \vert \lambda _j(\textbf{A}) \vert$, 
where $\lambda _j(\textbf{A})$ are the eigenvalues of the matrix $\textbf{A}$,  
and  
$\Vert\textbf{A} \Vert_{\infty}=\max_{i,j=1,...,p} \vert a_{ij}  \vert$. 
A diagonal matrix is denoted by $\textbf{D}$. For any two sequences $a_n$ 
and $b_n$, we write $a_n\lesssim b_n$ or $a_n=O(b_n)$, if for 
sufficiently large $n$ there exists a constant $C$  such that $a_n \leq C~ b_n$. We write  
$a_n\sim b_n$,  if $a_n/b_n \to 1$, as $n\to \infty$. 
And $a_n \asymp b_n$, if $a_n=O(b_n)$ 
and $b_n=O(a_n)$.  
Also, $a_n=o(b_n)$, when $a_n/b_n\rightarrow 0$ 
as $n\rightarrow \infty$. 
The notations $o_p$ and $O_p$ are  respectively used to 
indicate convergence and boundedness in probability.
An indicator function is denoted by $\textbf{1} \{\cdot \}$. 

\section{The LDA}
\label{sec:Impact}

In this section, we first describe the 
setting of the binary classification problem 
under our consideration. We then study the effect of dimension and 
imbalanced class sizes 
on the LDA, which motives the topics of the remaining sections. 

\subsection{Overview }
\label{sec:problem}

We consider 
the class labels 
$Y\in \{ 1,2 \}$, class prior probabilities $\pi_k = \Pr(Y = k)$, 
and a $p$-dimensional feature vector 
$\XX = (X_1, X_2, \ldots, X_p)^\top$ such that 
\(
\XX \vert Y=k \sim N_p(\bmu_k , \bSigma), k= 1, 2.
\)
The LDA is a well-known classification technique for this setting. 
More specifically, given the parameter vector $\btheta= (\bmu_1, \bmu_2, \bSigma)$ and 
assuming $\pi_1=\pi_2$, the optimal rule classifies a subject with an observed feature vector 
$\xx^* = (x^*_1, \ldots, x^*_p)^\top$ to Class 1 if and only if 
\begin{equation}
	\label{optimal bayes}
	\delta^{\text{opt}} (\xx^*; \btheta) = \bmu_d^{\top} \bSigma^{-1}(\xx^* - \bmu_a) < 0,
\end{equation}
where $\bmu_d = \bmu_2 - \bmu_1 \neq \bzero$, 
$\bmu_a=(\bmu_2 + \bmu_1)/2$. 

The misclassification rate (MCR) of a classifier is typically used to quantity its performance. 
The classifier in \eqref{optimal bayes} which is the Bayes' rule, is referred to 
as the optimal rule since it has the smallest average MCR, $\Pi^{\text{opt}}$ in \eqref{mcrk} below,  
among all classifiers. For $\delta^{\text{opt}}$, the class-specific MCRs are equal and given by 
\begin{eqnarray}
	\label{mcrk}
	\Pi_k=
	\Pr \bigg( (-1)^k\delta^{\text{opt}}(\XX^*; \btheta) < 0  \bigg\vert Y=k \bigg)  
	=
	\Phi \left( -\Delta_p/2\right)
	\equiv  
	\Pi^{\text{opt}}, \ \ k=1,2,
\end{eqnarray}
where $\Phi$ is the cumulative distribution function of the standard normal, 
and $\Delta_p^2=\bmu_d^{\top}\bSigma^{-1}\bmu_d$ 
is referred to as the discriminative power or signal value. 
It is seen that as $\Delta_p\to \infty$, 
high discriminative power, 
then $\Pi^{\text{opt}} \to 0$; and as $\Delta_p\to 0$, 
low discriminative power, then $\Pi^{\text{opt}} \to \frac{1}{2}$ 
implying that the 
classifier performs as a random guess. 
From now on, we assess the performance of other classifiers under 
consideration by comparing them with the optimal rule.  

In practice, the parameter vector $\btheta$ 
is unknown and 
needs to be estimated using a training data 
$\mathcal{D}_n = \{ \xx_{ik}, \ i=1,...,n_k , \ k=1,2 \}$, where
$\xx_{ik}$ is the $i$-th observed value of $\XX$ in Class $k$, and
the $n_k$ are the class sample sizes with 
the total sample size $n=n_1+n_2$. 
For a new feature vector $\xx^*$, 
a so-called plug-in discriminant
function based on the parameter estimates 
is given by 
\begin{equation}
	\label{rhat}
	\delta^{\text{\sc lda}}(\xx^*;\hat{\btheta}_n)
	= 
	\hat{\bmu}_d^{\top}~ \widehat{\bSigma}_n^{-1} (\xx^*-\hat{\bmu}_a),
\end{equation}
where 
$
\hat{\btheta}_n =
(\hat{\bmu}_{n,1},\hat{\bmu}_{n,2},\widehat{\bSigma}_n)$ 
and 
\begin{eqnarray}
	\hat{\bmu}_{n,k}  \equiv \hat{\bmu}_{k} 
	& = & 
	\frac{1}{n_k}\sum _{i=1}^{n_k} \xx_{ik}, \  \ k=1,2, \label{eq:hat_mu} \\
	\widehat{\bSigma}_n 
	& = & 
	\frac{1}{n-2}\sum_{k=1}^2\sum_{i=1}^{n_k} (\xx_{ik} 
	- 
	\hat{\bmu}_k)(\xx_{ik}-\hat{\bmu}_k)^{\top}. 
	\label{eq:hat_Sigma}
\end{eqnarray}
The matrix $\widehat{\bSigma}^{-1}_n$ in \eqref{rhat} is a generalized inverse when 
$\widehat{\bSigma}_n$ is not invertible. 
Given $\mathcal{D}_n$, the conditional MCR of the  
plug-in linear 
discriminant rule based on (\ref{rhat}) corresponding to 
Class $k\in \lbrace 1,2\rbrace$, 
is given by
\begin{equation}
	\label{ccrk}
	\Pi _k^{\text{\sc lda}} (\calD_n)  
	= 
	\Pr \left((-1)^k\delta^{\text{\sc lda}}(\XX^*;\hat{\btheta}_n)<0 \bigg\vert \ Y=k, \  \mathcal{D}_n \right) 
	=
	\Phi \bigg( \frac{ \Psi_k^{\text{\sc lda}}(\hat{\btheta}_n)}{\sqrt{\Upsilon^{\text{\sc lda}} (\hat{\btheta}_n)}} \bigg),
\end{equation}
where $\Psi_k^{\text{\sc lda}}(\hat{\btheta}_n) = 
(-1)^{k}\hat{\bmu}_d^{\top}\widehat{\bSigma}_n^{-1}(\hat{\bmu}_a-\bmu_k)$, 
and $\Upsilon^{\text{\sc lda}}(\hat{\btheta}_n)
=\hat{\bmu}_d^{\top}\widehat{\bSigma}_n^{-1}\bSigma\widehat{\bSigma}_n^{-1}\hat{\bmu}_d$.
As is common in the literature, we study large-sample 
properties of a classifier through its conditional MCR. 

\subsection{Impact of the dimension and imbalanced class sizes}
\label{sec:impact2}

The effect of the dimension $p$ 
on the LDA's performance is well studied in the literature. 
\cite{shao2011sparse} showed that when 
$p$ 
is fixed or diverges to infinity at a slower rate than $\sqrt{n}$, 
the classifier is asymptotically optimal \citep[Definition 1]{shao2011sparse}.  
When $p \rightarrow \infty$ 
such that $p/n\rightarrow\infty$, \cite{bickel2004some}, 
\cite{fan2008high}, and \cite{shao2011sparse} 
showed that this classifier performs no better than a random 
guess. Hence, 
feature selection is essential when 
$p$ is large compared to the sample size $n$. 

In the aforementioned works, the impact of dimensionality is studied 
under particular limiting settings on the class sizes $n_1$ and $n_2$. 
\cite{bickel2004some} and \cite{shao2011sparse} respectively considered
equal class sizes ($n_1=n_2$) and unequal sizes where 
$n_1,\ n_2\rightarrow \infty$ such that $\frac{n_2}{n}\rightarrow \pi$, $0<\pi <1$. 
\cite{fan2008high} developed their results by considering  
compatible class sizes, 
such that
$c_1\leq \frac{n_1}{n_2}\leq c_2$, with $0< c_1  \leq c_2 < \infty$. 
\cite{bak2016high} investigated the case where $n_1,\ n_2\rightarrow \infty$, 
such that $\frac{n_1-n_2}{n_1+n_2}=\rho>0$ is fixed.
All in all, it is seen that the sizes of the two classes grow 
similarly and proportional to the total sample size $n$, 
that is $n_k = O(n), k=1, 2$. We refer to these settings as a 
{\it balanced} classification problem. \cite{owen2007infinitely} 
analyzed the binary logistic regression models with fixed dimension $p$ 
in a so-called infinitely imbalanced case in which $n_1\rightarrow \infty$ 
but the class size $n_2$ is fixed. 
In this paper, we study 
imbalanced 
classification in which $n_1$ and $n_2$ grow to infinity such that $n_2= o(n_1)$, implying 
a different growth rate of the class sizes. 

In the balanced classification, 
typically average 
(over the two classes) MCR  \citep{shao2011sparse, fan2012road} or 
the MCR of one arbitrary class \citep{bickel2004some, fan2008high} 
is used as a performance measure of a classifier. However, in imbalanced 
situations due to data scarcity in the minority class, 
classification results have a tendency to favour the majority class. Thus, 
the average MCR is not an appropriate  performance measure for a classifier ${\cal T}$. 
This motivated us to adapt the optimality definition of a classifier 
from \cite{shao2011sparse} to our setting as follows. 

\begin{dfn}
	\label{defshao}
	Suppose ${\cal T}$ is a classifier in a binary classification problem.  
	The misclassification rates of ${\cal T}$, given the training data ${\cal D}_n$, 
	are denoted by $\Pi_k^{{\cal T}}({\cal D}_n), k=1,2$. Then, 
	\begin{enumerate}
		\item[(i)]
		$\cal T$ is asymptotically-strong optimal if 
		$\Pi_k^{\cal T}({\cal D}_n)/\Pi^{\text{opt}}\overset{p}{\longrightarrow} 1$, $k=1,2$,
		\item[(ii)]
		$\cal T$ is asymptotically-strong sub-optimal if 
		$\Pi_k^{\cal T}({\cal D}_n)-\Pi^{\text{opt}} \overset{p}{\longrightarrow} 0$, $k=1,2$,
		\item[(iii)] $\cal T$ is asymptotically-strong worst if 
		$\Pi_k^{\cal T}({\cal D}_n)\overset{p}{\longrightarrow} \frac{1}{2}$, $k=1,2$,
		\item[(iv)] $\cal T$ is asymptotically ignorant if 
		$\min_{k=1,2}  \Pi_k^{\cal T}({\cal D}_n)\overset{p}{\longrightarrow} 0$ 
		and \\ $\max_{k=1,2} \Pi_k^{\cal T}({\cal D}_n)\overset{p}{\longrightarrow} 1$.
	\end{enumerate}
\end{dfn}
Note that any classifier 
${\cal T}$ satisfying either of the properties 
in parts (i)-(iii) of the above definition also satisfies the 
properties discussed in the corresponding parts of 
Definition 1 of \cite{shao2011sparse} for a balanced case, 
but not vice versa. Part (iv) of the above definition occurs when a classifier completely ignores 
one of the classes, and more specifically the minority class. 
We now state our first result.

\begin{thm}
	\label{thrm:LDA}
	Suppose that the estimator $\widehat{\bSigma}_n^{-1}$ in $\delta^{\text{\sc lda}}$ in \eqref{rhat}
	is replaced by $\bSigma^{-1}$, and $\bSigma$ is known. When 
	$n_2 = o(n_1)$, such that $p/n_2 \to \infty$ 
	and $\sqrt{\frac{n_2}{p}}\Delta_p^2=o(1)$, 
	as $n_1,n_2 \to \infty$, then the LDA is asymptotically ignorant, that is, 
	\[
	\Pi _1^{\text{\sc lda}} (\calD_n) \overset{p}{\longrightarrow}\ 0~~,~~ 
	\Pi _2^{\text{\sc lda}}  (\calD_n) \overset{p}{\longrightarrow} \ 1.
	\]
\end{thm}
This result implies that in the high-dimensional 
imbalanced cases, 
the MCR of the majority class  
tends to 0 which will be even better than the optimal value 
$\Pi^{\text{opt}}$, but the MCR of the minority class approaches 1
which is worse than a random guess.  
Note that the above result also holds in the case of $p/n_1 \to c$, 
for some finite constant $c \ge 0$. 
\citep{hall2005geometric, nakayama2017support} showed that, 
under certain conditions, the {\sc svm} ignores the minority class  
in high-dimensional imbalanced problems.

\begin{rmk}
	\label{remark:prop1}
	When $p$ is fixed and $n_2=o(n_1)$, then the LDA 
	is asymptotically-strong optimal. 
\end{rmk}
Remark \ref{remark:prop1} illustrates that in the fixed-dimensional case, 
the impact of imbalanced class sizes asymptotically vanishes and  
$\Pi_k^{\text{\sc lda}}(\calD_n),~ k=1,2$, 
converge to the optimal value $\Pi^{\text{opt}}$. Hence,  
by Theorem \ref{thrm:LDA}, Shao's results, 
and Remark \ref{remark:prop1}, the effects of both dimension
and imbalanced class sizes are responsible for ignorance of the minority class.

\section{Proposed Method: Msplit hard-thresholding rule (Msplit-{\sc hr})}
\label{sec:DAC}

A common approach to deal with high-dimensionality 
in the LDA is to incorporate feature selection using a hard-thresholding rule ({\sc hr}) 
based on a two-sample t-statistic as in \cite{fan2008high}. More specifically, 
by ignoring the correlation among features, 
$\bSigma$ 
is estimated by the diagonal matrix 
$\widehat{\DD}_n = diag\lbrace \hsigma_1^2,...,\hsigma_p^2\rbrace$, 
and the discriminant function is given by 
\begin{eqnarray}
	\label{HLDF}
	\delta^{\text{\sc hr}} (\xx^*;\hbtheta_n) 
	=
	\sum_{j=1}^p r_j(\xx^*;\hbtheta_n)~ h_j(\hbtheta_n),
\end{eqnarray} 
where $\hbtheta_n=(\hbmu_1,\hbmu_2,\widehat{\DD}_n)$,    
$r_j(\xx^*;\hbtheta_n)
=(\hmu_{dj}/\hsigma_j^2)(x_j^*-\hmu_{aj})$, and 
$h_j(\hbtheta_n)=\textbf{1}\lbrace \vert t_j \vert > \tau_n \rbrace$ 
is the thresholding operator based on the t-statistic 
\begin{equation}
	\label{t-stat}
	t_j=\frac{\hmu_{j2}-\hmu_{j1}}{\hsigma_j\sqrt{n/n_1n_2}}.
\end{equation}
Here $\hmu_{jk}$'s and $\hsigma_j^2$'s are the entries of $\hbmu_k$ 
and $\widehat{\bSigma}_n$ in \eqref{eq:hat_mu} and \eqref{eq:hat_Sigma}, 
respectively. 
The discriminant function of 
FAIR proposed by \cite{fan2008high} for balanced problems 
belongs to the class of functions in \eqref{HLDF}. 
The authors select an optimal number of statistically 
most significant features, or equivalently the threshold value $\tau_n$ 
of t-statistic, by minimizing a common upper bound on its corresponding MCRs. 
However, 
for the case of general $\bSigma$, 
such choice of $\tau_n$ does not necessarily result in an asymptotically optimal classifier 
\cite{shao2011sparse}. Thus, for generality, 
in the rest of the paper, for any given sequence of $\tau_n$, 
we refer to a classifier based on \eqref{HLDF} as an HR unless otherwise 
is specified. 

If indeed 
$\bSigma=\DD$, \cite{bak2016high} showed that  
the {\sc hr} in \eqref{HLDF} based on a fixed threshold $\tau_n = \tau$, 
is asymptotically ignorant when 
$\rho=(n_1-n_2)/(n_1+n_2) > 0$ is fixed, 
as $n_1,n_2 \to \infty$. As stated after Theorem \ref{thrm:DACH} below, 
it is interesting to note that under 
the imbalanced setting $n_2=o(n_1)$ and by an appropriate choice of 
$\tau_n$, 
the {\sc hr} is indeed asymptotically-strong optimal.
However, our simulations in Section \ref{sec:sim}
show an unsatisfactory finite-sample performance of the 
{\sc hr} in the sense of both the MCR  
in the minority class and large difference between the two MCRs.
We propose a new construction of the {\sc hr} 
which outperforms 
\eqref{HLDF} in finite-samples, 
while maintaining the same desirable large-sample properties, to be discussed below. 
To fix ideas, 
we first consider the 
imbalanced problem with a diagonal 
$\bSigma=\DD$. 
The general case of a non-diagonal $\bSigma$ is discussed in 
Subsection \ref{subsec:general_sigma}, which is 
based on a feature screening technique. Note that under this case, 
the {\sc hr} based on \eqref{HLDF} 
is not optimal, as it ignores the correlation among the features. 

\subsection{Msplit-HR under a diagonal $\bSigma$ }
\label{subsec:indep}
As discussed in Section \ref{sec:Impact}, 
the class specific MCRs of the optimal rule are 
equal, and are given in \eqref{mcrk}.   
Our numerical experiments  
show that, due to the imbalanced class sizes, 
{\sc hr} performs well in majority class   
but underperforms in minority class, though it  
has large-sample optimal property as discussed 
after 
Theorem \ref{thrm:DACH} below. Thus, the idea in our work 
is to reduce the difference between two conditional 
MCRs of {\sc hr} toward that of the optimal rule which is zero. 
More specifically, 
our main goal is to propose a new discriminant 
function aiming to reduce the 
difference between the MCRs of the {\sc hr}, 
\begin{eqnarray*}
	\vert 
	\Pi_1^{\text{\sc hr}}(\calD_n)
	-
	\Pi_2^{\text{\sc hr}}(\calD_n)  
	\vert
	=
	\vert 
	\Phi ( {\psi}_{1,n} )
	-
	\Phi ( {\psi}_{2,n} )
	\vert ,
\end{eqnarray*}
where $\psi_{k,n}=\Psi_k^{\text{\sc hr}}(\hbtheta_n)/\surd\Upsilon^{\text{\sc hr}}(\hbtheta_n)$ 
and  
\begin{eqnarray*}
	\begin{matrix}
		\Psi_k^{\text{\sc hr}}(\hbtheta_n)=
		(-1)^{k+1} \sum_{j=1}^p r_j(\bmu_k;\hbtheta_n)h_j(\hbtheta_n)~,~
		\Upsilon^{\text{\sc hr}}(\hbtheta_n)=
		\sum_{j=1}^p (\hmu_{dj}/\hsigma_j^2)^2 \sigma_j^2 h_j(\hbtheta_n)
	\end{matrix}
\end{eqnarray*}
for $k=1,2$. To understand the above difference, 
\cite{bak2016high} studied distributional properties of the quantities 
$\psi_{k,n}, k=1,2$. They focused on reducing the so-called bias 
\begin{equation}
	\label{bias-HLDA}
	B_n^{\text{\sc hr}} 
	=
	\mathbb{E} \big \lbrace \Psi_1^{\text{\sc hr}}(\hbtheta_n)  - 
	\Psi_2^{\text{\sc hr}}(\hbtheta_n) \big \rbrace
\end{equation} 
to zero, which results in decreasing the bias between $\psi_{1,n}$  
and $\psi_{2,n}$ and consequently of that between MCRs. 
However, it turns out that due to the dependency between 
the random variables $r_j$ and $h_j$, 
computing 
$B_n^{\text{\sc hr}} $ is not an easy task. 
\cite{bak2016high} studied the origin of the bias and proposed 
methods for its correction. 
We instead propose a new construction of 
the {\sc hr} that facilities the computation of such bias
by adapting a sample-splitting strategy as follows.

The training sample of each class is randomly partitioned into 
two sub-samples of 
sizes $n '_k=\lfloor n_k/2 \rfloor$. The two sub-samples are  
used for computing two quantities similar to the $r_j$ and  
$h_j$ in \eqref{HLDF}, for each $j= 1, \ldots, p$, 
and then the results are merged. 
To reduce the effect of randomness due to the data splitting, 
this process is repeated, say, $\calL$ times. 
Our new discrimination function is then constructed by averaging over the {\sc hr}-type 
discriminant functions based on each splitting.  
Thus, we chose the name Msplit-{\sc hr} for our method.  
More specifically, at the $\ell$-th data splitting, for each $\ell=1,...,\calL$, 
the entire training data $\calD_n$ is partitioned into two parts  
$\calD_{n,\ell}^{(1)}$ and $\calD_{n,\ell}^{(2)}$. 
The parameter estimates based on 
each sub-sample are distinguished by the superscripts 
$^{(1)}$ and $^{(2)}$, that is, 
$\hbtheta_{n,\ell}^{(1)}$ 
and 
$\hbtheta_{n,\ell}^{(2)}$.
A new observation with a feature vector $\xx^*$ is then classified 
using the discriminant function 
\begin{equation}
	\label{DACH0}
	\delta_0^{\text{Msplit-{\sc hr}}}(\xx^*;\hbtheta_n)
	=
	\frac{1}{\calL}  \sum_{\ell=1}^{\calL}  \sum_{j=1}^p 
	r_j(\xx^*;\hbtheta_{n,\ell}^{(2)})~ h_j(\hbtheta_{n,\ell}^{(1)}),
\end{equation}
where 
$\hbtheta_n=\lbrace (\hbtheta_{n,\ell}^{(1)},\hbtheta_{n,\ell}^{(2)}): ~
\text{for } \ell=1,\dots ,\calL  \rbrace$. 
Due to the statistical independence of the two random functions 
$h_j$ and $r_j$ in \eqref{DACH0}, for all $j= 1, \ldots, p$, 
calculation of the bias $B_n$ 
for $\delta_0^{\text{Msplit-{\sc hr}}}$ 
is straight forward, which is shown below. 
Recall $n '_k=\lfloor n_k/2 \rfloor$, and let $n'=n'_1+n'_2 $ 
and $f_{n'}= n'/2-1$.
\begin{prn}
	\label{prop:DACH0}   
	The bias $B_n$ in \eqref{bias-HLDA} corresponding to
	$\delta_0^{\text{Msplit-{\sc hr}}}$ is given by
	\begin{equation*}
		B_{0,n}^{\text{Msplit-{\sc hr}}} =
		\mathbb{E}
		\big \lbrace
		\Psi_{0,1}^{\text{Msplit-{\sc hr}}}(\hbtheta_n)
		-
		\Psi_{0,2}^{\text{Msplit-{\sc hr}}}(\hbtheta_n)
		\big \rbrace =
		\frac{\bar r_n}{\calL}
		\sum_{\ell=1}^{\calL}\sum_{j=1}^p
		\mathbb{E}\{ h_j(\hbtheta_{n,\ell}^{(1)}) \},
	\end{equation*}
	where 
	$\bar{r}_n=
	f_{n'} ( \frac{1}{n'_1}-\frac{1}{n'_2})
	\frac{\Gamma ( f_{n'}-1 )}
	{\Gamma (f_{n'})}$, 
	and
	$\Gamma(\cdot)$ is the gamma function. 
\end{prn}

Finally, using the above result, we propose the bias-corrected 
discriminant function
\begin{equation}
	\label{DACH}
	\delta^{\text{Msplit-{\sc hr}}}(\xx^*;\hbtheta_n)
	=
	\frac{1}{\calL}  \sum_{\ell=1}^{\calL}  \sum_{j=1}^p 
	\bigg \lbrace r_j(\xx^*;\hbtheta_{n,\ell}^{(2)})-\frac{\bar{r}_n}{2} \bigg \rbrace
	h_j(\hbtheta_{n,\ell}^{(1)})
\end{equation}
which has its bias $B_n^{\text{Msplit-{\sc hr}}}=0$. 
The term $\bar{r}_n$ is a function of $(n'_2-n'_1)$
which is negative since $n_2<n_1$.
Hence, for any new feature vector $\xx^*$, 
the resulting discriminant function \eqref{DACH} 
tends to be more positive compared 
to the rule in \eqref{DACH0}. 
This increases the chance (or probability) 
of classifying a new observation to the minority class, 
and hence improving the classification results 
for this class. 
In our simulations and the real-data analysis, we evaluate the performance 
of Msplit-{\sc hr} based on the bias corrected function 
$\delta^{\text{Msplit-{\sc hr}}}$. 
We now describe Algorithm 
\ref{algo:DAC_indep} that summarizes the steps for computing \eqref{DACH}.
\begin{algorithm}
	\caption{: Computing the discriminant function $\delta^{\text{Msplit-{\sc hr}}}$.} 
	\begin{algorithmic}[1]\label{algo:DAC_indep}
		\REQUIRE Input $n'_1=\lfloor n_1/2\rfloor, 
		n'_2=\lfloor n_2/2\rfloor,  \xx^*, 
		\calL, \bar{r}_n$ and $\tau_n$. 
		\FOR{$\ell=1,\dots , \calL$}
		\STATE Split $\calD_n$ into $\calD_{n,\ell}^{(1)}$ and $\calD_{n,\ell}^{(2)}$
		\FOR{$j=1,\dots , p$}
		\STATE Step1:  Using $\calD_{n,\ell}^{(1)}$ compute $h_j(\hbtheta_{n,\ell}^{(1)})$
		\STATE Step2: Using $\calD_{n,\ell}^{(2)}$ compute  $r_j(\xx^*;\hbtheta_{n,\ell}^{(2)})-\frac{\bar{r}_n}{2}$ 
		\ENDFOR
		\ENDFOR
		\RETURN $\delta^{\text{Msplit-HR}}(\xx^*;\hbtheta_n)
		=
		\frac{1}{\calL}  \sum_{\ell=1}^{\calL}  \sum_{j=1}^p 
		\bigg \{ r_j(\xx^*;\hbtheta_{n,\ell}^{(2)})-\frac{\bar{r}_n}{2} \bigg \}
		h_j(\hbtheta_{n,\ell}^{(1)})$.
	\end{algorithmic}
\end{algorithm}

In practice, a value of $\calL$ is required to compute \eqref{DACH}. 
Figure \ref{fig:effect_T_1} shows the 
class-specific MCRs of \eqref{DACH} as a function of $\calL$, 
corresponding to scenario \textbf{(i)} in our simulations in 
Section \ref{subsec:indep_sim}.
It can be seen that a value of $\calL$ between 20 to 30 
provides a satisfactory performance of Msplit-{\sc hr}. 
We used $\calL=30$ in our numerical experiments.  


\begin{figure}
	\centering
	\begin{subfigure}[b]{0.4\textwidth}
		\includegraphics[width=65mm,height=70mm]{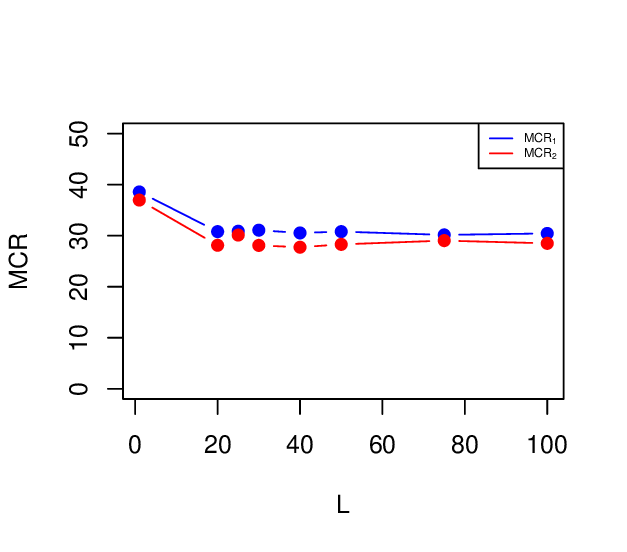}
		\caption{$n_1=50$, $n_2=10$}
		\label{fig:effect_T_50}
	\end{subfigure}%
	\qquad \qquad 
	\begin{subfigure}[b]{0.4\textwidth}
		\includegraphics[width=65mm,height=70mm]{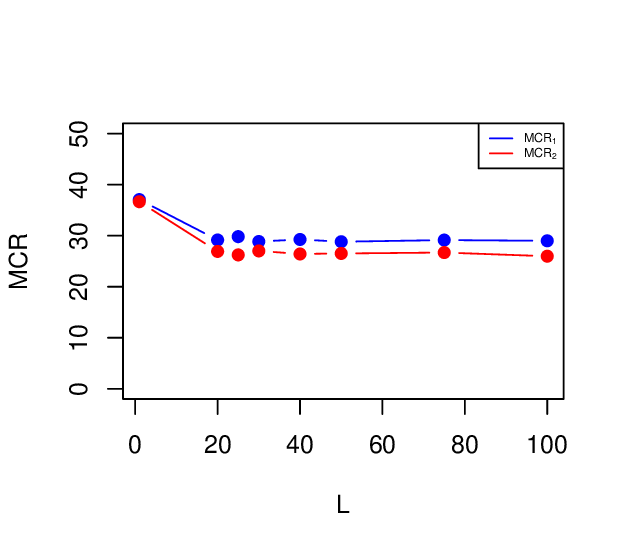}
		\caption{$n_1=100$, $n_2=10$}
		\label{fig:effect_T_100}
	\end{subfigure}
	\caption{
		Effect of the number of sample-splits $\calL$ on Msplit-{\sc hr} performance for the Simulation setting \textbf{(i)} and $p=1000$. 
	}
	\label{fig:effect_T_1}
\end{figure}

The following results show the asymptotic behaviour of 
$\delta^{\text{Msplit-{\sc hr}}}$.  
First, we state Lemma \ref{lem:true_actives} that 
provides conditions under which 
the t-statistic \eqref{t-stat} 
used in the thresholding operator $h_j$  in 
$\delta^{\text{Msplit-{\sc hr}}}$ 
selects all the important 
features. Since $\calL$ is fixed, 
the result of the lemma holds for all $\ell=1,\dots,\calL$. 

\begin{lem}
	\label{lem:true_actives}
	Assume that the mean difference vector 
	$\bmu_d = \bmu_2 - \bmu_1$ 
	is sparse. Let $\calS=\lbrace j: \mu_{dj}\neq 0 \rbrace$
	be the the corresponding active set 
	with the cardinality $s=\vert \calS \vert$, and define 
	$d_{0,n}
	=
	\min_{j\in \calS}
	\vert \mu_{dj} \vert$. 
	Under Conditions (C1) and (C2) in Appendix \ref{app:lemma}, 
	if $\tau_n=O(\sqrt{n_2}d_{0,n})$, $\log s =o(n_2d_{0,n}^2)$,  
	$\log (p-s)=o(\tau_n ^2)$, $n_2=o(n_1)$, and $\sqrt{n_2}d_{0,n}\to \infty$, 
	as $n_1,n_2\to \infty$, then
	\begin{eqnarray*}
		&(a)  &
		\Pr\bigg(  
		\bigcap _{j \not\in \calS} \lbrace \vert t_j \vert \leq \tau_n \rbrace 
		\bigg)   
		\to 1; ~~
		(b)~~
		\Pr\bigg(  
		\bigcap _{j \in \calS} \lbrace \vert t_j  \vert > \tau_n \rbrace 
		\bigg)  
		\to 1. 
	\end{eqnarray*}
\end{lem}

In the above Lemma, if $d_{0,n}=d_0$, for some constant $d_0>0$, 
then $\tau_n=O(\sqrt{n_2})$ and 
$\log p=o(n_2)$. On the other hand, 
if $d_{0,n}\sim n_2^{-\gamma}\alpha_{n_2}$, for $0<\gamma<1$ 
and some $\alpha_{n_2}\to \infty$, such that 
$d_{0,n}$ declines to zero and $\sqrt{n_2}d_{0,n}\to \infty$, 
then we have $\tau_n=O(n_2^{1/2-\gamma}\alpha_{n_2})$ and 
$\log p=o(n_2^{1-2\gamma}\alpha_{n_2}^2)$. 
Therefore, in both cases the divergence rate of the dimension $p$ 
is smaller than that of the minority class size $n_2$, 
as opposed to the balanced case 
where $\log p = o(n)$, that is, a larger dimension $p$ allowance. 

\begin{thm}
	\label{thrm:DACH}
	Suppose that the conditions of Lemma \ref{lem:true_actives} 
	are satisfied. 
	\newline
	 Let $\kappa_n=\max \lbrace \Delta_p^{-1}\sqrt{s/n_2} 
	~, ~ 
	\sqrt{\log p/n_1} \rbrace$.   
	For any fixed $\calL$,  
	\begin{itemize}
		\item[(a)]	 
		the MCRs of Msplit-{\sc hr} are given by  
		\begin{equation*}
			\Pi_{k}^{\text{Msplit-{\sc hr}}}(\calD_n)=
			\Phi \bigg(-\frac{1}{2}\Delta_p 
			( 1+O_p(\kappa_n) )
			\bigg), ~~ k=1,2
		\end{equation*}
		
		\item[(b)]
		if $s\Delta_p^2=o(n_2)$ and $\Delta_p^2\sqrt{\log p/n_1}=o(1)$, the 
		Msplit-{\sc hr} is asymptotically-strong optimal.   
	\end{itemize}
\end{thm}

Note that the result of Theorem \ref{thrm:DACH} 
also holds for the {\sc hr}. Part (b) of the theorem 
implies that the growth rates of both the sparsity size $s$ 
and the discriminative power $\Delta_p$ are controlled 
by the minority class size $n_2$. 

\subsection{Msplit-HR under a general $\bSigma$}
\label{subsec:general_sigma}

When the dimension $p$ is large compared to the sample size $n$,
the sample covariance matrix in \eqref{eq:hat_Sigma} is ill-conditioned. 
To deal with the singularity issue, many existing methods in the literature 
involve a feature selection strategy. 
In what follows, we use a variable screening method 
\citep{fan2008sure, pan2016ultrahigh} to select a subset of features $x_j$'s 
that have the highest discriminative power. 

At the $\ell$-th data splitting stage of Msplit-{\sc hr}, 
we consider the mean difference estimators 
$\hbmu_{d,\ell}^{(1)} =  \hat{\bmu}_{2,\ell}^{(1)}-\hat{\bmu}_{1,\ell}^{(1)}$, 
which are computed based on the training sub-samples $\calD_{n,\ell}^{(1)}$, 
for $\ell=1,...,\calL$. For a given 
threshold parameter $\tau_n$, we select those features $x_j$ whose indices belong to the 
set $\calS^{(1)}_{n,\ell} =\lbrace 1 \le j \le p: \vert \hmu_{dj,\ell}^{(1)} \vert>\tau_n  \rbrace$, where 
$\hmu_{dj,\ell}^{(1)}$ is the $j$-th entry of $\hbmu_{d,\ell}^{(1)}$.  

For any $p$-dimensional feature vector $\xx^*$, we define 
the discriminant function 
\begin{eqnarray}
	\label{DACH0-general}
	\delta_0^{\text{Msplit-{\sc hr}}}(\xx^*;\hbtheta_n)
	=
	\frac{1}{\calL} \sum_{\ell=1}^{\calL} \tbmu_{d,\ell}^{\top} 
	\widetilde{\bSigma}^{-1}_{n,\ell}~ (\xx_{\ell}^*-\tbmu_{a,\ell}),
\end{eqnarray}
where $\hbtheta_n$ is the vector of corresponding parameter estimates, and
$\xx_{\ell}^*=(x_j^*: j\in \calS^{(1)}_{n,\ell} )^{\top}$ are sub-vectors of 
the full feature vector $\xx^*$. Furthermore, 
for all $\ell= 1, \ldots, \calL$, we have   
$\tilde{\bmu}_{d,\ell} = \tilde{\bmu}_{2,\ell}-\tilde{\bmu}_{1,\ell}$, 
$\tilde{\bmu}_{a,\ell} = (\tilde{\bmu}_{1,\ell}+\tilde{\bmu}_{2,\ell})/2$, 
such that   
$
\tilde{\bmu}_{k,\ell}
=
(\hmu_{jk,\ell}^{(2)}: j\in \calS^{(1)}_{n,\ell} )^{\top}
$ 
for $k=1,2$, 
and 
$
\widetilde{\bSigma}_{n,\ell}
=
[ {\hsigma_{jj',\ell}^{(2)}}: j,j'\in \calS^{(1)}_{n,\ell} ]
$ 
are respectively the sub-vectors and sub-matrices of the sample means 
and covariance matrix given in \eqref{eq:hat_mu} and \eqref{eq:hat_Sigma}.
Note that for the existence of 
$\widetilde{\bSigma}^{-1}_{n,\ell}$, 
for all $\ell=1, 2, \ldots, \calL$, 
we include 
at most $(n'-2)$ features in each $\calS_{n,\ell}^{(1)}$.

As discussed in Subsection \ref{subsec:indep}, 
the data splitting technique facilitates  
computation of the bias $B_n$ \eqref{bias-HLDA} corresponding to 
\eqref{DACH0-general}. 
\begin{prn}
	\label{prop:gap_fill_general}
	If $\vert \calS_{n,\ell}^{(1)}\vert <n'-3$, 
	for all $\ell=1,..,\calL$, then 
	\begin{equation*}
		B_{0,n}^{\text{Msplit-{\sc hr}}} 
		=
		\mathbb{E} \lbrace 
		\Psi_{0,1}^{\text{Msplit-{\sc hr}}}(\hbtheta_{n})
		-
		\Psi_{0,2}^{\text{Msplit-{\sc hr}}} (\hbtheta_n)
		\rbrace
		=
		\frac{1}{\calL}\sum_{\ell=1}^{\calL} \mathbb{E}\{ \bar{r}_{n,\ell}\},
	\end{equation*}
	where 
	\begin{equation}
		\label{rbar}
		\bar{r}_{n,\ell}
		=
		\bigg ( \frac{1}{n'_1}-\frac{1}{n'_2} \bigg)
		\frac{n'-2}
		{n'-3- \vert \calS_{n,\ell}^{(1)} \vert } \times 
		\vert \calS_{n,\ell}^{(1)} \vert .
	\end{equation}
\end{prn}

\noindent	
Finally, our bias-corrected discriminant function is 
\begin{eqnarray}
	\label{DACH-general}
	\delta^{\text{Msplit-{\sc hr}}}(\xx^*;\hbtheta_n)
	=
	\frac{1}{\calL}\sum_{\ell=1}^{\calL} 
	\lbrace
	\tbmu_{d,\ell}^{\top} \tbSigma_{n,\ell}^{-1} 
	(\xx_{\ell}^*-\tbmu_{a,\ell})
	-\frac{\bar{r}_{n,\ell}}{2} 
	\rbrace
\end{eqnarray}
which has its bias $B_n^{\text{Msplit-{\sc hr}}} =0$. 
The term $\bar{r}_{n,\ell}$ as a function 
of $(n'_2-n'_1)$ makes the corrected 
discriminant function \eqref{DACH-general} more positive 
compared to the rule in \eqref{DACH0-general}. This increases 
the probability of classifying a new observation to the minority class, 
and hence improving the results for this class. 
Algorithm \ref{algo:DAC_corr} below summarize the steps for computing 
in \eqref{DACH-general}. 

Figure \ref{fig:effect_T_2} shows the 
class-specific MCRs of \eqref{DACH-general} as a function of $\calL$, 
corresponding to scenario \textbf{(iv)} in our simulations in 
Section \ref{subsec:corr_sim}. Based on these results, 
we used $\calL=30$ in our numerical experiments. 

\begin{algorithm}
	\caption{Computing the discriminant function in \eqref{DACH-general}}
	\begin{algorithmic}[1]\label{algo:DAC_corr}
		\REQUIRE Input $n'_1=\lfloor n_1/2 \rfloor, n'_2=\lfloor n_2/2 \rfloor, 
		\xx^*, \calL$, and $\tau_n$.
		\FOR{$\ell=1,\dots , \calL$}
		\STATE Split $\calD_n$ into $\calD_{n,\ell}^{(1)}$ and $\calD_{n,\ell}^{(2)}$
		\STATE Using $\calD_{n,\ell}^{(1)}$, obtain 
		$\calS^{(1)}_{n,\ell} =\lbrace 1 \le j \le p: \vert\hmu_{dj,\ell}^{(1)} \vert>\tau_n  \rbrace$ and compute $\bar{r}_{n,\ell}$ in \eqref{rbar}
		
		\IF{$\vert \calS_{n,\ell}^{(1)}\vert <n'_1+n'_2-3$}
		\STATE Using $\calS^{(1)}_{n,\ell}$ and $\calD_{n,\ell}^{(2)}$, 
		compute $\tbmu_{d,\ell}^{\top} \tbSigma_{n,\ell}^{-1} (\xx_{\ell}^*-\tbmu_{a,\ell})$  
		\ELSE 
		\STATE Step 1: Select the first $(n'_1+n'_2-4)$ features in 
		$\calS_{n,\ell}^{(1)}$ with highest value of $\vert\hmu_{dj,\ell}^{(1)} \vert$  
		\STATE Step 2: Using $\calD_{n,\ell}^{(2)}$ and 
		the selected features in Step 1, 
		compute $\tbmu_{d,\ell}^{\top} \tbSigma_{n,\ell}^{-1} (\xx_{\ell}^*-\tbmu_{a,\ell})$  
		\ENDIF
		\ENDFOR
		\RETURN $\delta^{\text{Msplit-HR}}(\xx^*;\hbtheta_n)
		=
		\frac{1}{\calL}\sum_{\ell=1}^{\calL}
		\lbrace
		\tbmu_{d,\ell}^{\top} \tbSigma_{n,\ell}^{-1} 
		(\xx_{\ell}^*-\tbmu_{a,\ell})
		-\frac{\bar{r}_{n,\ell}}{2}
		\rbrace$.
	\end{algorithmic}
\end{algorithm}


\begin{figure}
	\centering
	\begin{subfigure}[b]{0.4\textwidth}
		\includegraphics[width=65mm,height=70mm]{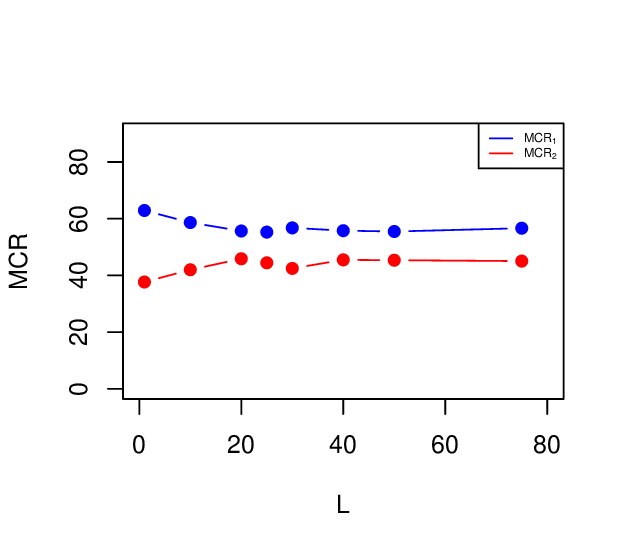}
		\caption{$n_1=50$, $n_2=10$}
		\label{fig:effect_T_general_50}
	\end{subfigure}%
	\qquad \qquad 
	\begin{subfigure}[b]{0.4\textwidth}
		\includegraphics[width=65mm,height=70mm]{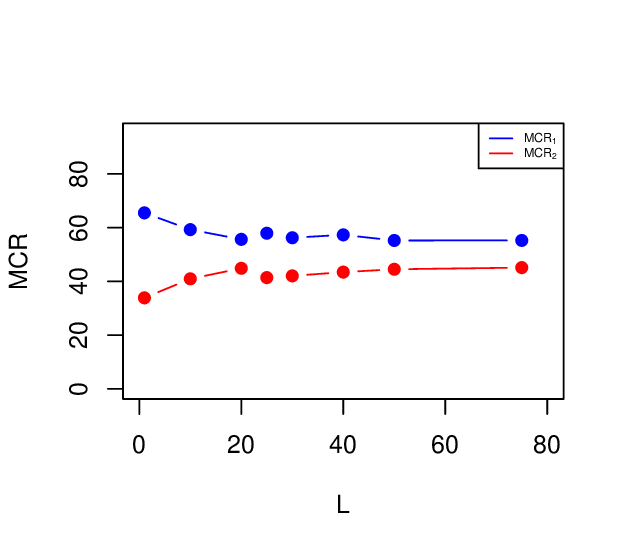}
		\caption{$n_1=100$, $n_2=10$}
		\label{fig:effect_T_general_100}
	\end{subfigure}
	\caption{
		Effect of the number of sample-splits $\calL$ on Msplit-{\sc hr} 
	performance for the Simulation setting \textbf{(iv)} and $p=500$. 
	}
	\label{fig:effect_T_2}
\end{figure}

The following lemma shows that 
the variable screening method used to obtain the selection  
sets ${\cal S}^{(1)}_{n,t}$ have a so-called
strong screening consistency property, 
as discussed in \cite{pan2016ultrahigh}.
We then establish the asymptotic optimality of  
$\delta^{\text{Msplit-{\sc hr}}}$ in Theorem \ref{thrm:geneal_optimal}.

\begin{lem}
	\label{lem:general_DACH}
	Let $\bbeta=\bSigma^{-1}\bmu_d$, and define the active set 
	$\calS=\lbrace 1 \le j \le p: \beta_j\neq 0  \rbrace$ with its cardinality denoted by 
	$|\calS|$. Furthermore, let $d_{0,n}=\min _{j\in \calS}\vert \mu_{dj}\vert$ 
	and $m_{max}=c_1 ( \max_{j\in \calS}\beta_j^2)|\calS|/d_{0,n}^2$, 
	for some constant $c_1>0$ such that $m_{max}\geq |\calS|$. 
	Under Condition 
	(C2) in 
	Appendix \ref{app:lemma}, if $\tau_n \asymp d_{0,n}$, 
	$\log p=o(n_2d_{0,n}^2)$, $n_2= o(n_1)$, and $\sqrt{n_2}d_{0,n}\to \infty$, 
	as $n_1,n_2\to \infty$, for any $\ell=1,...,\calL$, we have that 
	\[
	\text{(a)}~\Pr\bigg(\calS^{(1)}_{n,\ell} \supset \calS\bigg)\to 1~~;~~
	\text{(b)}~\Pr\bigg(\vert \calS^{(1)}_{n,\ell} \vert \leq m_{max}\bigg)\to 1.
	\]
\end{lem}
Part \text{\it (a)} 
implies that 
that for large sample sizes $n$, with probability tending to one, 
all the active features will be included in the selection 
sets $\calS_{n,\ell}^{(1)}$, for each $\ell=1, 2, \ldots, \calL$. 
Part \text{\it (b)} shows that the size of each set $\calS_{n,\ell}^{(1)}$
is of order $m_{max}$. 
These properties are obtained 
under the conditions that  
the divergence rate of the dimension $p$ is 
lower than that of the minority class size $n_2$. 
\begin{thm}
	\label{thrm:geneal_optimal} 
	Suppose that the conditions of Lemma \ref{lem:general_DACH} 
	are satisfied.  
	\newline
	Let $\kappa'_n= \max \lbrace \Delta_p^{-1}\sqrt{m_{max}/n_2}  
	~,~
	m_{max}\sqrt{\log p/n_1}  \rbrace$.	
	If $m_{max}\sqrt{\log p/n_1}=o(1)$, then for any fixed $\calL$, 
	\begin{itemize}
		\item[(a)]
		the MCRs of Msplit-{\sc hr} 
		are given by 
		\begin{equation*}
			\Pi_{k}^{\text{Msplit-{\sc hr}}}(\calD_n)=
			\Phi\bigg(-\frac{1}{2}\Delta_p(1+O_p(\kappa'_n)) \bigg), ~~ k=1,2
		\end{equation*}
		\item[(b)] 
		if 
		$\Delta_p^2m_{max}=o(n_2)$
		and 
		$\Delta_p^2m_{max}\sqrt{\log p /n_1}=o(1)$, 
		then the Msplit-{\sc hr} is asymptotically-strong optimal.
	\end{itemize}
\end{thm}
Condition $\Delta_p^2m_{max}=o(n_2)$ in the 
above theorem implies that the 
maximum size of the selection sets $\calS_{n,\ell}$, 
that is $m_{max}$, 
is affected by the minority class size $n_2$. 
Note that the results of the theorem also holds 
for the pairwise sure independence screening of \cite{pan2016ultrahigh} 
in the imbalanced binary cases, as well as in the balanced 
cases which was not studied before.  

\section{Two existing high-dimensional variants of LDA}
\label{sec:variants}

In this section, we investigate conditions under which 
two well-known 
sparse variants of the LDA 
obtain certain optimality properties under the imbalanced setting.  

\subsection{Sparse LDA {\sc (slda)}}
\label{subsec:SLDA}

This method, proposed by \cite{shao2011sparse}, 
uses thresholding-type estimators for both 
the mean-difference vector $\bmu_d = \bmu_2 - \bmu_1$ 
and 
$\bSigma$. 
In {\sc slda}, a new feature vector $\xx^*$ is allocated to Class 1 if and only if 
\begin{eqnarray*}
	\delta^{\text{\sc slda}}(\xx^*; \hat{\btheta}_n)
	=
	\tilde{\bmu}_d^{\top}\widetilde{\bSigma}_n^{-1}(\xx^*-\hat{\bmu}_a)< 0,
\end{eqnarray*}
where 
$\hbmu_a=(\hbmu_1+\hbmu_2)/2$, and       
$(\widetilde{\bSigma}_n, \tilde{\bmu}_d) $ are 
thresholded estimates of $\bSigma$ and $\bmu_d$, respectively, 
with the entries, 
\begin{eqnarray*}
	\tilde{\sigma}_{ij} & = & (1-2/n)~ \hsigma_{ij}~\textbf{1}\lbrace (1-2/n)\vert\hsigma_{ij} \vert >t_n \rbrace,
	~i, j= 1, \ldots, p\\ \\
	\tilde{\mu}_{dj} & = & 
	\hmu _{dj}~\textbf{1} \lbrace\vert \hmu _{dj}\vert >a_n\rbrace,~j= 1, \ldots, p,
\end{eqnarray*}
where $\hsigma_{ij}$ is $(i,j)$-th element of $\widehat{\bSigma}_n$ in \eqref{eq:hat_Sigma}, 
and $\hmu_{dj}$ is the $j$-th entry of $\hbmu_d$ in \eqref{eq:hat_mu}.
Further, $t_n=M_1\sqrt{\log p/n}$ with $M_1>0$, and 
$a_n=M_2(\log p/n)^{\alpha}$, $0<\alpha<1/2, M_2>0$.

\cite{shao2011sparse} derived conditions under which 
the {\sc slda} is 
optimal according to their Definition 
\ref{defshao}, when $p/n \to \infty$ and $n_1/n\rightarrow \pi$ 
with $0< \pi <1$, as $n\rightarrow \infty$. 
It turns out that their conditions do not yield an optimal {\sc slda} 
in the imbalanced case.   
In Theorem \ref{thrm:slda} below, we investigate conditions under 
which the {\sc slda} is asymptotically-strong optimal 
under the imbalanced case.
We then discuss and compare these conditions with those 
of \cite{shao2011sparse} under the balanced case.

First, for ease of comparison, we recall some notations 
introduced in \cite{shao2011sparse}. 
Let $\hat{q}_n$ be the number of features 
for which the value $\vert\hmu _{dj}\vert$ is greater than $a_n$. Further, let 
$q_{n0}$ and $q_n$ be the number of features for which the 
value of $\vert \mu_{dj}\vert$ is greater than $ra_n$ and 
$a_n/r$, respectively, for some fixed constant $r>1$. Also let    
$D_{g,p}=\sum_{j=1}^p\mu_{dj}^{2g}$, $0\leq g < 1$, and 
$C_{h,p}=\max _{1\leq i\leq p}\sum_{j=1}^p \vert \sigma_{ij} \vert ^{h}$, 
$ 0\leq h < 1$, 
be the sparsity measures corresponding to 
$\bmu_d$ and $\bSigma$, respectively. Here, $0^0$ is defined to be $0$. Furthermore, let   
$d_{n_1}=C_{h,p}({n_1}^{-1}\log p)^{(1-h)/2}$, and 
\[
b_{n_1}=\Delta_p^{-1}\max 
\left\lbrace  
\Delta_pd_{n_1}, ~
\sqrt{a_n^{2(1-g)}D_{g,p}} , ~
\sqrt{q_n/n_2} ,~
\sqrt{C_{h,p}q_n/n_1} 
\right\rbrace, 
\]
\[
b_{n_2}=\Delta_p^{-1}\max 
\left\lbrace  
\Delta_pd_{n_1}, ~
\sqrt{a_n^{2(1-g)}D_{g,p}} , ~
\sqrt{C_{h,p}q_n/n_2}
\right\rbrace, 
\]
where $\Delta_p^2=\bmu_d^{\top}\bSigma^{-1}\bmu_d$.
Note that under the imbalanced setting $n_2 = o(n_1)$, we have 
$d_{n_1} \sim d_n$, where $d_n=C_{h,p}(n^{-1}\log p)^{(1-h)/2}$. 

The following Lemma shows that the 
set $\{ 1 \le j \le p: \vert \hmu _{dj} \vert  > a_n \}$
has indeed a sure screening property, which is essential 
in Theorem \ref{thrm:slda} 
for the assessment of {\sc slda}. 
\begin{lem}\label{lem:shao}
	Suppose that,  
	\begin{eqnarray}\label{cond_lem2_shao}
		(\log p) \left( n_1/\log p\right) ^{2\alpha}=o(n_2), 
	\end{eqnarray}
	and $n_2=o(n_1)$, then as $n_1,n_2 \to \infty$, 
	
	\noindent
	\(
	(a) \ \Pr \bigg( \bigcap _{ j: \vert \mu_{dj} \vert > ra_n }
	\left\lbrace \vert \hmu _{dj} \vert > a_n \right\rbrace 
	\bigg) \to 1, \\    
	(b) \  \Pr \bigg( \bigcap _{ j: \vert \mu_{dj} \vert \leq a_n/r} 
	\left\lbrace \vert \hmu _{dj} \vert \leq a_n \right\rbrace
	\bigg)  \to 1,\\
	(c) \ \Pr \bigg( q_{n0} \leq \hat{q}_n \leq q_{n} \bigg)  \to 1.
	\)
\end{lem}
Condition \eqref{cond_lem2_shao} replaces 
the condition $\log p/n =o(1)$ in \cite{shao2011sparse}. 
One implication of 
\eqref{cond_lem2_shao} is $\log p/n_2=o(1)$, 
which shows the impact of the minority class size 
$n_2$ on the dimension allowance $p$.

\begin{thm}
	\label{thrm:slda}
	Suppose that the conditions of Lemma \ref{lem:shao}, 
	and Conditions (C2) and (C3) in Appendix \ref{app:lemma} are satisfied.  
	Then, as $n_1, n_2 \to \infty$, 
	\begin{itemize}
		\item[(a)]
		the MCRs 
		of {\sc slda} are given by 
		\begin{eqnarray*}
			\Pi_k^{\text{\sc slda}}(\calD_n)
			=
			\Phi \bigg(
			-\frac{1}{2}\Delta_p 
			\left\lbrace 1+O_p(b_{n_k}) \right\rbrace 
			\bigg),~~ k= 1, 2.  
		\end{eqnarray*}
		
		\item[(b)]
		the {\sc slda} is asymptotically-strong optimal if
		
		i. $\Delta_p^2$ is bounded, and $b_{n_2}=o(1)$, or 
		
		ii. $\Delta_p^2 \rightarrow \infty$, such that $\Delta_p^2b_{n_2}=o(1)$ holds.
	\end{itemize}
\end{thm}

The difference between the above theorem and Theorem 
3 of \cite{shao2011sparse} 
appears in $b_{n_2}$. 
To simplify the comparison in this case as in \cite{shao2011sparse}, 
suppose that $\bSigma$ is a diagonal matrix ($C_{0,p}=1$), and let $s$ 
be the number of nonzero (active) entries of the mean difference vector $\bmu_d$. If 
there are two constant $c_1,c_2>0$, such that $c_1\leq \vert \mu_{dj} \vert \leq c_2$,  
for the active $j$'s, then we have $q_n= s$. This implies that, by 
the Conditions (C2) and (C3),  
$\Delta_p^2$ and $D_{0,p}$ are of order $s$. Now, in this case, 
if $s \to \infty$, according to Theorem \ref{thrm:slda}-(b)-ii above, 
under condition \eqref{cond_lem2_shao}, $\Delta_p^2b_{n_2}=o(1)$ 
is equivalent to $s=o((n_1/\log p)^{\alpha})$. 
This implies that under the imbalanced setting, the growth rate of the sparsity factor  
$s$ is smaller than 
$\sqrt{n_2}$ and consequently is 
smaller than the growth rate of $s$ in the balanced setting. Therefore, 
due to the data scarcity in the minority class ($n_2$) in the imbalanced setting, 
in order for the {\sc slda} to be asymptotically-strong optimal 
more restrictive conditions are required on both the dimension $p$ 
and the sparsity size $s$ compared to the balanced case.

Next, we compare the optimality conditions of Msplit-{\sc hr} 
and {\sc slda}. The relation between these conditions for 
a general $\bSigma$ is not straightforward, and thus to get some insight 
we consider a diagonal case. 
Suppose that $\bSigma$ is diagonal ($C_{0,p}=1$), and  
$g=0$ such that $D_{0,p}= s =\vert \calS \vert$, where 
$\calS=\{1 \le j \le p: ~ \mu_{dj}\neq 0 \}$.  
By condition \eqref{cond_lem2_shao}, 
we have $\log p=o(n_2)$ which implies   
the necessary conditions of Lemma \ref{lem:true_actives} on ($s,p$), if 
$d_{0,n}=\min_{j\in \calS} \vert \mu_{dj} \vert=d_0>0$ and  
$\tau_n=M\sqrt{n_2}$, for some constant $M>0$. 
On the other hand, if $d_{0,n}$ decays, the same conclusion  
holds when $a_n=O(d_{0,n})$ and $\tau_n=M\sqrt{n_2}d_{0,n}$.  
Furthermore,  by \eqref{cond_lem2_shao} the conditions of 
Theorem \ref{thrm:slda}-(b) are equivalent to 
$s\Delta_p^2(\log p/n_1)^{2\alpha}=o(1)$ implying 
$s\Delta_p^2=o(n_2)$ which is required for 
the optimality of Msplit-{\sc hr}.  Therefore, 
the conditions of Theorem \ref{thrm:slda} for {\sc slda} 
on the dimension $p$ and the sparsity size 
$s$ are more restrictive than those in Theorem \ref{thrm:DACH}
for Msplit-{\sc hr}. 
In terms of feature selection,  
Lemma \ref{lem:general_DACH}-(b) 
provides an upper bound $m_{max}=o(n_2\Delta_p^{-2})$ on  
the size of the set of selected features by Msplit-{\sc hr},  
whereas the {\sc slda} allows the number of
nonzero estimators of $\mu_{dj}$'s or $\sigma_{l,j}$'s   
to be much larger than the class sizes to ensure 
optimality of the classifier, see \cite{shao2011sparse}. 
Therefore, the number of selected features by 
{\sc slda} could be potentially larger than the class sizes 
which we have also observed in 
our numerical study in Section \ref{sec:sim}. 

\subsection{Regularized optimal Affine discriminant {\sc (road)}}
\label{subsec:ROAD}
This method, proposed by \cite{fan2012road}, is 
constructed based on a 
sparse estimate of $\ww = \bSigma^{-1} \bmu_d$, 
unlike the {\sc slda} which uses sparse estimates 
of $\bmu_d$ and $\bSigma$, separately. The {\sc road} 
assigns $\xx^*$ to Class 1 if and only if 
\begin{eqnarray}\label{ROADF}
	\delta^{\text{\sc road}}(\xx^* ; \hbtheta_{n},c)
	=
	\hat{\ww}_c^{\top}(\xx^*-\hbmu_a) < 0,
\end{eqnarray}
where $\hbtheta_{n} = (\hbmu_1,\hbmu_2,\widehat{\bSigma}_n)$, 
$\hbmu_a = (\hbmu_1+\hbmu_2)/2$, 
and 
\begin{eqnarray}
	\label{ROAD}
	\hat{\ww}_c
	\in
	arg \min _{\Vert \ww \Vert _1\leq c, \ \ww^{\top}\hbmu_d=1}
	\ww^{\top}\widehat{\bSigma}_n\ww
\end{eqnarray} 
with $\hbmu_d=\hbmu_2-\hbmu_1$, and $(\hbmu_k, \widehat{\bSigma}_n)$ 
are the estimates in \eqref{eq:hat_mu}-\eqref{eq:hat_Sigma}. 
Note that in (\ref{ROAD}) the smaller the $c$, 
the sparser the solution $\hat{\ww}_c$, and as 
$c \to \infty$ the solution is equivalent to the regular 
weight $\ww_c\propto \bSigma^{-1}\bmu_d$. 
\cite{fan2012road} studied the asymptotic difference 
between the average MCR of the {\sc road} and its oracle version for which 
the true values of $(\bmu_1, \bmu_2, \bSigma)$ are used in (\ref{ROAD}). 
However, as discussed in Section \ref{sec:impact2}, 
under the imbalanced setting the average 
MCR is not an appropriate performance measure for a classifier. Therefore, 
in the following theorem, 
we study the class-wise MCRs of 
the {\sc road}.  
\begin{thm}
	\label{thrm:ROAD}
	Let $s_c=\Vert \ww_c\Vert _0$, $s_c^{(1)}=\Vert\ww_c^{(1)} \Vert_0$ 
	and $\hat{s}_c=\Vert \hat{\ww}_c \Vert_0$,  
	where $\ww_c$, $\ww_c^{(1)}$, and 
	$\hat{\ww}_c$ are respectively the solutions of 
	\eqref{ROAD} when $(\bmu_d, \bSigma)$, $(\hbmu_d, \bSigma)$ 
	and $(\hbmu_d, \widehat{\bSigma}_n)$ are used. 
	Furthermore, let $\Pi_k^{\text{\sc road}}(\calD_{n}; c)$
	be the MCR of Class $k=1,2$, associated with {\sc road}, and 
	$\Pi_k^{\text{orc}}(c)$ denotes its oracle value. 
	Under Condition (C2) in Appendix \ref{app:lemma},   
	if  $n_2=o(n_1)$  and $\log p = o(n_2)$,  
	then as $n_1,n_2\to \infty$,  
	\begin{equation}\label{eq:opt_road}
		\Pi_k^{\text{\sc road}}(\calD_{n}; c)-\Pi_k^{\text{orc}}( c) = O_p(e_{n}), \ 
		~	k=1,2,
	\end{equation}
	where $e_{n}=\max\bigg\lbrace c^2 (\log p)/n_1 ~,~  
	\sqrt{(\log p)/n_2}\times
	\sqrt{\max \lbrace s_c, \ s_c^{(1)}, \ \hat{s}_{c} \rbrace }	
	\bigg\rbrace$.
\end{thm}

By Theorem \ref{thrm:ROAD}, a necessary condition for 
convergency of the MCRs of {\sc road} to their oracle values 
is that the sparsity size $s_c$ of the vector $\ww_c$ and 
the dimension $p$ are controlled by the minority class size $n_2$ 
(similar to the {\sc slda}), which in turn shows the effect of imbalanced class sizes on 
the performance of {\sc road}. 

In general, the conditions of Theorem \ref{thrm:ROAD} 
do not guarantee the optimality of {\sc road} 
according to Definition \ref{defshao}. 
\cite{fan2012road} showed that when the 
penalty parameter $c$ is chosen as  
$c\geq \Delta_p^{-2}\parallel \bSigma^{-1}\bmu_d 
\parallel_1 $, then $\ww_c\propto \bSigma^{-1}\bmu_d$ and 
the oracle MCRs $\Pi_k^{orc}(c)$ reduce to those of the 
optimal rule in \eqref{mcrk}. Hence, by Definition \ref{defshao}, 
for such $c$'s, Theorem \ref{thrm:ROAD} shows that {\sc road} is 
asymptotically-strong sub-optimal as long as $e_n\to 0$. 
Furthermore, {\sc road} becomes asymptotically-strong optimal  
if $\Delta_p$ is bounded. 
The condition $e_n\to 0$ 
is the same as $\log p=o(n_1/c^2)$ and 
$\log p=o(n_2/s_{max})$, 
where $s_{max}=\max \lbrace s_c, \ s_c^{(1)}, \ \hat{s}_{c} \rbrace$.
Note that, the larger the $c$, the larger the quantities 
$s_c$, $\hat{s}_c$ and $s_c^{(1)}$, 
and hence more restrictions on $(n_1,n_2,p)$  
compared to those in Theorem \ref{thrm:ROAD}, and  
the conditions of Msplit-{\sc hr}. 
In our numerical study, we observe that the performance of 
{\sc road} in terms of MCR$_2$ improves for lower dimensions.

\section{Simulation Study}
\label{sec:sim}
In this section, we assess the finite-sample performance 
of Msplit-{\sc hr} and several binary classification methods 
using simulations. We consider two settings of diagonal and general 
covariance matrix $\bSigma$ under the model 
$\XX|Y=k \sim N_p(\bmu_k,\bSigma)$, $k=1,2$.

\subsection{Diagonal $\bSigma$}
\label{subsec:indep_sim}
We compare 
the following methods:  
the bias adjusted independence ({\sc bai}) and 
leave-one-out independence rules ({\sc loui}) \cite{bak2016high}, 
diagonal {\sc road} method ({\sc droad}) \cite{fan2012road}, the 
bias corrected LDA ({\sc blda}) \cite{huang2010bias}, 
the {\sc hr} and its under-sampling version ({\sc us-hr}),  
and our proposed method Msplit-{\sc hr}. 
Note that the aforementioned methods 
use the knowledge of a diagonal 
$\bSigma$. 
In our comparison, we also include a bias-corrected support vector machines  
proposed by \citep{nakayama2017support} coupled with 
an under-sampling method ({\sc us-bcsvm}).   
In regards to over-sampling techniques such as the {\sc somte},  
\cite{bak2016high} and \cite{Blagus2013} showed that such  
techniques deduce larger differences between the MCRs in 
high-dimensional imbalanced problems. 
For example, we examined the performance of {\sc hr} and 
{\sc bcsvm} coupled with {\sc smote} (under both diagonal and general 
$\bSigma$) and since their performances were not satisfactory, 
we did not report the results here. 

We implemented the methods using R software.
The {\sc droad} results are based on 
the authors' MATLAB codes available on their website 
\footnote{https://github.com/statcodes/ROAD}.  
Our computations are carried out on a computer with an AMD Opteron(tm) 
Processor 6174 CPU 2.2GHz. 

The above methods involve certain tuning (threshold) parameters 
that need to be chosen using data-driven methods. We chose best threshold 
parameters in {\sc blda}, {\sc bai} and 
{\sc loui} by a grid search  
using the techniques 
outlined by the authors. 
As in  \cite{huang2010bias}, an F-statistic is used to select the important features 
in {\sc blda} method. In both {\sc hr} and Msplit-{\sc hr}, 
we choose the tuning parameter $\tau$ 
by minimizing MCR of the minority class based on a leave-one-out cross validation.

We consider the binary classification problem 
$\XX \vert (Y=k) \sim N_p(\bmu_k, \DD), k= 1, 2$, and 
$\DD=\text{diag}\{\sigma_1^2, ..., \sigma_p^2\}$. We generated 
training data with different class sizes $n_1$ and $n_2$, and 
test data sets of size $50$ in both classes. 
We considered two dimensions $p=1000, 3000$, and class-wise 
sample sizes  $(n_1,n_2)=(25,5)$, $(50,10)$, 
$(100,10)$ for the training data. 
The simulation results are based 
on $100$ randomly generated data sets, and  
the two parameter settings: 

\begin{itemize}
	\item[\textbf{(i)}] 
	$\bmu_1=(1, 1,   \textbf{0}_{p-2})^{\top}$,  
	$\bmu_2=(2, 2.2, \textbf{0}_{p-2})^{\top}$, 
	$\sigma_1^2=1.5^2 $,  $\sigma_2^2=0.75^2$, and  
	$\sigma_j^2=1$, for $j=3,...,p$.
	
	\item[\textbf{(ii)}] 
	$\bmu_1=(\textbf{1}_9 ,   \textbf{0}_{p-9})^{\top}$,  
	$\bmu_2=(2*\textbf{1}_4, 2.5*\textbf{1}_3, 3*\textbf{1}_2, \textbf{0}_{p-9})^{\top}$,   
	$\sigma_j^2=10$, for $j=1,...,4$,  
	$\sigma_j^2=2.25^2 $, for $j=5,6,7$, 
	$\sigma_j^2=1.5^2$, $j=8,9$, and  
	$\sigma_j^2=1$, for $j=10,...,p$.
\end{itemize}
The number ($s$) of active features 
$x_j$'s that distinguish the two classes, and also the value 
of $\Delta_p$  in the two settings are respectively $s= 2, \Delta_p^2=3$
and $s= 9, \Delta_p^2=8.7$. 
Since the signal strength is measured by 
$\Delta_p$, setting {\bf (i)} has a weaker signal than {\bf (ii)}. Under these settings, 
the value of the optimal MCR, 
$\Pi^{\text{opt}}$ in \eqref{mcrk}, 
are respectively $19.32\%$ and $7\%$.
Also, the active features have different marginal signal values 
$\vert\mu_{dj}\vert/\sigma_j$, in each of the settings. 

The performance measures used to compare different methods are: 
per-class misclassification rates (MCR$_1$, MCR$_2$), and the 
geometric mean ($GM$) of the MCRs. 
The results reported in the tables are average and standard 
deviations (in parentheses) of the 
measures over $100$ generated samples. 
We also reported median number of true selected features, denoted by 
$A$, and falsely selected features denoted by 
$N$, respectively.  For the new method Msplit-{\sc hr}, 
similar to the stability selection technique of \cite{meinshausen2010stability},
the selected features for each simulated sample are those with a relative 
frequency more than $50\%$, that is the set $\mathcal{S}_n=\{ j: \frac{f_j}{\calL}\geq 0.5 \}$, 
where $f_j$ is selection frequency of $j$-th feature among $\calL$ splits.  

\subsubsection{Discussion of the results}

The 
results for the cases $(n_1,n_2, p)=$$(25,5,1000)$, $(50,10,1000)$ and $(100,10,1000)$ 
are given in Table \ref{tab:1000-50&100-10}. 
The results corresponding to dimension $p=3000$ 
are given in Table \ref{tab:3000-50&100-10}.

From Table \ref{tab:1000-50&100-10}, under both settings {\bf (i)} and {\bf (ii)},  
we can see that {\sc droad}, {\sc hr}, and {\sc blda} have smaller error rates 
in the majority class (MCR$_1$) compared to the other methods, 
but the differences between their MCR$_1$ and MCR$_2$ are larger. 
The class-wise error rates corresponding to  
{\sc us-hr} and {\sc us-bcsvm}  
have smaller differences than those of {\sc droad, hr}, and {\sc blda}. 
Furthermore, the {\sc us-hr} 
outperforms {\sc us-bcsvm}, {\sc droad, hr}, 
and {\sc blda} in terms of MCR$_2$. 
Under setting \textbf{(i)}, Msplit-{\sc hr} outperforms all the other 
methods in terms of MCR$_2$; for example, its MCR$_2$ is better 
than the next best method {\sc loui} up to about $8\%$, 
depending on class sizes $(n_1, n_2)$ and dimension $p$, 
while having balanced results for both classes.  
In setting \textbf{(ii)}, Msplit-{\sc hr} behaves similarly 
to {\sc loui} and {\sc bai}, 
with its MCR$_2$ better than {\sc loui} and {\sc bai} respectively 
up to about $3\%$ and $7\%$. 
Note that in {\bf (i)}, we have a weaker signal strength ($\Delta_p^2$) 
and fewer number of active features ($s$) than {\bf (ii)}, 
which matches the conditions of Theorem \ref{thrm:DACH} 
for Msplit-{\sc hr} on controlling the size of $s\Delta_p^2$. In other words, 
we can see that the weaker the signal, the better the performance 
of Msplit-{\sc hr} in terms of MCRs in both classes. 
On the other hand, from the columns $A$ and $N$ of 
Table \ref{tab:1000-50&100-10}, 
Msplit-{\sc hr} tends to select fewer number of inactive (noise) 
features compared to the two its competitors  
{\sc bai} and {\sc loui}. 
In {\sc bcsvm}, the bias caused by dimension is corrected 
by using all features in the model and therefore 
this method does not perform any feature selection. 

\begin{table*}
	\caption{Classification results for the simulation settings {\bf (i)}-{\bf(ii)} 
		with a diagonal $\bSigma$ and $p=1000$.}
	\label{tab:1000-50&100-10}
	\begin{center}
		\resizebox{8cm}{!}{
		\begin{tabular}{c|c|lrrrrr}
			\hline
			$(n_1,n_2)$ &
			Setting & 
			Methods & \multicolumn{1}{c}{$MCR_1\%$} & \multicolumn{1}{c}{$MCR_2\%$} & 
			\multicolumn{1}{c}{$GM\%$} 
			& \multicolumn{1}{c}{$A$} 
			& \multicolumn{1}{c}{$N$} \\
			\hline
			\multirow{8}{*}{(25,5)} &
			\multirow{8}{*}{\textbf{(i)}}
			&{\sc us-bcsvm}  &  48.96$_{(13.99)}$    &  47.46$_{(14.38)}$    &   46.33$_{(5.99)}$  & 2  & 998
			\\
			&&DROAD   & 2.62$_{(7.45)}$   &  93.14$_{(15.91)}$   &  5.45$_{(11.37)}$    
			&  2      
			& 364  \\
			&&HR       &    15.96$_{(13.02)}$   &     67.3$_{(26.43)}$    & 26.25$_{(15.45)}$       
			&   1     
			&  2 \\
			&&US-HR  &    44.34$_{(16.45)}$   &    45.14$_{(16.49)}$    &    42.73$_{(8.91)}$      
			&  1       
			&  143.5  \\
			&&BLDA   &     14.72$_{(11.05)}$    &   70.16$_{(22.47)}$    &  28.04$_{(11.75)}$   
			&  1    
			&   5   \\
			&&BAI   &     38.86$_{(15.16)}$    &   48.2$_{(17.23)}$    &  41.15$_{(10.06)}$   
			&   1   
			&  75.5    \\
			&&LOUI    &     41.46$_{(18.14)}$    &   43.9$_{(19.43)}$    &  39.05$_{(11.75)}$   
			&    1  
			&   20.5   \\
			&&Msplit-HR   &     42.66$_{(16.97)}$    &  40.04$_{(16.65)}$    &  39.12$_{(11.01)}$   
			&  1    
			&   2  \\
			\hline
			\multirow{8}{*}{(25,5)} &
			\multirow{8}{*}{\textbf{(ii)}}
			&{\sc us-bcsvm}  &  46.42$_{(14.04)}$    &  41.22$_{(12.27)}$    &   41.99$_{(5.75)}$  & 9  & 991
			\\
			&&DROAD    &  5.46$_{(8.63)}$    &  58.48$_{(30.27)}$   &   9.84$_{(10.24)}$    
			&      6
			&   17  \\
			&&HR   &  12.38$_{(10.52)}$    &  55.78$_{(27.55)}$   &   20.80$_{(12.40)}$    
			&      1
			&       2\\
			&&US-HR&  39.34$_{(14.75)}$    &  35.48$_{(15.43)}$   &   35.13$_{(9.11)}$    
			&      4
			&   124 \\
			&&BLDA   &  11.06$_{(8.74)}$    &  57.48$_{(26.60)}$   &   20.85$_{(11.30)}$    
			&      2
			&   3 \\
			&&BAI   &  30.72$_{(13.94)}$    &  35.06$_{(16.32)}$   &   30.53$_{(10.12)}$    
			&      3
			&   33 \\
			&&LOUI    &  29.86$_{(14.54)}$    &  31.24$_{(16.37)}$   &   27.95$_{(10.83)}$    
			& 3     
			&  36.5    \\
			&&Msplit-HR  &  32.2$_{(15.57)}$    &  28.22$_{(15.44)}$   &   27.61$_{(9.91)}$    
			&   1   
			&   3.5 \\
			\hline
			\hline
			\multirow{8}{*}{(50,10)} &
			\multirow{8}{*}{\textbf{(i)}}
			&{\sc us-bcsvm}  &  47.88$_{(10.19)}$    &  44.56$_{(10.78)}$    &   45.25$_{(5.71)}$  & 2  & 998
			\\
			&&DROAD   & 6.30$_{(9.10)}$   &  75.28$_{(30.25)}$   &  11.23$_{(11.61)}$    
			&   2     
			&  68 \\
			&&HR       &    19.36$_{(7.47)}$   &    40.82 $_{(20.99)}$    & 26.15$_{(7.68)}$       
			&    1     
			& 1 \\
			&&US-HR  &    34.22$_{(14.05)}$   &     32.66$_{(14.34)}$    &    32.12$_{(10.72)}$      
			&    1     
			&   0 \\
			&&BLDA   &     18.26$_{(8.82)}$    &   48.68$_{(21.24)}$    &  25.27$_{(8.81)}$   
			&  1    
			&   3   \\
			&&BAI   &    31.94$_{(14.39)}$       &   36.92$_{(15.19)}$    &    32.71$_{(10.52)}$       
			&   1        
			&   11\\
			&&LOUI    &     29.28$_{(12.21)}$    &    34.12$_{(17.04)}$     &   29.99$_{(10.39)}$     
			&    1    
			&  8.5      \\
			&&Msplit-HR   &     30.22$_{(12.66)}$    &    26.68$_{(13.42)}$     &   26.99$_{(9.75)}$  
			& 1    
			&  0 \\
			\hline
			\multirow{8}{*}{(50,10)} &
			\multirow{8}{*}{\textbf{(ii)}}
			&{\sc us-bcsvm}  &  41.72$_{(9.03)}$    &  38.02$_{(10.21)}$    &   38.93$_{(5.23)}$  & 9  & 991
			\\
			&&DROAD    &  5.60$_{(6.04)}$    &  30.72$_{(19.67)}$   &   9.39$_{(6.22)}$   
			& 7    
			&  17.5  \\
			&&HR     &  11.02$_{(7.04)}$    &  25.42$_{(15.69)}$   &  14.59$_{(6.84)}$    
			&   2      
			&  0     \\
			&&US-HR    &   22.84$_{(10.14)}$       &     19.04$_{(8.49)}$    &   19.74$_{(7.23)}$    
			&   1  
			&  0  \\
			&&BLDA    &     11.72$_{(6.70)}$    &   24.36$_{(15.71)}$    &  14.80$_{(6.13)}$   
			&  2    
			&  0   \\
			&&BAI       &    17.6$_{(8.76)}$       &   19.8$_{(11.79)}$    &    17.18$_{(8.07)}$     
			&  3       
			&   3   \\
			&&LOUI     &  16.72$_{(8.55)}$    &  19.16$_{(10.99)}$   &  16.55$_{(7.39)}$      
			&   3   
			&   3.5    \\
			&&Msplit-HR    &  19.22$_{(9.58)}$    &  17.82$_{(9.07)}$   &  17.08$_{(6.84)}$     
			&  2   
			&   0 \\
			\hline
			\hline
			\multirow{8}{*}{(100,10)} &
			\multirow{8}{*}{\textbf{(i)}}
			&{\sc us-bcsvm}  &  47.96$_{(10.08)}$    &  44.1$_{(10.50)}$    &   45.09$_{(5.42)}$  & 2  & 998
			\\
			&&DROAD    &  2.60$_{(5.25)}$       &   85.74$_{(22.67)}$   &  6.28$_{(9.67)}$   
			&  2      
			&  494   \\
			&&HR        &     19.96$_{(8.53)}$     &   34.82$_{(19.40)}$   &   24.31$_{(8.09)}$  
			&    1        
			&   0    \\
			&&US-HR   &    34.08$_{(13.17)}$  &     30.52$_{(13.07)}$    &  31.14$_{(9.29)}$     
			& 1      
			&  0 \\
			&&BLDA    &     16.84$_{(7.60)}$    &   45.48$_{(22.81)}$    &  25.08$_{(8.01)}$   
			&  1   
			&   2  \\
			&&BAI       &    28.86$_{(12.15)}$       &   33.64$_{(16.62)}$    &     29.72$_{(9.85)}$       
			&    1       
			&   7 \\
			&&LOUI     &     26.26$_{(11.03)}$   &   32.48$_{(16.76)}$   &   27.81$_{(9.26)}$   
			&    1      
			&   6    \\
			&&Msplit-HR    &     27.94$_{(12.09)}$   &   24.84$_{(13.11)}$   &    24.95$_{(8.93)}$     
			& 1   
			&   0   \\
			\hline
			\multirow{8}{*}{(100,10)} &
			\multirow{8}{*}{\textbf{(ii)}}
			&{\sc us-bcsvm}  &  41.66$_{(9.67)}$    &  37.38$_{(10.88)}$    &   38.51$_{(6.02)}$  & 9  & 991
			\\
			&&DROAD    &  3.22$_{(4.23)}$    &  37.96$_{(20.38)}$   &  6.57$_{(6.07)}$    
			&  8    
			&  31.5   \\
			&&HR    &  10.02$_{(6.20)}$    &  22.14$_{(12.49)}$   &  13.11$_{(5.77)}$      
			&  3     
			&    0   \\
			&&US-HR   &    20.64$_{(10.65)}$       &     18.98$_{(10.47)}$    &   18.30$_{(7.76)}$      
			&   1    
			&   0 \\
			&&BLDA    &     10.44$_{(6.26)}$    &   22.28$_{(14.29)}$    &   12.96$_{(5.83)}$   
			&  3   
			&   0  \\
			&&BAI &    16.44$_{(9.70)}$       &   17.08$_{(10.19)}$    &   15.49$_{(7.89)}$    
			&     3.5      
			&   2   \\
			&&LOUI     &  15.02$_{(8.37)}$    &  15.94$_{(9.32)}$   &  14.13$_{(6.42)}$      
			&  3      
			&   2    \\
			&&Msplit-HR    &  16.56$_{(8.98)}$    &  14.38$_{(7.88)}$   &   14.04$_{(5.36)}$    
			&  3    
			&  0   \\
			\hline
		\end{tabular}
	}
	\end{center}
\end{table*}

\begin{table*}
	\caption{Classification results for Simulation settings {\bf (i)}-{\bf(ii)} 
		with a diagonal $\bSigma$ and $p=3000$.}
	\label{tab:3000-50&100-10}
	\begin{center}
		\resizebox{8cm}{!}{%
		\begin{tabular}{c|c|lrrrrr}
			\hline
			$(n_1,n_2)$ &
			Setting & 
			Methods & \multicolumn{1}{c}{$MCR_1\%$} & \multicolumn{1}{c}{$MCR_2\%$} & 
			\multicolumn{1}{c}{$GM\%$} 
			& \multicolumn{1}{c}{$A$} 
			& \multicolumn{1}{c}{$N$} \\
			\hline
			\multirow{8}{*}{(25,5)} &
			\multirow{8}{*}{\textbf{(i)}}
			&{\sc us-bcsvm}  &  50.84$_{(13.78)}$    &  47.32$_{(13.75)}$    &   47.32$_{(5.03)}$  &  2 & 2998
			\\
			&&DROAD    &  3.14$_{(7.66)}$    &  94.40$_{(12.91)}$   &   6.71$_{(13.44)}$   
			&   1  
			&  529  \\
			&&HR     &  14.92$_{(12.68)}$    &  73.64$_{(24.84)}$   &   26.10$_{(16.85)}$   
			&   0  
			&   1.5 \\
			&&US-HR    &  46.38$_{(16.21)}$    &  46.86$_{(16.10)}$   &   44.25$_{(7.23)}$   
			&   1  
			&   255 \\
			&&BLDA    &  13.6$_{(11.63)}$    &  76.44$_{(23.64)}$   &   26.03$_{(14.53)}$   
			& 1    
			&  5  \\
			&&BAI   &  38.04$_{(15.86)}$    &  50.92$_{(17.61)}$   &   41.77$_{(9.48)}$   
			& 1    
			&  107.5  \\
			&&LOUI      &  41.02$_{(18.00)}$    &  47.04$_{(18.60)}$   &   41.16$_{(10.20)}$   
			&   1  
			&   42 \\
			&&Msplit-HR   &  43.06$_{(19.61)}$    &  44.08$_{(19.25)}$   &   40.09$_{(9.18)}$   
			&  1   
			&  3  \\		
			\hline
			\multirow{8}{*}{(25,5)} &
			\multirow{8}{*}{\textbf{(ii)}}
			&{\sc us-bcsvm}  &  47.12$_{(13.88)}$    &  45.32$_{(13.29)}$    &   44.37$_{(5.11)}$  &  9 & 2991
			\\
			&&DROAD  &  5.54$_{(8.59)}$    &  60.58$_{(28.94)}$   &   10.24$_{(11.52)}$    
			&      6
			&   16   \\
			&&HR    &  14.04$_{(12.40)}$    &  62.44$_{(27.65)}$   &   23.41$_{(14.78)}$    
			&     1 
			&    1 \\
			&&US-HR &  43.78$_{(15.90)}$    &  41.52$_{(15.33)}$   &   40.35$_{(8.22)}$    
			&      3
			&   251.5 \\
			&&BLDA  &  11.04$_{(10.41)}$    &  65.64$_{(27.20)}$   &   20.50$_{(13.70)}$    
			&    1  
			&   3 \\
			&&BAI   &  33.04$_{(15.35)}$    &  41.68$_{(17.60)}$   &   34.84$_{(10.61)}$    
			&      3
			&  79.5  \\
			&&LOUI     &  31.64$_{(16.28)}$    &  41.84$_{(18.91)}$   &   33.63$_{(10.90)}$    
			&      3
			&    78 \\
			&&Msplit-HR    &  37.78$_{(17.62)}$    &  36.32$_{(17.68)}$   &   34.18$_{(10.36)}$    
			&      1
			&    3 \\	
			\hline
			\hline
			\multirow{8}{*}{(50,10)} &
			\multirow{8}{*}{\textbf{(i)}}
			&{\sc us-bcsvm}  &  48.46$_{(11.56)}$    &  47.98$_{(11.55)}$    &   47.04$_{(5.29)}$  & 2  & 998
			\\
			&&DROAD  &  5.7$_{(9.27)}$     &    81.02 $_{(24.78)}$    &  11.01$_{(13.54)}$     
			&  2        
			& 77  \\
			&&HR   &  18.68$_{(8.29)}$    &  40.88 $_{(24.32)}$    &    24.61$_{(9.61)}$    
			&    1           
			&     0    \\
			&&US-HR  &    35.62$_{(13.25)}$       &     36.66$_{(14.57)}$    &   34.87$_{(10.01)}$     
			&   1    
			&  0  \\
			&&BLDA   &     17.16$_{(8.35)}$    &   48.18$_{(24.08)}$    &   25.74$_{(8.52)}$     
			&   1    
			&  2   \\
			&&BAI   &  32.08$_{(11.86)}$   &   37.42$_{(17.28)}$    &    33.32$_{(11.03)}$    
			&    1        
			&   12.5   \\
			&&LOUI   &   31.1$_{(12.33)}$    &   34.82$_{(17.41)}$    &   31.48$_{(11.18)}$    
			&    1     
			&  9.5    \\
			&&Msplit-HR   &  32.3$_{(12.81)}$    &   30.98$_{(16.05)}$    &    30.35$_{(11.34)}$     
			&  1    
			&   0     \\
			\hline
			\multirow{8}{*}{(50,10)} &
			\multirow{8}{*}{\textbf{(ii)}}
			&{\sc us-bcsvm}  &  44.08$_{(10.75)}$    &  44.16$_{(10.42)}$    &   43.04$_{(4.85)}$  & 9  & 991
			\\
			&&DROAD  &    5.12$_{(5.21)}$    &     32.40$_{(17.42)}$    &  9.24$_{(6.25)}$     
			&    1   
			&  25.50 \\
			&&HR        &  12.98$_{(8.14)}$    & 28.7 $_{(18.25)}$   &  17.14$_{(8.18)}$       
			& 2      
			&    0   \\
			&&US-HR   &   26.8$_{(12.13)}$       &     24.72$_{(12.17)}$    &   24.48$_{(8.81)}$    
			&   1     
			& 0   \\
			&&BLDA    &     12.7$_{(6.88)}$    &   29.1$_{(19.32)}$    &   16.70$_{(7.20)}$     
			&  2     
			&  1   \\
			&&BAI       &    20.24$_{(10.71)}$       &   22.44$_{(13.68)}$    &   19.55$_{(9.60)}$     
			&    3    
			&  4 \\
			&&LOUI     &  19.02$_{(10.22)}$    &  22.74$_{(13.48)}$   &  19.45$_{(9.35)}$    
			&   3   
			&    9   \\
			&&Msplit-HR    &  21.4$_{(11.07)}$    &  19.2$_{(10.14)}$   &  18.93$_{(7.47)}$      
			&  2      
			&   0 \\
			\hline
			\hline
			\multirow{8}{*}{(100,10)} &
			\multirow{8}{*}{\textbf{(i)}}
			&{\sc us-bcsvm}  &  48.3$_{(10.85)}$    &  48.58$_{(11.83)}$    &   47.32$_{(5.50)}$  & 2  & 998
			\\
			&&DROAD  &  1.80$_{(4.23)}$       &     88.66$_{(20.70)}$    &  4.49$_{(8.31)}$    
			&   2       
			& 861.50  \\
			&&HR  &     18.42$_{(8.24)}$    &   42.38$_{(24.65)}$    &   25.08$_{(9.02)}$     
			&  1     
			& 0.50     \\
			&&US-HR  &    36.94$_{(14.20)}$       &     37.1$_{(13.93)}$    &  35.91$_{(10.47)}$    
			&   1    
			&  0  \\
			&&BLDA   &     15.64$_{(8.56)}$    &   50.18$_{(26.11)}$    &   24.00$_{(9.47)}$     
			&   1     
			&  2   \\
			&&BAI  &    29.92$_{(10.65)}$  &   39.64$_{(16.46)}$    &   33.37$_{(10.74)}$     
			&    1        
			&  14.5    \\
			&&LOUI    &  27.04$_{(10.72)}$   &   36.04$_{(17.31)}$    &   29.95$_{(10.71)}$      
			&   1      
			&   7        \\
			&&Msplit-HR   &     31.46$_{(11.87)}$        &   29.1$_{(15.09)}$    &   29.14$_{(11.13)}$  
			&  1  
			&   0     \\
			\hline
			\multirow{8}{*}{(100,10)} &
			\multirow{8}{*}{\textbf{(ii)}}
			&{\sc us-bcsvm}  &  44.52$_{(10.96)}$    &  44.74$_{(10.91)}$    &   43.52$_{(5.09)}$  &  9  & 991
			\\
			&&DROAD  &    3.28$_{(4.47)}$       &     38.90$_{(20.50)}$    &   6.74$_{(6.38)}$   
			&     1   
			&  31.5 \\
			&&HR        &  10.18$_{(6.08)}$    &  27.96$_{(18.17)}$   &  14.07$_{(6.17)}$     
			&  2      
			&   0    \\
			&&US-HR   &    24.28$_{(11.97)}$   &     24.48$_{(12.20)}$    &  22.97$_{(8.76)}$     
			&   1    
			& 0 \\
			&&BLDA    &     10.06$_{(6.06)}$    &   28.26$_{(18.13)}$    &    14.25$_{(6.17)}$    
			&   2    
			&  1   \\
			&&BAI       &    17.32$_{(9.01)}$       &   20.94$_{(14.02)}$    & 17.39$_{(8.37)}$      
			&  3      
			&  3 \\
			&&LOUI     & 16.08$_{(8.97)}$    &  21.32$_{(13.90)}$   &  17.06$_{(8.32)}$  
			&  3     
			&   3    \\
			&&Msplit-HR    &  18.64$_{(9.41)}$    &  18.04$_{(11.09)}$   &  16.84$_{(8.28)}$   
			& 2  
			&  0   \\
			\hline			
		\end{tabular}
		}
	\end{center}
\end{table*}

Table \ref{tab:3000-50&100-10} consists of 
the results for dimension $p=3000$. 
As expected, the class-specific MCRs  
of all the methods increase compared to  
$p=1000$.  
Msplit-{\sc hr} outperforms all the other techniques   
in terms of MCR$_2$ while 
having balanced misclassification rates.  
For example, the MCR$_2$ of Msplit-{\sc hr} is smaller than 
the next best method {\sc loui} up to about $7\%$.  
In addition, we observe that Msplit-{\sc hr} 
has better performance 
than {\sc bai} and {\sc loui} even in setting 
{\bf (ii)} in which they have comparable performance for $p=1000$.    

We now assess the computational efficiency of the different methods. 
For a fixed threshold, the computational 
complexity of {\sc bai} and {\sc loui} is 
$O(n^2p)$ and that of all the other methods is $O(np)$. 
In our simulations, the threshold (or tuning) parameter in each method 
was chosen using a cross validation criterion. 
Table \ref{tab:Time_diag} provides the average computational time 
(in seconds) taken by each method to complete per-sample results.  
Note that since {\sc us-bcsvm} does not involve any feature 
selection step, as expected, this method is among the faster methods 
discussed here.
It can be seen that the {\sc hr} and {\sc blda}, followed 
by {\sc us-hr} and {\sc us-bcsvm},
are the fastest among all the methods we considered, but they are outperformed
by the other methods in terms of the error rate in the minority class. 
In addition, while {\sc bai} and 
{\sc loui}'s performances in terms of the error rates in the minority class
are comparable to our proposed method Msplit-{\sc hr}; the former are slower in terms 
of computational time. 

\begin{table*}[h]
	\caption{Average computational time (in seconds) taken by  
		a method to complete per-sample results: Simulation setting {\bf(i)}.}
	\label{tab:Time_diag}
	\begin{center}
		\resizebox{12cm}{!}{
			\begin{tabular}{crrrrrrrc}
				\hline
				$(n_1,n_2,p)$ & \multicolumn{1}{c}{{\sc us-bcsvm}} 
				&\multicolumn{1}{c}{DROAD}
				&  \multicolumn{1}{c}{HR}  &    \multicolumn{1}{c}{US-HR}
				&  \multicolumn{1}{c}{BLDA} & \multicolumn{1}{c}{BAI} 
				& \multicolumn{1}{c}{LOUI} & \multicolumn{1}{c}{Msplit-HR} \\
				\hline
				{(25,5,1000)}  & 2.8 & 21.73 & 0.9  &  4.66  & 1.05  &  6   &   6.39   &  9.27   \\			
				(50,10,1000)  & 5.12 &30.77  & 1.47  & 19.98   & 3.53  &   58   &   260   &   92  \\   
				(100,10,1000) & 4.76 &35.00  &  5.43 &  42.22  & 11.20  &   421   &   365   &   185  \\ 
				{(25,5,3000)} & 7.5 &  97.58 &  1.13 & 9.05 & 1.75  &  14.72  &  13.83   & 19.63   \\ 			
				(50,10,3000) & 12.38 &146.17 &  4.40  &  62.54 &  12.24 &   225  &    219  &    282  \\ 
				(100,10,3000) & 10.67 &141.82  &  19.90  &  169.97& 29.34  &   1517   &   2294   &  1200    \\    
				\hline
			\end{tabular}
		}
	\end{center}
\end{table*}

\subsection{General $\bSigma$}
\label{subsec:corr_sim}
We considered the same binary classification problem as in Section 
\ref{subsec:indep_sim}, 
i.e.\ $\XX \vert (Y=k) \sim N_p(\bmu_k, \bSigma), k= 1, 2$, 
but with a general non-diagonal $\bSigma$. We generated 
training data with different class sizes $n_1$ and $n_2$, and 
test data sets of sizes $50$ in both classes. 
The simulation results are based on $100$ randomly generated data sets. 
The parameter settings are: 
\begin{itemize}
	\item [\textbf{(iii)}] 
	$\bmu_{1}=\textbf{0}_p$, 
	$\bmu_{2}^{\top}=(1,0.5*\textbf{1}_5^{\top},0.1*\textbf{1}_5^{\top},\textbf{0}_{p-11}^{\top})$, 
	$(\bSigma)_{ij}=0.8$, for $i\neq j$, $(\bSigma)_{ii}=4$, for $i=1,...,p$ and $\Delta_p^2=0.71$.  
	
	\item[\textbf{(iv)}] 
	$\bmu_{1}=\textbf{0}_p$, 
	$\bmu_{2}^{\top}=(1,\textbf{0}_4^{\top},0.1,\textbf{0}_{p-6}^{\top})$, 
	$
	\bSigma = 
	\left[ 
	\begin{smallmatrix}
		\bSigma_1     &   & \textbf{0}  \\
		&   \bSigma_2  &   \\
		\textbf{0}  &    & \ddots 
	\end{smallmatrix}
	\right] 
	$, 
	where $(\bSigma_1)_{ij}=0.3$, and $(\bSigma_2)_{ij}=0.8$, 
	for $i\neq j$, $(\bSigma_1)_{ii}=(\bSigma_2)_{ii}=1$, for 
	$i=1,...,5$ and $\Delta_p^2=1.27$.
\end{itemize}
In what follows, using the same performance measures 
described in Section \ref{subsec:indep_sim}, we compare  
these methods: {\sc fair}, {\sc slda}, {\sc road}, Msplit-{\sc hr}, a binary version of 
the pairwise sure independent screening ({\sc psis}) method by \cite{pan2016ultrahigh},  
bias adjusted {\sc road} ({\sc ba-road}) and 
leave-one-out {\sc road} ({\sc lou-road}) by \cite{bak2016high},  
and {\sc us-bcsvm} mentioned in Section 5.1.
For the {\sc fair}, {\sc road}, {\sc ba-road}, {\sc lou-road}, 
we used the techniques
based on cross-validation described in the related papers for selecting tuning parameters.  
We applied the bi-section method of 
\cite{li2015sparse} for tuning parameter selection 
in {\sc slda} by minimizing the MCR of the minority class 
(called {\sc slda}$_{\text{\sc mcr}_2}$, in the tables). 

All the aforementioned methods provide sparse estimates, say $\hat \bbeta$, of the vector 
$\bbeta = (\beta_j: 1 \le j\le p)^{\top} = \bSigma^{-1} \bmu_d$ by either plugging in particular 
sparse estimates of $\bmu_d$ and $\bSigma$, or by directly 
finding sparse estimate of $\bbeta$. Thus, in our simulation 
results for each method, we also report the number of $j$'s 
for which 
$\hat \beta_j \not=0$, denoted by $S$ in the tables. 
For Msplit-{\sc hr}, we report the cardinality of the set 
$\mathcal{S}_n = \{ 1 \le j \le p: \frac{f_j}{\calL} \geq 0.5 \}$, where $f_j$ is 
the selection frequency corresponding to index $j$ over the splits $\ell=1,...,\calL$. 
Table \ref{tab:sig-200-50&100-10} contains the simulation results
for 
$(n_1, n_2, p) = (25,5,200)$, 
$(50,10,200)$  and $(100,10,200)$, 
and the results 
for the dimension 
$p=500$ are given in the Table \ref{tab:sig-500-50&100-10}.

\subsubsection{Discussion of the results}
\label{subsec:result2}

From Tables \ref{tab:sig-200-50&100-10} 
and \ref{tab:sig-500-50&100-10}, 
under both settings {\bf (iii)} and {\bf (iv)}, 
we can see that 
{\sc fair}, {\sc slda}, {\sc psis} and {\sc road} tend to classify 
more observations to the majority class, and resulting in large  
differences between the two MCRs. 
Overall, the techniques 
{\sc us-bcsvm}, {\sc ba-road}, {\sc lou-road} 
and Msplit-{\sc hr} perform better  
than {\sc fair}, {\sc slda}, {\sc psis} 
and {\sc road} in terms of MCR$_2$ and 
the geometric mean. 
For the setting {\bf (iii)}, in the case  
$(n_1,n_2)=(25,5)$,  
Msplit-{\sc hr} outperforms others, and 
in the cases, $(n_1,n_2)=(50,10)$ and $(100,10)$, 
the {\sc us-bcsvm} and {\sc lou-road} have 
better performance than others; 
for example, when $(n_1,n_2)=(100,10)$,  
{\sc lou-road} outperforms Msplit-{\sc hr} about $4\%$. 
For the setting {\bf (iv)}, 
Msplit-{\sc hr} outperforms all 
the other techniques in terms of MCR$_2$; for example 
outperforms {\sc bc-svm} and {\sc lou-road} 
respectively up to about $10\%$ and $12\%$ depending on the values of $(n_1, n_2, p)$.  
Moreover, this performance of Msplit-{\sc hr} is based 
on a much smaller set of selected features compared to its competitors.  
In summary, Msplit-{\sc hr} has better performance in  
the setting {\bf (iv)} which includes more features with weak signals than {\bf (iii)}.

\begin{table*}
	\caption{Classification results for the simulation settings {\bf (iii)}-{\bf(iv)} 
		with a general $\bSigma$ and $p=200$.}
	\label{tab:sig-200-50&100-10}
	\begin{center}
		\resizebox{7cm}{!}{
		\begin{tabular}{c|c|lrrrr}
			\hline
			$(n_1,n_2)$&
			Setting &Methods & \multicolumn{1}{c}{$MCR_1\%$} & 
			\multicolumn{1}{c}{$MCR_2\%$} & \multicolumn{1}{c}{$GM\%$} 
			& \multicolumn{1}{c}{$S$} \\
			\hline
			\multirow{8}{*}{(25,5)}&
			\multirow{8}{*}{\textbf{(iii)}}
			&{\sc us-bcsvm}  &  46.26$_{(15.97)}$    & 51.22$_{(15.50)}$    &   46.20$_{(6.30)}$  &   200
			\\
			&&FAIR        &  23.22$_{(9.64)}$    &  78.56$_{(10.26)}$    &   40.53$_{(8.04)}$  & 6.87 
			\\
			&&SLDA$_{\text{\sc mcr}_2}$      &  42.04$_{(14.38)}$    &  57.38$_{(14.87)}$    &   47.02$_{(6.55)}$  &  147.07
			\\
			&&PSIS        &  31.56$_{(9.44)}$    &  66.98$_{(10.51)}$    &   45.02$_{(6.21)}$  &  1
			\\
			&&ROAD     &  15.47$_{(8.03)}$    &  82.67$_{(8.87)}$    &   34.16$_{(6.89)}$  &  26.17
			\\
			&&BA-ROAD        & 48.20 $_{(12.51)}$    &  48.91$_{(12.36)}$    &   46.83$_{(4.55)}$  &  56.45
			\\
			&&LOU-ROAD     &  48.07$_{(12.27)}$    &  49.03$_{(12.40)}$    &   46.97$_{(2.55)}$  &  54.09
			\\
			&&Msplit-HR       &  53.58$_{(15.06)}$    &  45.76$_{(16.23)}$    &   47.12$_{(6.65)}$  &  4
			\\
			\hline
			\multirow{8}{*}{(25,5)}&
			\multirow{8}{*}{\textbf{(iv)}}
			&{\sc us-bcsvm}  &  49.96$_{(15.76)}$    & 46$_{(15.44)}$    &   45.39$_{(5.62)}$  &   200
			\\
			&&FAIR    &  20.96$_{(8.00)}$    &  76.38$_{(10.82)}$    &   38.83$_{(6.85)}$  &  8.07
			\\
			&&SLDA$_{\text{\sc mcr}_2}$     &  37.18$_{(14.30)}$    &  60.74$_{(14.84)}$    &   45.31$_{(7.84)}$  & 124.39 
			\\
			&&PSIS      &  30.02$_{(9.15)}$    &  62.52$_{(16.16)}$    &   42.17$_{(8.64)}$  &  1
			\\
			&&ROAD      &  15.77$_{(7.24)}$    &  78.93$_{(11.66)}$    &   33.94$_{(6.09)}$  &  24.38
			\\
			&&BA-ROAD     &  46.52$_{(15.06)}$    &  46.63$_{(16.65)}$    &   43.92$_{(7.17)}$  &  48.53
			\\
			&&LOU-ROAD       &  46.34$_{(15.79)}$    &  46.22$_{(17.97)}$    &   43.17$_{(7.63)}$  &  47.26
			\\
			&&Msplit-HR      &  52.32$_{(18.66)}$    &  43.02$_{(18.51)}$    &   43.77$_{(7.92)}$  & 4.5 
			\\
			\hline
			\hline
			\multirow{8}{*}{(50,10)}&
			\multirow{8}{*}{\textbf{(iii)}}
			&{\sc us-bcsvm}  &  46.34$_{(11.67)}$    & 48.88$_{(12.77)}$    &   46.27$_{(5.25)}$  &   200
			\\
			&&FAIR        &  28.98$_{(8.96)}$    &  69.1$_{(8.95)}$    &   43.96$_{(6.76)}$  &  
			6 \\
			&&SLDA$_{\text{\sc mcr}_2}$       &  44.5$_{(11.65)}$    &  55.84$_{(12.40)}$    &   48.57$_{(5.37)}$  &   
			195 \\
			&&PSIS                         &  37.96$_{(8.26)}$    &  60.72$_{(8.67)}$    &  47.47$_{(5.54)}$  & 
			1\\
			&&ROAD   &  19.98$_{(8.79)}$    &  77.54$_{(9.02)}$    &    38.07$_{(7.11)}$  &
			44 \\
			&&BA-ROAD                  &  48.36$_{(14.40)}$    &  48.50$_{(14.15)}$    &   45.99$_{(9.29)}$  &   
			53.50 \\
			&&LOU-ROAD                     &  47.38$_{(11.90)}$    &  49.14$_{(12.19)}$    &     46.99$_{(6.34)}$ & 
			53\\
			&&Msplit-HR       &  50.76$_{(13.85)}$    &  47.64$_{(12.15)}$    &  47.65$_{(6.11)}$     &
			3  \\
			\hline
			\multirow{8}{*}{(50,10)}&
			\multirow{8}{*}{\textbf{(iv)}}
			&{\sc us-bcsvm}  &  47.88$_{(11.98)}$    & 48.42$_{(12.55)}$    &   46.76$_{(5.53)}$  &   200
			\\
			&&FAIR        &  23.98$_{(8.65)}$    &  64.5$_{(12.53)}$    &   38.33$_{(7.57)}$  &
			6\\
			&&SLDA$_{\text{\sc mcr}_2}$       &  37.5$_{(13.23)}$    &  54.58$_{(16.77)}$    &    43.61$_{(9.53)}$  & 
			189.5 \\
			&&PSIS                         &  32.16$_{(8.79)}$    &  51.04$_{(17.23)}$    &   39.64$_{(9.48)}$ & 
			1 \\
			&&ROAD       &  21.68$_{(9.25)}$    &  64.44$_{(18.71)}$    &  35.45$_{(7.40)}$ & 
			19.50 \\    
			&&BA-ROAD                  &  37.82$_{(13.06)}$    &  44.74$_{(15.80)}$    &   39.00$_{(10.23)}$   & 
			26 \\
			&&LOU-ROAD                    &  39.74$_{(12.51)}$    &  42.46$_{(13.50)}$    &    39.80$_{(8.27)}$  & 
			27\\
			&&Msplit-HR      &  44.62$_{(13.96)}$    & 40.1$_{(14.25)}$    &    40.77$_{(8.48)}$  &
			1 \\
			\hline
			\hline
			\multirow{8}{*}{(100,10)}&
			\multirow{8}{*}{\textbf{(iii)}}
			&{\sc us-bcsvm}  &  46.54$_{(11.31)}$    & 48.46$_{(12.50)}$    &   46.19$_{(5.04)}$  &  200 
			\\
			&&FAIR    &  26.08$_{(7.63)}$    &  70.62$_{(7.94)}$    &   42.27$_{(6.27)}$  &  6.68 \\
			&&SLDA$_{\text{\sc mcr}_2}$    &  47.24$_{(13.75)}$    &  54.18$_{(12.48)}$    &   49.09$_{(6.46)}$  &  169.26
			\\
			&&PSIS     &  36.06$_{(8.41)}$    &  63$_{(8.95)}$    &   46.52$_{(6.62)}$  & 1.01 
			\\
			&&ROAD    &  11.16$_{(6.67)}$    &  86.04$_{(8.46)}$    &   29.49$_{(7.64)}$  &  71.70
			\\
			&&BA-ROAD    &  44.50$_{(13.57)}$    &  50.94$_{(13.67)}$    &   45.40$_{(8.71)}$  &  85.41
			\\
			&&LOU-ROAD    &  44.62$_{(10.18)}$    &  42.44$_{(9.37)}$    &   42.79$_{(6.12)}$  &  66.22
			\\
			&&Msplit-HR      &  50.42$_{(15.15)}$    &  46.46$_{(14.43)}$    &   46.22$_{(6.99)}$  &  11.45
			\\
			\hline
			\multirow{8}{*}{(100,10)}&
			\multirow{8}{*}{\textbf{(iv)}}
			&{\sc us-bcsvm}  &  47.76$_{(12.15)}$    & 47.26$_{(12.50)}$    &   46.03$_{(5.56)}$  &   200
			\\
			&&FAIR      &  22$_{(7.97)}$    &  67.54$_{(11.80)}$    &   37.63$_{(7.61)}$  &  8.03
			\\
			&&SLDA$_{\text{\sc mcr}_2}$    &  34.2$_{(11.71)}$    &  50.54$_{(17.12)}$    &   40.03$_{(8.67)}$  &  96.77
			\\
			&&PSIS    &  31.36$_{(8.13)}$    &  47.58$_{(18.24)}$    &   37.74$_{(9.75)}$  &  1
			\\
			&&ROAD     &  11.16$_{(6.67)}$    &  86.04$_{(8.46)}$    &   29.49$_{(7.64)}$  & 71.70 
			\\
			&&BA-ROAD     &  44.50$_{(13.57)}$    &  50.96$_{(13.67)}$    &   45.39$_{(8.71)}$  &  85.41
			\\
			&&LOU-ROAD   &  44.12$_{(12.27)}$    &  50.54$_{(11.72)}$    &   45.84$_{(5.82)}$  &  97
			\\
			&&Msplit-HR    &  45.86$_{(17.28)}$    &  37.6$_{(14.92)}$    &   39.19$_{(8.68)}$  &  9.25
			\\
			\hline
		\end{tabular}
	}
	\end{center}
\end{table*}

\begin{table*}
	\caption{Classification results for the simulation settings {\bf (iii)}-{\bf(iv)} 
		with a general $\bSigma$  and $p=500$.}
	\label{tab:sig-500-50&100-10}
	\begin{center}
		\resizebox{7cm}{!}{
		\begin{tabular}{c|c|lrrrr}
			\hline
			$(n_1,n_2)$ &
			Setting & Methods & \multicolumn{1}{c}{$MCR_1\%$} & 
			\multicolumn{1}{c}{$MCR_2\%$} & \multicolumn{1}{c}{$GM\%$} 
			& \multicolumn{1}{c}{$S$} \\
			\hline
			\multirow{8}{*}{(25,5)}&
			\multirow{8}{*}{\textbf{(iii)}}
			&{\sc us-bcsvm}  &  49.18$_{(17.81)}$    &  50.46$_{(17.69)}$    &   46.39$_{(8.47)}$ & 500
			\\
			&&FAIR       &  20.48$_{(8.72)}$    &  79$_{(8.91)}$    &   38.79$_{(8.29)}$  &  8.91
			\\
			&&SLDA$_{\text{\sc mcr}_2}$     &  44.84$_{(15.58)}$    &  54.56$_{(16.11)}$    &   46.97$_{(5.57)}$  & 350.18 
			\\
			&&PSIS       &  23.22$_{(9.64)}$    &  75.68$_{(10.26)}$    &   40.53$_{(8.05)}$  &  6
			\\
			&&ROAD       &  12.01$_{(8.27)}$    &  87.16$_{(8.70)}$    &   30.21$_{(8.08)}$  &  30.79
			\\
			&&BA-ROAD       &  44.63$_{(13.74)}$    &  53.71$_{(14.30)}$    &   45.01$_{(9.74)}$  &  57.49
			\\
			&&LOU-ROAD      &  45.82$_{(12.19)}$    &  52.32$_{(12.41)}$    &   47.40$_{(2.84)}$  &  69.18
			\\
			&&Msplit-HR      &  55.96$_{(16.92)}$    &  44.58$_{(17.43)}$    &   46.77$_{(7.11)}$  &  3
			\\
			\hline
			\multirow{8}{*}{(25,5)}&
			\multirow{8}{*}{\textbf{(iv)}}
			&{\sc us-bcsvm}  &  48.9$_{(15.18)}$    &  47.66$_{(13.96)}$    &   46.17$_{(5.45)}$ & 500
			\\
			&&FAIR     &  15.1$_{(8.91)}$    &  84.08$_{(8.87)}$    &   33.15$_{(10.97)}$  &  13.38
			\\
			&&SLDA$_{\text{\sc mcr}_2}$      &  41.04$_{(15.56)}$    &  58.3$_{(15.45)}$    &   46.52$_{(7.29)}$  & 315.96 
			\\
			&&PSIS    &  30.26$_{(9.52)}$    &  65.9$_{(13.01)}$    &   43.60$_{(7.70)}$  &  1
			\\
			&&ROAD       &  12.50$_{(8.10)}$    &  85.07$_{(10.66)}$    &   30.67$_{(7.51)}$  &  27.82
			\\
			&&BA-ROAD      &  48.07$_{(13.10)}$    &  48.40$_{(14.23)}$    &   46.20$_{(6.71)}$  &  60.06
			\\
			&&LOU-ROAD     &  48.47$_{(13.37)}$    &  47.48$_{(14.94)}$    &   45.97$_{(5.10)}$  &  59.19
			\\
			&&Msplit-HR       &  53.06$_{(17.66)}$    &  43.16$_{(17.54)}$    &   44.50$_{(8.83)}$  &  2
			\\
			\hline
			\hline
			\multirow{8}{*}{(50,10)}&
			\multirow{8}{*}{\textbf{(iii)}}
			&{\sc us-bcsvm}  &  48.46$_{(12.44)}$    &  48.64$_{(14.25)}$    &   46.94$_{(5.82)}$ & 500
			\\
			&&FAIR        &  26.1$_{(8.65)}$    &  72.12$_{(8.35)}$    &   42.48$_{(6.57)}$  &  
			9 \\
			&&SLDA$_{\text{\sc mcr}_2}$       &  43.76$_{(12.28)}$    &  55.18$_{(11.98)}$    &   47.78$_{(5.51)}$ &  
			493.5 \\
			&&PSIS                         &  35.44$_{(8.02)}$    &  63.02$_{(9.33)}$    &  46.67$_{(5.37)}$  &
			1 \\
			&&ROAD    &  14.10$_{(9.43)}$    &  83.70$_{(11.63)}$    &   31.80$_{(9.44)}$    &
			59.50 \\
			&&BA-ROAD                  &  48.32$_{(14.03)}$    &  49.66$_{(12.68)}$    &   47.15$_{(7.68)}$ &  
			64.50 \\
			&&LOU-ROAD                 &  48.18$_{(13.51)}$    &  48.26$_{(12.33)}$    &   46.71$_{(6.01)}$   &
			68 \\
			&&Msplit-HR       &  50.06$_{(15.67)}$    &  48.26$_{(14.52)}$    &   46.91$_{(5.78)}$   &
			1 \\  
			\hline
			\multirow{8}{*}{(50,10)}&
			\multirow{8}{*}{\textbf{(iv)}}
			&{\sc us-bcsvm}  &  49.04$_{(10.87)}$    &  49.28$_{(11.96)}$    &   48.01$_{(5.56)}$ & 500
			\\
			&&FAIR        &  17.38$_{(6.91)}$    &  76.14$_{(10.84)}$    &   35.41$_{(7.35)}$  &
			13.5 \\
			&&SLDA$_{\text{\sc mcr}_2}$       &  38.58$_{(13.19)}$    &  51.14$_{(14.79)}$    &   42.98$_{(9.05)}$   & 
			473   \\
			&&PSIS              &  32.02$_{(8.58)}$    &  54.56$_{(17.60)}$    &   40.97$_{(9.69)}$ &  
			1 \\
			&&ROAD                       &  15.46$_{(9.32)}$    &  73.56$_{(19.75)}$    &   30.97$_{(7.64)}$ &  
			38.50 \\
			&&BA-ROAD                  &  42.20$_{(13.08)}$    &  44.26$_{(15.04)}$    &   41.40$_{(8.97)}$  & 
			38  \\
			&&LOU-ROAD                     &  42.60$_{(13.97)}$    &  43.16$_{(14.41)}$    &  41.25$_{(8.55)}$  &   
			47  \\
			&&Msplit-HR       &  46.4$_{(14.65)}$    &  40.7$_{(15.10)}$    &   41.55$_{(8.81)}$    & 
			1 \\ 
			\hline
			\hline
			\multirow{8}{*}{(100,10)}&
			\multirow{8}{*}{\textbf{(iii)}}
			&{\sc us-bcsvm}  &  47.96$_{(11.92)}$    &  49.98$_{(13.98)}$    &   47.47$_{(5.67)}$ & 500
			\\
			&&FAIR        &  23.91$_{(8.31)}$    &  74.6$_{(7.48)}$    &   41.28$_{(7.55)}$  & 8.22
			\\
			&&SLDA$_{\text{\sc mcr}_2}$       &  44.28$_{(12.54)}$    &  54.34$_{(12.45)}$    &   47.61$_{(5.27)}$  & 403.99
			\\
			&&PSIS            &  33.32$_{(7.73)}$    &  65.48$_{(8.83)}$    &   46.24$_{(5.87)}$  & 1.01
			\\
			&&ROAD           &  4.60$_{(3.47)}$    &  94.60$_{(4.99)}$    &   19.02$_{(8.42)}$  & 96.49
			\\
			&&BA-ROAD        &  45.40$_{(13.39)}$    &  51.16$_{(14.24)}$    &   45.90$_{(8.21)}$  & 105.05
			\\
			&&LOU-ROAD       &  46.24$_{(8.52)}$    &  43.12$_{(10.44)}$    &   43.99$_{(6.11)}$  & 103.05
			\\
			&&Msplit-HR        &  51.76$_{(14.45)}$    &  46$_{(14.62)}$    &   46.61$_{(6.55)}$  &  5.67
			\\
			\hline
			\multirow{8}{*}{(100,10)}&
			\multirow{8}{*}{\textbf{(iv)}}
			&{\sc us-bcsvm}  &  48.72$_{(11.81)}$    &  48.94$_{(12.06)}$    &   47.65$_{(5.53)}$ &   500
			\\
			&&FAIR         &  15.1$_{(7.41)}$    &  79.42$_{(9.89)}$    &   33.09$_{(8.62)}$  &  17.18
			\\
			&&SLDA$_{\text{\sc mcr}_2}$       &  36.98$_{(12.42)}$    &  53.58$_{(14.94)}$    &   43.12$_{(9.01)}$  &  226.33
			\\
			&&PSIS        &  30.28$_{(7.57)}$    &  54.96$_{(18.29)}$    &   39.94$_{(9.12)}$  &  1.01
			\\
			&&ROAD      &  4.64$_{(3.47)}$    &  94.60$_{(4.99)}$    &   19.02$_{(8.42)}$  &  96.49
			\\
			&&BA-ROAD       &  45.40$_{(13.39)}$    &  51.16$_{(14.24)}$    &   45.90$_{(8.21)}$  & 105.05
			\\
			&&LOU-ROAD        &  45.64$_{(13.15)}$    &  49.18$_{(12.87)}$    &   45.79$_{(5.82)}$  & 107.80
			\\
			&&Msplit-HR       &  43.52$_{(12.21)}$    &  44.02$_{(14.34)}$    &   42.38$_{(8.26)}$  & 5.74
			\\
			\hline
		\end{tabular}
	}
	\end{center}
\end{table*}

Next, we assess the computational efficiency of different methods by studying 
the average computational time (in seconds) taken by each method to 
complete per-sample results, which are given in Table \ref{tab:Time_general}.
We can see that {\sc psis} is the fastest method 
followed by {\sc fair} and {\sc us-bcsvm}. 
However, as seen above, these methods do not perform well 
in terms of the MCRs.  As mentioned before, 
{\sc us-bcsvm} is computationally fast, since it 
does not involve any feature selection step.
The {\sc slda} is slower than the Msplit-{\sc hr} when the dimension $p$ 
is increased from $p=200$ to $500$. 
On the other hand, Msplit-{\sc hr} is computationally 
more efficient than its two competitors {\sc ba-road} and {\sc lou-road}. 
Note that, for a fixed value of tuning parameter,  
the computational complexity of {\sc ba-road} and {\sc lou-road} 
is $O(n^2p^2)$, and that of Msplit-{\sc hr} is $O(np^2)$. Therefore, even without a 
tuning selection procedure, our technique has lower computational cost. 

\begin{table*}
	\caption{Average computational time (in seconds) taken by  
		a method to complete per-sample results: Simulation setting {\bf(iv)}.}
	\label{tab:Time_general}
	\begin{center}
		\resizebox{12cm}{!}{
			\begin{tabular}{crrrrrrrr}
				\hline
				$(n_1,n_2,p)$ & \multicolumn{1}{c}{{\sc us-bcsvm}} 
				& \multicolumn{1}{c}{FAIR} 
				& \multicolumn{1}{c}{SLDA$_{\text{\sc mcr}_2}$}  & \multicolumn{1}{c}{PSIS} 
				& \multicolumn{1}{c}{ROAD}& \multicolumn{1}{c}{BA-ROAD} 
				& \multicolumn{1}{c}{LOU-ROAD} 
				& \multicolumn{1}{c}{Msplit-HR}  \\
				\hline
				{(25,5,200)} &  1.91 & 1.75  & 8.75  & 0.28 & 50.23  &  110.34  &  59.93  &   11.42    \\ 			
				(50,10,200) & 1.48  & 4.73  & 25.71 & 4.11 & 66.00  &  192.10  &  189.86   & 39.31      \\   
				(100,10,200) & 1.86  & 5.56 & 66.46 & 3.06 & 120.93 & 468.02  & 443.23  & 114.44     \\ 
				{(25,5,500)}  & 3.00 & 29.3 & 80.05 & 0.41 & 204.52 &   254.64  & 184.99 &  16.4  \\			
				(50,10,500)  & 3.01 & 27.09 & 234.35 & 3.00 & 272.15 &   483.82   & 477.45 & 62.23   \\ 
				(100,10,500) &  3.58  & 29.66 & 451.75 & 3.61 & 219.95 &  974.34   &  1084.75   & 176.65   \\
				\hline
			\end{tabular}
		}
	\end{center}
\end{table*}

In summary, given the difficulty of the imbalanced problem, 
our current simulation study shows that (considering all the three factors: misclassification 
rates, feature selection, and computational efficiency)
Msplit-{\sc hr} has a good performance  
compared to the methods discussed here, and is 
yet another reliable technique for high-dimensional imbalanced problems.

\section{Real-data analysis}
\label{sec:real}

We now demonstrate the performance of different methods on two real 
data sets. 
{\footnote{\label{real_data}Both data sets are publicly available from the R package 
		\textit{datamicroarray} 
		\citep{ramey2016datamicroarray}, and are available at https://github.com/.}
	
	\noindent
	The first data set, on breast cancer \citep{gravier2010prognostic}, consists of 
	the expression profiles of $2905$ genes for $168$ patients of whom 
	$111$ patients with no event after diagnosis were labelled as ``good'' and 
	the remaining $57$ patients with early metastasis were labelled as ``poor''. 
	In our analysis, we randomly split the data into training data of sizes  
	$56$ and $28$ of respectively good cases (the majority Class 1) and 
	poor cases (the minority Class 2). The rest of the data is used for testing. 
	The classification results, under the assumptions of (a) uncorrelated and 
	(b) correlated features, are given in Table \ref{tab:gravier_1500}. 
	Under (a), the results suggest that 
	{\sc bai}, {\sc loui}, Msplit-{\sc hr}, and {\sc us-hr} 
	have comparable performance, with {\sc bai} 
	and {\sc loui} performing slightly better than 
	the other two in terms of the MCR of the minority class (MCR$_2$). 
	Under (b),
	{\sc ba-road}, {\sc lou-road}, and Msplit-{\sc hr} perform similar in terms 
	of the MCRs. 
	{\sc us-bcsvm} has smaller MCRs compared 
	to the others but by using the set of all features as it is 
	not able to perform any feature selection.   
	Note that in both cases, Msplit-{\sc hr} selects a much smaller number of  
	features toward the classification task. 
	
	\begin{table*}
		\caption{Classification results for Breast Cancer data set.   
			$S$ denotes the median number of selected features. 
		}
		\label{tab:gravier_1500}
		\begin{center}
			\begin{tabular}{clrrrrrrr}
				\hline
				$\bSigma$ &
				Methods & \multicolumn{1}{c}{$MCR_{1}\%$} & \multicolumn{1}{c}
				{$MCR_{2}\%$} & 
				\multicolumn{1}{c}{$GM\%$} & \multicolumn{1}{c}{$S$}  
				\\
				\hline
				\multirow{7}{*}{Diagonal}&
				DROAD    &  19.62$_{(10.20)}$  &  46.31$_{(13.38)}$   &  28.72$_{(7.77)}$   
				&  201.50       
				\\
				&HR   &  16.82$_{(6.70)}$    &  46.72$_{(10.24)}$   &   27.08$_{(5.98)}$ 
				&   26          
				\\
				&US-HR  &  20.67$_{(7.58)}$   &  39.79$_{(9.78)}$    &  27.93$_{(6.10)}$  
				&  32       
				\\
				&BLDA   &  16.8$_{(6.14)}$    &  45.59$_{(10.86)}$   & 26.78$_{(5.63)}$   
				&  35     
				\\
				&BAI         &  22.24$_{(7.09)}$   &  37$_{(10.08)}$    &  27.83$_{(5.40)}$   
				&  99                    
				\\
				&LOUI    &  22.65$_{(7.05)}$   &  37.03$_{(10.82)}$   &  28.06$_{(5.20)}$   
				&  83.5      
				\\
				&Msplit-HR      &  20.96$_{(7.00)}$   &  39.56$_{(10.74)}$    &  27.83$_{(5.19)}$  
				&  6        
				\\
				\hline
				\hline
				\multirow{8}{*}{General}&
				{\sc us-bcsvm}    &  19.78$_{(5.74)}$  &  34.79$_{(10.01)}$   &  25.50$_{(4.24)}$   
				&    1500     
				\\
				&FAIR        &  16.24$_{(5.54)}$    &  45.41$_{(9.31)}$    &  26.43$_{(4.95)}$   &   
				22     \\
				&SLDA$_{\text{\sc mcr}_2}$       &  22.91$_{(12.32)}$    &  47.76$_{(12.19)}$    &   31.58$_{(9.17)}$   &
				1500 \\
				&PSIS          &  27.47$_{(14.62)}$    &  46.17$_{(15.30)}$    &   33.77$_{(9.85)}$   & 
				1 \\
				&ROAD                  &  19.51$_{(10.03)}$    &  47.41$_{(13.81)}$    &   28.96$_{(7.26)}$   &
				25 \\
				&BA-ROAD   &  22.16$_{(5.95)}$    &  38.83$_{(9.72)}$    &   28.62$_{(4.24)}$   &  
				51.50 \\
				&LOU-ROAD              &  22.16$_{(5.90)}$    &  38.10$_{(9.84)}$    &   28.37$_{(4.32)}$   &
				56.50  \\
				&Msplit-HR    &  24.11$_{(8.85)}$    &  40.55$_{(10.76)}$    &   30.35$_{(6.50)}$   &
				5 \\
				\hline
			\end{tabular}
		\end{center}
	\end{table*}

	\begin{table*}
		\caption{Classification results for Myeloma Cancer data set. 
			$S$ denotes the median number of selected features. 
		}
		\label{tab:tian_1500}
		\begin{center}
			\begin{tabular}{clrrrrrrr}
				\hline
				$\bSigma$ &
				Methods & \multicolumn{1}{c}{$MCR_{1}\%$} & \multicolumn{1}{c}
				{$MCR_{2}\%$} & 
				\multicolumn{1}{c}{$GM\%$} & \multicolumn{1}{c}{$S$}   
				\\
				\hline
				\multirow{7}{*}{Diagonal}&
				DROAD  &  26.03$_{(11.29)}$  &  49.33$_{(10.30)}$   &  34.93$_{(8.58)}$  
				&  5    
				\\
				&HR    &  25.94$_{(11.60)}$    &  57.78$_{(11.92)}$   & 37.43$_{(8.47)}$   
				&  19        
				\\
				&US-HR  &  41.6$_{(9.74)}$  &  41$_{(12.75)}$   & 40.17$_{(6.66)}$     
				&   92.5
				\\
				&BLDA    &  25.58$_{(9.10)}$  & 53.28$_{(11.23)}$   &  35.89$_{(6.36)}$  
				&    11 
				\\
				&BAI    &  34.31$_{(10.44)}$  &  44.17$_{(13.20)}$   &  37.50$_{(5.98)}$     
				&  30          
				\\
				&LOUI     &  35.14$_{(10.54)}$  &  44.39$_{(11.35)}$   &  38.26$_{(5.95)}$   
				&  27.5
				\\
				&Msplit-HR   &  38.18$_{(13.68)}$  &  41.94$_{(14.37)}$   &  37.89$_{(7.51)}$  
				&   7     
				\\
				\hline
				\hline
				\multirow{8}{*}{General}&
				{\sc us-bcsvm}  &  53.78$_{(27.56)}$  &  39.44$_{(28.32)}$   &  46.06$_{(18.47)}$  
				&     1500 
				\\
				&FAIR       &  27.92$_{(7.64)}$    &  49.56$_{(11.16)}$    &   36.50$_{(6.15)}$   & 
				14  \\
				&SLDA$_{\text{\sc mcr}_2}$       &  28.83$_{(9.79)}$    &  47.22$_{(10.34)}$    &   36.18$_{(7.59)}$   &
				13 \\
				&PSIS      &   31.42$_{(15.24)}$    &  50.11$_{(10.33)}$    &   38.43$_{(10.72)}$   &
				1   \\
				&ROAD                  &  26.01$_{(10.27)}$    &  53.22$_{(10.63)}$    &   36.47$_{(8.13)}$   &  
				7.50 \\
				&BA-ROAD               &  34.01$_{(13.75)}$    &  43.17$_{(13.92)}$    &   35.52$_{(9.38)}$   & 
				20 \\
				&LOU-ROAD              &  33.74$_{(9.63)}$    &  42.78$_{(10.67)}$    &   38.05$_{(6.43)}$   &
				23.50 \\
				&Msplit-HR     &  34.74$_{(11.84)}$    &  42.61$_{(11.66)}$    &   37.27$_{(7.16)}$   & 
				6 \\
				\hline
			\end{tabular}
		\end{center}
	\end{table*}	
	
	The second data set, on multiple-myeloma cancer \citep{tian2003role}, 
	consists of the expression profiles of $12,2625$ genes for $173$ patients 
	with newly diagnosed multiple-myeloma, of whom $137$ were with bone lytic lesions  
	and the remaining $36$ patients were without bone lytic lesions.  
	We randomly choose a training set containing $18$ observations from patients labelled 
	by MRI-no-lytic-lesion (the minority Class 2), and $72$ observations from patients labelled 
	by MRI-lytic-lesion (the majority Class 1). The rest of the data were used for testing.
	Table \ref{tab:tian_1500} contains the classification results 
	under the aforementioned assumptions (a) and (b). 
	Under (a), 
	the results show that Msplit-{\sc hr} and {\sc us-hr} outperform the 
	other methods in terms of 
	the error rate in the minority class, MCR$_2$. 
	In addition, Msplit-{\sc hr} outperforms {\sc us-hr} 
	in terms of the error rate in the majority class, MCR$_1$. 
	Under (b), 
	the three methods {\sc ba-road}, {\sc lou-road}, and Msplit-{\sc hr} 
	perform similar in terms of the MCRs. For this data set, the overall performances 
	of the aforementioned three methods are better than {\sc us-bcsvm}. 
	Note that in both cases, Msplit-{\sc hr} selects a smaller number of  
	features toward the classification task. 
	
	To reduce the computational cost of each method, and 
	by using a t-statistic, we screened the initial 
	number of features in each of the above data sets by 
	selecting a subset of $p=1500$ genes.

	\section{Conclusion}
	\label{sec:discuss}
	In this paper, we have studied linear discriminant analysis 
	(LDA) 
	in high-dimensional imbalanced binary classification.  
	To the best of our knowledge, this is the first work 
	that rigorously investigates such 
	problems which frequently arise 
	in a wide range of applications. 
	
	First, we showed that in the aforementioned settings the standard LDA asymptotically 
	ignores the so-called minority class. Second, using a multiple data splitting 
	technique, 
	we proposed a new method, 
	called Msplit-{\sc hr}, that 
	obtains desirable large-sample properties.  
	Third, we derived conditions under which two 
	well-known sparse versions of the LDA in our setting 
	obtain certain desirable  
	large-sample properties.   
	We then examined the finite-sample performance of different methods via 
	simulations and by analyzing two real data sets. 
	In our simulations, the Msplit-{\sc hr} either outperforms competing 
	methods or has comparable performance
	in terms of misclassification rate in the minority class,   
	while it has a lower computational cost.

	The methodology (Msplit-{\sc hr}) and theory developed in this paper are based on  
	normal distribution for the feature vector $\XX$.  
	The normality is used 
	for bias calculations in Propositions 
	\ref{prop:DACH0}-\ref{prop:gap_fill_general}, 
	and to establish 
	feature selection consistency  in Lemmas 
	\ref{lem:true_actives}-\ref{lem:general_DACH}. 
	On the other hand, \cite{delaigle2012effect} 
	showed that feature selection methods based on mean-differences are sensitive to 
	heavy-tailed distributions for $\XX$, and they suggested 
	transformation approaches in feature space which are more 
	resistant to extreme observations from heavy-tailed distributions. 
	Properties of such transformations with respect to our theoretical guidelines, 
	and in general, extension of our results to non-normal models require further investigation
	and is a topic of future research.

	If the covariance matrix differs between the two classes, i.e. \(
	\XX|(Y=k) \sim N(\bmu_k, \bSigma_k), k= 1, 2 
	\), the optimal (Bayes) rule is the quadratic discriminant analysis (QDA). 
	Our limited numerical experiment shows that the QDA 
	in imbalanced high-dimensional problems behaves similarly to the LDA 
	ignoring the minority class. A potential approach to 
	alleviate the impact of imbalanced class sizes 
	is to reduce the difference between MCRs of 
	an empirical QDA toward that of the optimal rule. 
	However, the main challenge is that none of the 
	aforementioned MCRs have workable closed 
	forms. \cite{li2015sparse} studied such differences for sparse QDA, and their results 
	might be useful toward imbalanced problems in the 
	context of QDA. This, however, requires a careful investigation and is a 
	subject of future work. 
	
	Another possible future research direction is to investigate the possibility of extending 
	the methodology and theory developed in this paper  
	to imbalanced multi-class classification problems.

\section*{Acknowledgements}
We would like to thank the editor, an associate editor, and two referees for 
their insightful comments and suggestions that improved the quality 
of this paper. We thank the National High Performance Computing Center 
(NHPCC) at Isfahan University of Technology for their computational support 
to conduct our numerical experiments. 
Arezou Mojiri is grateful to (late) Soroush Alimoradi and also Ali Rejali for their 
help and constant support during her graduate studies. 
Abbas Khalili was supported by the Natural Sciences and Engineering 
Research Council of Canada through Discovery 
Grants (NSERC RGPIN-2015-03805 and NSERC RGPIN-2020-05011).

\appendix


\section{Technical Lemmas}
\label{app:lemma}

\noindent
In this Appendix, we first state the technical conditions (C1)-(C3) 
required in our theoretical developments. 
Next, we
state several lemmas that are used in the proofs of our main results. 
Lemmas \ref{lem:A3bickel} and \ref{lem:lem1_shao} 
are from \cite{bickel2008regularized} and \cite{shao2011sparse}. 
Lemmas \ref{lem:bickel_cov}-\ref{lem:pan2016} 
are the results from other papers adapted to the 
imbalanced setting under our consideration. 
Lemma \ref{lem:T_bound} states an upper bound for 
the tail of Student's t-distribution.
}

\vspace{0.3 in}

\noindent
\textbf{Technical Conditions:}

\noindent
\begin{itemize}		
\item[(C1)] 
$\log p=o(n_1)$, where $n_1$ is the majority class size.

\item[(C2)]
$~0<c_0^{-1}<\lambda_{min}(\bSigma)\leq \lambda_{max}(\bSigma)<c_0<\infty$, for a 
constant $c_0>0$. 

\item[(C3)] 
$~0<c_0^{-1}<\max_{j=1,...,p}\mu_{dj}^2<c_0<\infty$, 
where $\bmu_d=\bmu_2-\bmu_1$.
\end{itemize}

\vspace{0.3 in}

\begin{lem} 
\label{lem:A3bickel}
\citep[Lemma A.3]{bickel2008regularized} 
Let $\textbf{Z}_i$ be independent and identically random variables from 
$N_p(\textbf{0},\bSigma)$ and 
$\lambda_{\max}(\bSigma) \leq \varepsilon _0 ^{-1}<\infty$. 
Then, 
\begin{equation*}
	P \left(  
	\mid \sum _{i=1}^n (Z_{ij}Z_{ik}-\sigma _{jk}) \mid 
	> n \nu 
	\right) 
	\leq 
	C_1 \exp (-C_2n \nu ^2) \quad 
	\text{for all} \ 
	\vert \nu \vert \leq \delta
\end{equation*}
where $\sigma_{jk}$'s are entries of $\bSigma$, and 
$C_1$, $C_2$, and 
$\delta$ depend on $\epsilon _0$ only.
\end{lem}

\noindent
\begin{lem}
\label{lem:lem1_shao}
\citep[Lemma 1]{shao2011sparse}
Let $\xi_n$ and $\nu_n$ be two sequence of positive numbers such that $\xi_n\rightarrow \infty$ and $\nu_n\rightarrow 0$ as $n\rightarrow \infty$. If $\lim _{n \rightarrow \infty}\xi_n \nu_n =\gamma$, 
where $\gamma$ may be $0$, positive or $\infty$, then
\begin{eqnarray*}
	\lim _{n\rightarrow \infty}\frac{\Phi (-\sqrt{\xi_n}(1-\nu_n))}{\Phi (-\sqrt{\xi_n})}&=& e^{\gamma}.\\
\end{eqnarray*}
\end{lem}

\noindent
\begin{lem}
\label{lem:bickel_cov}
Denote the sets 
\begin{eqnarray*} 
	U_{\tau}(h, c_0(p), M)
	&=&
	\bigg\lbrace  
	\bSigma: \ 
	\sigma_{ii}\leq M, 
	\sum_{j=1}^p\vert \sigma _{ij} \vert^h\leq c_0(p) , \
	\forall i, \ 
	0\leq h <1
	\bigg\rbrace , \\
	U_{\tau}(h, c_0(p), M,\epsilon_0)
	&=&
	\bigg\lbrace  
	\bSigma: \ 
	\bSigma \in U_{\tau}(h, c_0(p), M), 
	\lambda_{min}(\bSigma)\geq \epsilon_0 >0
	\bigg\rbrace.
\end{eqnarray*} 
Let $\widetilde{\bSigma}_n$ be a thresholded version of the pooled sample covariance matrix 
$\widehat{\bSigma}_n$ in \eqref{eq:hat_Sigma}, such that 
$\tilde{\sigma}_{ij}
=(1-2/n)\hat{\sigma}_{ij}\textbf{1}\lbrace (1-2/n)\vert \hat{\sigma}_{ij} \vert >t_{n}  \rbrace$, 
with $t_{n}=M_1\sqrt{\frac{\log p}{n}}$ and some positive constant $M_1$. 
Then uniformly on $U_{\tau}(h, c_0(p), M)$, 
and for sufficiently large $M_1$, 
under the Condition (C3) and 
$n_2=o(n_1)$, as 
$n_1,n_2\to \infty$, then
\begin{eqnarray*}
	\parallel \widetilde{\bSigma}_n-\bSigma \parallel
	=
	O_p 
	\bigg(  
	c_0(p)
	\left( \log p/n_1 \right)^{\frac{1-h}{2}}  
	\bigg), 
\end{eqnarray*}
and uniformly on $U_{\tau}(h, c_0(p), M,\epsilon_0)$, 
\begin{eqnarray*}
	\parallel \widetilde{\bSigma}_n^{-1}-\bSigma^{-1} \parallel
	=
	O_p 
	\bigg(  
	c_0(p)
	\left( \log p/n_1 \right)^{\frac{1-h}{2}}  
	\bigg) .
\end{eqnarray*}
\end{lem}
\noindent
\textbf{Proof}.
The proof is a straight forward extension of Theorem 1 of \cite{bickel2008covariance} 
to imbalanced case, and thus omitted here. $\blacksquare$\\

\noindent
\begin{lem}
\label{lem:Dhat}
Let $\XX_{ik}=(X_{i1k},...,X_{ipk})^{\top}$, for $i=1,...,n_k$, and $k=1, 2$,  
be random samples from $p$-variate normal distribution 
with mean vector $\bzero$ and diagonal covariance matrix 
$\DD=diag \lbrace \sigma_1^2,...,\sigma_p^2 \rbrace$.  
If the Conditions (C1) and (C2) are satisfied and 
$n_2=o(n_1)$, then as $n_1,n_2 \to \infty$, we have 
\begin{eqnarray*}
	\max_{1\leq j\leq p} 
	\mid \hat{\sigma}_j^2 - \sigma _j ^2\mid 
	=O_p(\sqrt{(\log p)/n_1}), 
\end{eqnarray*}
where $\hat{\sigma}_j^2, j= 1, \ldots, p$, are the diagonal elements of  
the pooled sample variance $\widehat \bSigma_n$ in \eqref{eq:hat_Sigma}.
\end{lem}
\noindent
\textbf{Proof}.
Let $\overline{X} _{jk}=\frac{1}{n_k}\sum_{i=1}^{n_k}X_{ijk}$, for $k=1,2$, $j=1,...,p$. 
We have, 
\begin{eqnarray*} 
&& \Pr \left( 
\max _{1\leq j \leq p} 
\mid \hat{\sigma}_j^2 -\sigma _j^2 \mid > \eta 
\right) 
\leq 
\sum_{j=1} ^p 
\Pr \left( 
\mid \hat{\sigma}_j^2-\sigma _j^2\mid > \eta 
\right) \\
&\leq & 
\sum_{k=1}^{2}\sum_{j=1} ^p 
\Pr \left( 
\frac{1}{\sqrt{n_k}} 
\mid \sum _{i=1}^{n_k}({X}_{ijk}^2
-\sigma _j^2)
\mid > \frac{1}{\sqrt{n_k}} \frac{\eta }{4}(n_1+n_2-2)
\right)\\
&+&
\sum_{k=1}^{2}\sum_{j=1} ^p 
\Pr \left( 
\mid n_k\overline{X}_{jk}^2
-\sigma _j^2
\mid > \frac{\eta }{4}(n_1+n_2-2)
\right)  \\
&\leq & 
\sum_{k=1}^2 
p C_1 \exp \bigg\lbrace -C_2 \frac{\eta  ^2}{16}\frac{(n-2)^2}{n_k} \bigg\rbrace
+ p C_3 \exp  \bigg\lbrace -C_4 \frac{\eta  ^2}{16}(n-2)^2 \bigg\rbrace 
\end{eqnarray*}
for $\vert \eta  \vert <\delta$, 
where $C_1,C_2,C_3,C_4$, 
and $\delta$ 
are constants depending only on $c_0$.  
The last inequality follows from 
Lemma \ref{lem:A3bickel}. 
By taking $\eta =M\sqrt{\log p/n_1}$,  
for sufficiently large $M>0$,  
under the imbalanced setting and  
the Condition (C1),   
the result holds. $\blacksquare$\\


\noindent
\begin{lem}
\label{lem:pan2016}  
Under conditions of Lemma \ref{lem:general_DACH} and 
the imbalanced setting $n_2=o(n_1)$, 
assume that $m_{max}\sqrt{\log p /n_1}=o(1)$. 
Then for $\ell=1,\dots,\calL$, as long as $n_1,n_2\to \infty$,  
\[
\parallel  \widetilde{\bSigma}_{n,\ell}-\bSigma_{\ell} \parallel 
= O_p\bigg(m_{max}\sqrt{\log p/n_1}\bigg).
\]
where $\widetilde{\bSigma}_{n,\ell}=[\hsigma_{jj',\ell}^{(2)}:~ j,j'\in \calS_{n,\ell}^{(1)}]$
and 
$\bSigma_{\ell}=[\sigma_{jj'}: ~j,j'\in \calS_{n,\ell}^{(1)}]$.
\end{lem}
\textbf{Proof}. 
Note that  
if $\textbf{A}=[a_{jj'}]$ be a symmetric $p\times p$ matrix  
then  
$
\parallel \textbf{A} \parallel \leq \max_{j'} \sum_{j=1}^p \vert  a_{jj'} \vert 
$.
Thus, the result is implied by 
\begin{equation}
\Pr \bigg( 
\max_{j\in \calS_{n,\ell}^{(1)}} \sum_{j'\in \calS_{n,\ell}^{(1)} } \vert 
\hsigma_{jj',\ell}^{(2)}-\sigma_{jj'}  \vert > \eta
\bigg)
\leq 
\sum_{j,j'\in \calS_{n,\ell}^{(1)} } 
\Pr \bigg( 
\vert \hsigma_{jj',\ell}^{(2)}-\sigma_{jj'}  \vert 
> 
\frac{\eta}{m_{max}}
\bigg)
\hspace{.5in}
\label{sigm_t}
\end{equation}
where $m_{max}=c_1 \vert  \calS \vert ( \max_{j\in \calS}\beta_j^2)/d_{0,n}^2$.   
The inequality follows from part (ii) of Lemma \ref{lem:general_DACH}.  
Let $\bmu_{k,\ell}=[\mu_{jk}:~ j\in\calS_{n,\ell}^{(1)}]$,  
$Z_{ijk,\ell}=X_{ijk,\ell}-\mu_{jk,\ell}$, 
and 
$\bar{Z}_{jk,\ell}=\sum_{i=1}^{n'_k}X_{ijk,\ell}/n'_k$, 
where $X_{ijk,\ell}\in\calD_{n,\ell}^{(2)}$, 
for $i=1,...,n'_k$, $j=1,...,p$,  
$k=1,2$, and $\ell=1,...,\calL$, 
where $\XX_{ik,\ell} \sim N_p(\bmu_{k,\ell}, \bSigma_{\ell})$.  
For the first probability term in \eqref{sigm_t}, we have 
\begin{eqnarray*}
&&
\Pr \bigg( 
\vert \hsigma_{jj',\ell}^{(2)}-\sigma_{jj'}  \vert 
> 
\frac{\eta}{m_{max}}
\bigg) \leq\\
&&
\sum_{k=1}^2
\Pr \bigg(
\bigg\vert \sum_{i=1}^{n'_k}Z_{ijk,\ell}Z_{ij'k,\ell}
-n'_k\bar{Z}_{jk,\ell}\bar{Z}_{j'k,\ell}
-(n'_k-1)\sigma_{jj'}  \bigg\vert
>
\frac{(n'-2)\eta}{m_{max}}
\bigg)\\
&\leq &
\sum_{k=1}^2
\bigg \lbrace 
\Pr \bigg(
\bigg\vert \sum_{i=1}^{n'_k}Z_{ijk,\ell}Z_{ij'k,\ell}
-n'_k\sigma_{jj'}  \bigg\vert
>
\frac{(n'-2)\eta}{m_{max}}
\bigg)\\
&& +  
\Pr \bigg(
\vert n'_k\bar{Z}_{jk,\ell}\bar{Z}_{j'k,\ell}
-\sigma_{jj'}  \vert
>
\frac{(n'-2)\eta}{m_{max}}
\bigg) 
\bigg \rbrace .
\end{eqnarray*}
Finally, using Lemma \ref{lem:A3bickel}, 
\begin{eqnarray*}
&&	\sum_{j,j'\in \calS_{n,t}^{(2)} } 
\Pr \bigg( 
\vert \hsigma_{jj',\ell}^{(2)}-\sigma_{jj'}  \vert 
> 
\frac{\eta}{m_{max}}
\bigg)\\
&\leq & 
\sum_{k=1}^2
C_1p^2 
\exp \bigg \lbrace  -C_2 \frac{(n-2)^2\eta^2}{m_{max}^2n_k}  \bigg \rbrace
+
C'_1p^2
\exp \bigg \lbrace  -C'_2 \frac{(n-2)^2\eta^2}{m_{max}^2}  
\bigg \rbrace ,
\end{eqnarray*} 
where $C_1,C'_1,C_2,C'_2$ are some positive constants. 
If $m_{max}\sqrt{\log p / n_1}=o(1)$  
and by taking 
$\eta=M\times m_{max}\sqrt{\log p/ n_1}$, 
for sufficiently large $M>0$, 
the desired result is obtained. 
$\blacksquare$\\

\noindent
\begin{lem}
\label{lem:T_bound}
Suppose that $T$ has the Student's t-distribution with 
$n>1$ degrees of freedom. 
Then, for any large constant $\tau>0$, we have
\begin{eqnarray*}
	\Pr (T>\tau) 
	\leq 
	\frac{c_n}{\tau} 
	\frac{n}{n-1}
	\left( 1+\frac{1}{n}\tau^2  \right) ^{-\frac{n-1}{2}}, 
\end{eqnarray*}
where $c_n=\frac{\Gamma(\frac{n+1}{2})}{\Gamma(\frac{n}{2})\sqrt{n\pi}}$, 
and $\Gamma(.)$ is the gamma function. 
\end{lem}
\noindent
\textbf{Proof}.
For any $\tau>0$, 
\begin{eqnarray*}
\Pr (T>\tau)
&=&
\int _{\tau}^\infty 
\frac{c_n}
{(1+\frac{x^2}{n})^{\frac{n+1}{2}}} dx 
< 
\int _{\tau}^\infty 
\frac{x}{\tau}
\frac{c_n}{(1+\frac{x^2}{n})^{\frac{n+1}{2}}} dx
\\
& =& 
\frac{c_n}{\tau} 
\frac{n}{n-1}
\left( 1+\frac{1}{n}\tau^2  \right) ^{-\frac{n-1}{2}}~.
\end{eqnarray*}
The result follows from the facts that 
$\tau>0$ and $\tau<x<\infty$. $\blacksquare$


\section{Proofs of the main results}
\label{app:proofs}
In this Appendix, we provide the proofs of Theorems \ref{thrm:LDA}-\ref{thrm:ROAD}. \\

\noindent
\textbf{Proof of Theorem \ref{thrm:LDA}}.
Let $\pmb{\epsilon}_{ik}=\XX_{ik}-\bmu_k$, 
for $i=1,...,n_k$, and $k=1,2$, 
where $\XX_{ik} = (\XX_i | Y_i = k) \sim N_p(\bmu_k, \bSigma)$, 
and the vectors 
$\bbeps_k=(\bar{\epsilon}_{1k},\bar{\epsilon}_{2k},...,\bar{\epsilon}_{pk}) ^{\top}$ 
with entires 
$\bar{\epsilon}_{jk}=\frac{1}{n_k}\sum _{i=1}^{n_k}{\epsilon} _{ijk}$. 
Also, recall $\Delta_p^2= \bmu_d^{\top} \bSigma^{-1} \bmu_d$ and 
$\bmu_d = \bmu_2-\bmu_1$. 
The quantities $\Psi_1^{\text{\sc lda}}(\hbtheta_n)$, $\Psi_2^{\text{\sc lda}}(\hbtheta_n)$, and 
$\Upsilon^{\text{\sc lda}}(\hbtheta_n)$ in \eqref{ccrk} can be decomposed as
\begin{eqnarray*}
\Psi _1^{\text{\sc lda}}(\hbtheta_n) 
&=& 
(\bmu_1-\hbmu_a)^{\top} {\bSigma}^{-1}(\hbmu_2-\hbmu_1) 
\\
&=&
\frac{1}{2}
(-\bbeps_2-\bbeps_1-\bmu_d)^{\top}
\bSigma^{-1}
(\bbeps_2-\bbeps_1+\bmu_d) \\
&=&  
\frac{1}{2} 
\left\lbrace \bbeps_1^{\top}\bSigma^{-1}\bbeps_1
-\bbeps_2^{\top}\bSigma^{-1}\bbeps_2
-2\bbeps_2^{\top}\bSigma^{-1}\bmu_d
-\bmu_d^{\top}\bSigma^{-1}\bmu_d \right\rbrace \\
&=&  
\frac{1}{2} 
\left\lbrace 
\calI_1 - \calI_2 - 2\calI_3
-\bmu_d^{\top}\bSigma^{-1}\bmu_d  
\right\rbrace ,
\end{eqnarray*}
\begin{eqnarray*}
\Psi _2^{\text{\sc lda}}(\hbtheta_n) 
&=& 
-(\bmu_2-\hbmu_a)^{\top} \bSigma^{-1}(\hbmu_2-\hbmu_1)
\\
&=&
-\frac{1}{2}
(-\bbeps_2-\bbeps_1+\bmu_d)^{\top}
\bSigma^{-1}
(\bbeps_2-\bbeps_1+\bmu_d)\\
&=& 
-\frac{1}{2} 
\left\lbrace 
-\bbeps_2^{\top}\bSigma^{-1}\bbeps_2
+\bbeps_1^{\top}\bSigma^{-1}\bbeps_1
-2\bbeps_1^{\top}\bSigma^{-1}\bmu_d
+\bmu_d^{\top}\bSigma^{-1}\bmu_d 
\right\rbrace \\
&=&  
\frac{1}{2} 
\left\lbrace 
\calI_2-\calI_1+2\calI_4
-\bmu_d^{\top}\bSigma^{-1}\bmu_d  
\right\rbrace ,
\end{eqnarray*}
and
\begin{eqnarray*}
\Upsilon ^{\text{\sc lda}}(\hbtheta_n) 
&=& 
(\hbmu_{2}-\hbmu_{1})^{\top} 
\bSigma^{-1}\bSigma \bSigma^{-1}
(\hbmu_{2}-\hbmu_{1}) 
\\
&=&
(\bbeps_2-\bbeps_1+\bmu_d)^{\top}  
\bSigma^{-1} 
(\bbeps_2-\bbeps_1+\bmu_d) \\
&=&  
(\bbeps_2-\bbeps_1)^{\top}
\bSigma^{-1}(\bbeps_2-\bbeps_1)
+2(\bbeps_2-\bbeps_1)^{\top}\bSigma^{-1}\bmu_d
+\bmu_d^{\top}\bSigma^{-1}\bmu_d   \\
&=&  
\calI_5+2\calI_6
+\bmu_d^{\top}\bSigma^{-1}\bmu_d 
.
\end{eqnarray*}
We first show that
\begin{eqnarray*}
\calI_1
=
\bbeps_1^{\top}\bSigma^{-1}\bbeps_1
=p/n_1+o_p(\sqrt{p/n_1}).
\end{eqnarray*}
Note that $\bbeps_1\sim N_p(\textbf{0}, n_1^{-1}\bSigma) $. 
By Chebyshev's inequality, for any $\tau >0$, 
\begin{eqnarray*}
\Pr\bigg(
\sqrt{\frac{n_1}{p}} \mid \calI_1-\frac{p}{n_1} \mid >\tau
\bigg)
\leq 
\frac{1}{\tau ^2}Var \lbrace\calI_1.\sqrt{n_1/p}\rbrace .
\end{eqnarray*}
This together with the fact that 
$Var \lbrace\calI_1.\sqrt{n_1/p}\rbrace \to 0$, 
when $n_1,n_2\to \infty$ such that $n_2=o(n_1)$, implies that   
$\calI_1=p/n_1+o_p(\sqrt{p/n_1})$. 
Similarly, we have 
\[
\calI_2 =  p/n_2+o_p(\sqrt{p/n_2})~~,~~
\calI_3=\bbeps_2^{\top}\bSigma^{-1}\bmu_d
=O_p\bigg( \sqrt{\Delta_p^2/n_2}\bigg),
\]
\[
\calI_4=\bbeps_1^{\top}\bSigma^{-1}\bmu_d
=O_p\bigg(\sqrt{\Delta_p^2/n_1}\bigg),\] 
\[
\calI_5=(\bbeps_2-\bbeps_1)^{\top}\bSigma^{-1}(\bbeps_2-\bbeps_1)
= \sqrt{\frac{np}{n_1n_2}} o_p(1)+\frac{np}{n_1n_2},
\]
and 
\[
\calI_6=(\bbeps_2-\bbeps_1)^{\top}\bSigma ^{-1}\bmu_d
=O_p\bigg ( \sqrt{\frac{n}{n_1n_2}\Delta_p^2}~ \bigg).
\]

By combining the above results, we have
\begin{eqnarray*}
\frac{\Psi _1^{\text{\sc lda}} (\hbtheta_n)}{\sqrt{\Upsilon^{\text{\sc lda}}  (\hbtheta_n)}}
&=&
\frac{ \calI_1-\calI_2-2\calI_3-\Delta_p^2 }
{ 2\left\lbrace \calI_5+2\calI_6+\Delta_p^2 \right\rbrace^{1/2}}\\
&=& 
\frac{  \frac{p}{n_1}
	+o_p(\sqrt{p/n_1})
	-\frac{p}{n_2}
	+o_p(\sqrt{p/n_2})
	+O_p(\sqrt{\Delta_p^2 /n_2})
	-\Delta_p^2}
{ 2\left\lbrace  \sqrt{\frac{np}{n_1n_2}}o_p(1)
	+\frac{np}{n_1n_2}
	+O_p( \sqrt{n\Delta_p^2/n_1n_2}) 
	+\Delta_p^2 \right\rbrace ^{1/2}
}\\
&=&
\frac{  -\sqrt{\frac{p}{n_2}}(1-\frac{n_2}{n_1})
	+o_p(\sqrt{n_2/n_1})
	+O_p(\sqrt{\Delta_p^2 /p})
	-\sqrt{\frac{n_2}{p}}\Delta_p^2}
{ 2\left\lbrace 1
	+o_p(\sqrt{n_2/p})
	+O_p(\sqrt{n_2\Delta_p^2}/p)
	+n_2\Delta_p^2/p\right\rbrace ^{1/2}
}\\
\end{eqnarray*}
and
\begin{eqnarray*}
\frac{ \Psi _2^{\text{\sc lda}} (\hbtheta_n)}{\sqrt{\Upsilon ^{\text{\sc lda}} (\hbtheta_n)}}
&=&
\frac{\calI_2-\calI_1+2\calI_4-\Delta_p^2 }
{2\left\lbrace     \calI_5+2\calI_6+\Delta_p^2  \right\rbrace  ^{1/2}}\\
&=& 
\frac{  \frac{p}{n_2}
	-\frac{p}{n_1}
	+o_p(\sqrt{p/n_2})
	+o_p(\sqrt{p/n_1})
	+O_p(\sqrt{\Delta_p^2 /n_1})
	-\Delta_p^2 }
{  2\left\lbrace  \sqrt{\frac{np}{n_1n_2}}o_p(1)
	+\frac{np}{n_1n_2}
	+O_p( \sqrt{n\Delta_p^2/n_1n_2}) 
	+\Delta_p^2 \right\rbrace ^{1/2}
}\\
&=& 
\frac{    \sqrt{\frac{p}{n_2}}(1-\frac{n_2}{n_1})
	+o_p(\sqrt{n_2/n_1})
	+O_p(\sqrt{n_2\Delta_p^2/pn_1})
	-\sqrt{\frac{n_2}{p}}\Delta_p^2}
{   2\left\lbrace 1
	+o_p(\sqrt{n_2/p})
	+O_p(\sqrt{n_2\Delta_p^2}/p)
	+n_2\Delta_p^2/p\right\rbrace ^{1/2}
}.
\end{eqnarray*}
Since $\sqrt{\frac{n_2}{p}}\Delta_p^2=o(1)$, 
as long as $n_1,n_2 \to \infty$, thus we obtain 
\begin{eqnarray*}
\begin{matrix}
	\frac{\Psi _1^{\text{\sc lda}} (\hbtheta_n)}{\sqrt{\Upsilon^{\text{\sc lda}}  (\hbtheta_n)}}\overset{p}{\longrightarrow} -\infty , 
	&&
	\frac{\Psi _2^{\text{\sc lda}} (\hbtheta_n)}{\sqrt{\Upsilon^{\text{\sc lda}}  (\hbtheta_n)}}\overset{p}{\longrightarrow} +\infty . 
\end{matrix}
\end{eqnarray*}
Hence, $\Pi_1^{\text{\sc lda}}(\calD_n)\overset{p}{\longrightarrow} 0$ 
and $\Pi_2^{\text{\sc lda}}(\calD_n)\overset{p}{\longrightarrow} 1$, which completes the proof. 
$\blacksquare$\\ 

\noindent
\textbf{Proof of Lemma \ref{lem:true_actives}.}	
\textbf{(a)} 
Note that 
\begin{eqnarray*}
\Pr \bigg( 
\bigcap _{j\not\in \calS} 
\lbrace \vert t_j \vert \leq \tau_n \rbrace 
\bigg) 
=
1- \Pr\bigg( \smash{\displaystyle \max _{j\not\in \calS}} 
\vert t_j \vert ~> \tau_n \bigg). 
\end{eqnarray*}
By Lemma \ref{lem:T_bound} of the Appendix \ref{app:lemma}, 
with 
$c_n=\frac{\Gamma(\frac{n-1}{2})}{\Gamma(\frac{n-2}{2})\sqrt{(n-2)\pi}}$, 
we have 
\begin{eqnarray*}
\Pr \bigg( 
\smash{\displaystyle \max _{j\not\in \calS}} 
\vert t_j \vert ~ > \tau_n
\bigg)   
&\leq &  
\sum _{j\not\in \calS}  
\Pr \bigg( \vert t_j \vert > \tau_n \bigg) \\
&\leq & 
(p-s)\frac{n-2}{n-3}
\frac{c_n}{\tau_n} 
\bigg(1+\frac{1}{n-2}\tau_n ^2 \bigg)^{-\frac{n-3}{2}}\\
&:=& 
u(n_1,n_2,p-s,\tau_n),
\end{eqnarray*}
where $n=n_1+n_2$. The last inequality follows from 
the upper bound described in Lemma \ref{lem:T_bound}, 
for the tail of a Student's t-distributed 
random variable, with $n-2$ degrees of freedom.
Since $n_2=o(n_1)$ as $n_1,n_2\to \infty $, we then obtain 
\begin{eqnarray*}
u(n_1,n_2,p-s,\tau_n)
\sim 
\frac{p-s}{\tau_n}
\left( 1+\frac{1}{n_1}\tau_n ^2  \right)^{-n_1}
\end{eqnarray*}
and hence, as $n_1 \to \infty$,  
\begin{eqnarray*}
u(n_1,n_2,p-s,\tau_n)
\sim 
\frac{p-s}{\tau_n}e^{ -\tau_n ^2 }.
\end{eqnarray*}
Since $\log( p-s)=o(\tau_n ^2)$, therefore 
$\Pr(
\smash{\displaystyle \max _{j \not\in \calS}} 
\vert t_j \vert ~> \tau_n
)\to 0$, and this completes the proof. 

\textbf{(b)} Note that 
\begin{eqnarray*}
\Pr \bigg( 
\bigcap _{j\in \calS} 
\lbrace \vert t_j \vert > \tau_n \rbrace 
\bigg)
=
\Pr \bigg(
\smash{\displaystyle \min _{j\in \calS}} 
\vert t_j \vert ~> \tau_n
\bigg) 
=
1- 
\Pr\bigg( 
\smash{\displaystyle \min _{j\in \calS}} 
\vert t_j \vert ~\leq \tau_n
\bigg).
\end{eqnarray*}
Let $ \widetilde{t}_j 
=
t_j 
- \frac{\mu_{dj}}
{\hat{\sigma}_j\sqrt{n/n_1n_2}}$. We have  
\begin{eqnarray*}
\Pr \bigg(
\smash{\displaystyle \min _{j\in \calS}} 
\vert t_j \vert \leq \tau_n
\bigg)
&=&
\Pr \bigg( 
\smash{\displaystyle \max _{j\in \calS}} 
\vert  \tilde{t}_j \vert
\geq 
\smash{\displaystyle \min _{j\in \calS}} ~
\frac{\vert \mu_{dj}
	\vert}{\hat{\sigma}_j\sqrt{n/(n_1n_2)}} - \tau_n
\bigg)
\\
&\leq & 
\sum _{j\in \calS}    
\Pr \bigg( 
\vert  \tilde{t}_j \vert 
\geq  
\smash{\displaystyle \min _{j\in \calS}}~ 
\frac{ \vert \mu_{dj}
	\vert}{\hat{\sigma}_j\sqrt{n/(n_1n_2)}}  
-\tau_n 
\bigg). 
\end{eqnarray*}
Also by Lemma \ref{lem:Dhat} and under the Condition (C2),  
\begin{eqnarray*}
\smash{\displaystyle \min _{j\in \calS}}~
\frac{ \vert \mu_{dj}
	\vert}{\hat{\sigma}_j\sqrt{n/(n_1n_2)}}
=
d_{0,n}(1+o_p(1)).
\end{eqnarray*}
Hence, 
\begin{eqnarray*}
&&\Pr \bigg(
\smash{\displaystyle \min _{j\in \calS}} 
\vert t_j \vert ~\leq \tau_n
\bigg)
\leq 
\sum_{j\in \calS}
\Pr \bigg( 
\vert \tilde{t}_j \vert ~> \frac{d_{0,n}}
{\sqrt{n/n_1n_2}}(1+o_p(1))-\tau_n 
\bigg)\\
& \leq & 
s~
\frac{n-2}{n-3}~
\frac{c_n}
{\frac{d_{0,n}(1+o_p(1))}{\sqrt{n/n_1n_2}}-\tau_n}
\bigg( 1+
\frac{1}{n-2}\bigg[\frac{d_{0,n}(1+o_p(1))}{\sqrt{n/n_1n_2}}-\tau_n\bigg]^2 \bigg) ^{-\frac{n-3}{2}}
\\
& := & 
u(n_1,n_2,s,d_{0,n},\tau_n),
\end{eqnarray*}
where the last inequality follows from Lemma \ref{lem:T_bound}, 
when $\tau_n=O(\sqrt{n_2}d_{0,n})$.
Since $\sqrt{n_2}d_{0,n}\to \infty$, $\log s=o(n_2d_{0,n}^2)$, and $n_2=o(n_1)$, then 
as $n_1,n_2 \to \infty$, we have 
$\sqrt{\frac{n}{n_1n_2}}
\sim 
\frac{1}{\sqrt{n_2}} $, 
$ \frac{n-2}{n-3}\sim 1$ 
and $c_n=\frac{\Gamma(\frac{n-1}{2})}
{\Gamma(\frac{n-2}{2})\sqrt{(n-2)\pi}} 
\to \frac{1}{\sqrt{2\pi}}$. Therefore,   
\begin{eqnarray*}
u(n_1,n_2,s,d_{0,n},\tau_n) \to 0
\end{eqnarray*}
and it completes the proof. $\blacksquare$ \\

\noindent
\textbf{Proof of Theroem \ref{thrm:DACH}}.
\textbf{(a)} The class-specific misclassification rates of Msplit-{\sc hr} in \eqref{DACH} are given by 
\begin{eqnarray*}
\Pi_k^{\text{Msplit-{\sc hr}}}(\calD_n)
=
\Phi \bigg(  
\frac{ \Psi_k^{\text{Msplit-{\sc hr}}}(\hbtheta_n)}
{\sqrt{\Upsilon^{\text{Msplit-{\sc hr}}}(\hbtheta_n)}}
\bigg), ~~k=1,2, 
\end{eqnarray*}
where 
\begin{eqnarray*}
\Psi_k^{\text{Msplit-{\sc hr}}}(\hbtheta_n) 
&=&
\frac{(-1)^{k+1}}{\calL}\sum_{\ell=1}^{\calL}\sum_{j=1}^p
\bigg\{ r_j(\bmu_k;\hbtheta_{n,\ell}^{(2)}) -\frac{\bar{r}_n}{2} \bigg\} h_j(\hbtheta_{n,\ell}^{(1)}), 
\\
\Upsilon^{\text{Msplit-{\sc hr}}}(\hbtheta_n)  
&=&
\frac{1}{\calL^2}\sum_{\ell=1}^{\calL}\sum_{j=1}^p
\sigma_j^2\bigg(\hmu_{dj,\ell}^{(2)}/\hsigma_{j,\ell}^{(2),2}\bigg)^2 h_j(\hbtheta_{n,\ell}^{(1)}).
\end{eqnarray*}

By Lemma \ref{lem:true_actives}, 
if $\sqrt{n_2}d_{0,n}\to 0$, 
$\tau_n=O(\sqrt{n_2}d_{0,n})$, 
$\log (p-s)=o(\tau_n^2)$, and 
$\log s=o(n_2d_{0,n})$, 
as $n_1,n_2\to \infty$, then  
\[
\max_{j\in\calS} \bigg\vert h_j(\hbtheta_{n,\ell}^{(1)})-1\bigg\vert \overset{p}{\longrightarrow} 0~~, ~~
\max_{j\not\in\calS} h_j(\hbtheta_{n,\ell}^{(1)}) \overset{p}{\longrightarrow} 0. 
\]
Using these results, for any $\epsilon >0$, 
we have, for $k=1,2$, 
\begin{equation*}
\Pr\left( 
\bigg\vert 
\sum_{j\not\in\calS} 
r_j(\bmu_k;\hbtheta_{n,\ell}^{(2)})
h_j(\hbtheta_{n,\ell}^{(1)}) 
\bigg\vert >\epsilon 
\right) 
\leq   
\Pr \bigg(  \max_{j\not\in\calS}  h_j(\hbtheta_{n,\ell}^{(1)}) >\epsilon  \bigg) \overset{p}{\longrightarrow}  0 ,
\end{equation*}
and consequently, 
\begin{eqnarray*}
\Psi_k^{\text{Msplit-{\sc hr}}}(\hbtheta_{n}) 
=
\frac{(-1)^{k+1}}{\calL}\sum_{\ell=1}^{\calL}\sum_{j\in\calS}
\bigg\lbrace r_j(\bmu_k;\hbtheta_{n,\ell}^{(2)}) -\frac{\bar{r}_n}{2} \bigg\rbrace 
(1+o_p(1)), ~~ k=1,2.
\end{eqnarray*} 
Similarly, we have   
\begin{eqnarray*}
\Upsilon^{\text{Msplit-{\sc hr}}}(\hbtheta_{n})  
&=&
\frac{1}{\calL^2}\sum_{\ell=1}^{\calL}\sum_{j\in\calS}
\sigma_j^2 \bigg(\hmu_{dj,\ell}^{(2)}/\hsigma_{j,\ell}^{(2),2}\bigg)^2 
(1+o_p(1)).
\end{eqnarray*} 

Let  
$\bar{\epsilon}_{jk,\ell}^{(2)}=\hmu_{jk,\ell}^{(2)}-\mu_{jk}$,  
$\calI_{k,\ell}=\sum_{j\in\calS}(\bar{\epsilon}_{jk,\ell}^{(2)}/\sigma_{j})^2$, 
for $k=1,2$, and 
$\calI_{3,\ell}=\sum_{j\in\calS}(\bar{\epsilon}_{j2,\ell}^{(2)}\mu_{dj}/\sigma_{j}^2)$, 
for each $\ell=1,..,\calL$. 
By the result of Lemma \ref{lem:Dhat} in 
the Appendix \ref{app:lemma}, we have  
\begin{equation}
\sum_{j\in\calS} 
r_j(\bmu_{j1},\hbtheta_{n,\ell}^{(2)}) h_j(\hbtheta_{n,\ell}^{(1)}) 
=
\frac{1}{2}\bigg\lbrace
\calI_{1,\ell}-\calI_{2,\ell}-2\calI_{3,\ell}-\Delta_p^2  
\bigg\rbrace \bigg( 1+O_p(\sqrt{\log p/n_1}) \bigg),
\label{dec_1}
\end{equation}
\begin{equation}
\sum_{j\in\calS} 
r_j(\bmu_{j2},\hbtheta_{n,\ell}^{(2)}) h_j(\hbtheta_{n,\ell}^{(1)}) 
=
\frac{1}{2}\bigg\lbrace
\calI_{1,\ell}-\calI_{2,\ell}-2\calI_{4,\ell}+\Delta_p^2  
\bigg\rbrace \bigg( 1+O_p(\sqrt{\log p/n_1}) \bigg),
\label{dec_2}
\end{equation}
where $\Delta_p^2=\sum_{j\in\calS}(\mu_{dj}^2/\sigma_j^2)$. 
Now, for $\eta>0$, and $k=1,2$
\begin{equation}
\label{prob_1}
\Pr\bigg( \vert \calI_{k,\ell} \vert >\eta \bigg) \leq \frac{s}{n_k\eta}, 
\end{equation}
by taking $\eta=M.s/n_k$, for sufficiently large $M>0$, 
then $\calI_{k,\ell}=O_p(s/n_k)$, for $k=1,2$.. 
By Cauchy-Schwartz inequality, we have 
$\calI_{3,\ell}=O_p(\Delta_p\sqrt{s/n_2})$ and 
$\calI_{4,\ell}=O_p(\Delta_p\sqrt{s/n_1})$. 
In addition,  we have 
$\sum_{j=1}^p\bar{r}_nh_j(\hbtheta_{n,\ell}^{(2)})=o_p(s/n_2)$.
By combining these results in 
\eqref{dec_1}-\eqref{dec_2}, we arrive at
\begin{equation}
\Psi_k^{\text{Msplit-{\sc hr}}}(\hbtheta_n)
= 
O_p(s/n_2)+O_p\bigg(\Delta_p\sqrt{s/n_2}\bigg)-\frac{1}{2}\Delta_p^2+
O_p\bigg(\Delta_p^2\sqrt{\log p/n_1}\bigg), 
\label{ine3}
\end{equation}
for $k=1,2$. 
Let 
$\calI_{5,\ell}=\sum_{j\in\calS}(\bar{\epsilon}_{j2,\ell}-\bar{\epsilon}_{j1,\ell})^2/\sigma_j^2$, 
$\calI_{6,\ell}=\sum_{j\in\calS}(\bar{\epsilon}_{j2,\ell}-\bar{\epsilon}_{j1,\ell})\mu_{dj}/\sigma_j^2$ for each $\ell=1,..,\calL$. 
Similar to \eqref{prob_1}, we result 
$\calI_{5,\ell}=O_p(s/n_2)$ and also 
$\calI_{6,\ell}=O_p(\Delta_p\sqrt{s/n_2})$. 
Therefore 
\begin{eqnarray}
&&\Upsilon ^{\text{Msplit-{\sc hr}}}(\hbtheta_n)
=
\frac{1}{\calL^2}
\sum_{\ell=1}^{\calL} \bigg\lbrace
\calI_{5,\ell}+2\calI_{6,\ell}+\Delta_p^2
\bigg\rbrace
\nonumber\\
&=&
O_p(s/n_2)+O_p\bigg(\Delta_p\sqrt{s/n_2}\bigg) 
+\Delta_p^2 +O_p\bigg(\Delta_p^2\sqrt{\log p/n_1}\bigg).
\label{ine5}
\end{eqnarray}
By combining \eqref{ine3} and \eqref{ine5}, 
we have, for $k=1,2$,  
\begin{eqnarray*}
&&\Pi_{k}^{\text{Msplit-{\sc hr}}}(\calD_n)
\\
&=&\Phi \left(   
\frac{O_p(s/n_2)+O_p(\Delta_p\sqrt{s/n_2})-\frac{1}{2}\Delta_p^2+
	O_p\bigg(\Delta_p^2\sqrt{\frac{\log p}{n_1}}\bigg)}
{\bigg\lbrace O_p(s/n_2)+O_p(\Delta_p\sqrt{s/n_2}) 
	+\Delta_p^2 +O_p\bigg(\Delta_p^2\sqrt{\frac{\log p}{n_1}}\bigg) 
	\bigg\rbrace ^{1/2}}
\right) \nonumber\\
& =&
\Phi \bigg( -\frac{1}{2}\Delta_p
\lbrace 1+O_p(\kappa_n) \rbrace  \bigg),
\label{kapa_1}
\end{eqnarray*}
where 
$\kappa_n=\max \lbrace \Delta_p^{-1}\sqrt{s/n_2},~
\sqrt{\log p/n_1} \rbrace$. 

\textbf{(b)}
When $\Delta_p\to \infty$, 
by Lemma \ref{lem:lem1_shao}, 
if $\Delta_p^2\kappa_n=o(1)$, then 
Msplit-{\sc hr} is asymptotically-strong optimal 
and the result follows.  
The condition $\Delta_p^2\kappa_n=o(1)$ 
is equivalent to $\Delta_p^2\sqrt{\log p/n_1}=o(1)$, 
and 
$\Delta_p^2s=o(n_2)$.
$\blacksquare$ \\

\noindent 
\textbf{Proof of Lemma \ref{lem:general_DACH}.}
We follow a similar line of proof as in 
\citep[Theorem 1]{pan2016ultrahigh}, 
to show the results of both parts (a) and (b), 
under the imbalanced setting.

\textbf{(a)} 
It is enough to show that for any 
$\ell=1,...,\calL$, as $n_1,n_2\to \infty$, 
\begin{eqnarray*}
\Pr \bigg( \calS \not\subset \calS_{n,\ell}^{(1)} \bigg)\to 0.
\end{eqnarray*}
Suppose that there exist an index $j$ in $\calS$ for which $j\not\in\calS_{n,\ell}^{(1)}$. 
Thus, $\vert \mu_{dj} \vert \geq d_{0,n}$ and $\vert  \hat{\mu}_{dj,\ell}^{(1)} \vert < \tau_n$, 
where $d_{0,n}=\min _{j\in \calS} \vert  \mu_{dj} \vert$. 
It results in 
$\vert \hat{\mu}_{dj,\ell}^{(1)}-\mu_{dj}\vert >d_{0,n}-\tau_n$. 
By conditions $\tau_n\asymp d_{0,n}$ and $\lambda_{\max}(\bSigma)<c_0$, and for some 
constants $C_1,C_2>0$, we have 
\begin{eqnarray*}
&&\Pr \bigg( \calS \not\subset \calS_{n,\ell}^{(1)} \bigg) 
\leq 
\sum_{j=1}^p\Pr \bigg(  \vert  \hat{\mu}_{dj,\ell}^{(1)}-\mu_{dj} \vert >d_{0,n}-\tau_n\bigg)\\
&\leq &
C_1 \bigg (\frac{p}{d_{0,n}-\tau_n} \bigg) \sqrt{\frac{n'_1+n'_2}{n'_1n'_2}}~
\exp \bigg\lbrace  -C_2 \frac{n'_1n'_2(d_{0,n}-\tau_n)^2}{n'_1+n'_2} \bigg\rbrace.
\end{eqnarray*}
The last term tends to zero, since $\log p =o(n_2d_{0,n}^2)$ and $n_2=o(n_1)$,  
and thus the result follows. 

\noindent
\textbf{(b)}
By condition $\lambda_{\max}(\bSigma)<c_0$, 
we have 
\begin{eqnarray}
\bmu_d^{\top}\bmu_d = 
\bmu_d^{\top} \bSigma^{-1} \bSigma\bSigma^{\top} \bSigma^{-1}\bmu_d 
\leq 
\lambda_{\max}(\bSigma\bSigma^{\top})\bbeta^{\top}\bbeta
\leq
c_0 \times \vert  \calS \vert  \times \max_{j\in \calS} \beta_j^2 \label{ge1}.
\end{eqnarray}
Let $\calS^{*}=\lbrace j: \vert  \mu_{dj} \vert >d_{0,n}/r  \rbrace$, 
for some constant $r>1$. Thus, 
$\bmu_d^{\top}\bmu_d 
\geq \vert  \calS^{*}\vert d_{0,n}^2/r^2 $.  
This together with \eqref{ge1}, result in 
$\vert \calS^{*} \vert \leq C_3 
\vert \calS \vert\max_{j\in \calS} \beta_j^2 /d_{0,n}^2 \doteq m_{\max}$, 
for constant $C_3>0$. 
The result in part (b) follows by proving that, 
$\vert \calS_{n,\ell}^{(1)}\vert <\vert  \calS^{*} \vert $, 
with probability tending to one, 
for any $\ell=1,...,\calL$.  
If there exists an index $j$ in $\calS_{n,\ell}^{(1)}$ for which 
$j\not\in \calS^{*} $, thus $\vert  \hmu_{dj,\ell}^{(1)}\vert >\tau_n$ 
and $\vert \mu_{dj}\vert <d_{0,n}/r$ and consequently, 
$\vert  \hmu_{dj,\ell}^{(1)}-\mu_{dj} \vert  > \tau_n-d_{0,n}/r$. 
Therefore, by condition $\tau_n\asymp d_{0,n}$ and for constants $C_4,C_5>0$
\begin{eqnarray*}
\Pr \bigg( \vert \calS_{n,\ell}^{(1)}\vert \geq \vert  \calS^{*} \vert   \bigg) 
&\leq &
\Pr \bigg( \calS_{n,\ell}^{(1)} \not\subset  \calS^{*} \bigg) 
\leq 
\sum_{j=1}^p
\Pr \bigg(  \vert  \hmu_{dj,\ell}^{(1)}-\mu_{dj} \vert  > \tau_n-d_{0,n}/r\bigg) \\
&\leq &
C_4\frac{p}{\tau_n-d_{0,n}}\sqrt{\frac{n'_1+n'_2}{n'_1n'_2}} 
\exp \bigg\lbrace -C_5 \frac{(\tau_n-d_{0,n})^2n'_1n'_2}{n'_1+n'_2}  \bigg\rbrace.
\end{eqnarray*}
The last term tends to zero, as $\log p=o(n_2d_{0,n}^2)$ and  
$\sqrt{n_2}d_{0,n}\to \infty$. $\blacksquare$ \\	

\noindent
\textbf{Proof of Theroem \ref{thrm:geneal_optimal}}.
\textbf{(a)}
The misclassification rates of Msplit-{\sc hr} in \eqref{DACH-general}, 
are given as 
\begin{eqnarray*}
\Pi_k^{\text{Msplit-{\sc hr}}}(\calD_n)
=
\Phi \bigg( 
\frac{\Psi_k^{\text{Msplit-{\sc hr}}}(\hbtheta_n)}
{\sqrt{\Upsilon^{\text{Msplit-{\sc hr}}}(\hbtheta_n)}}
\bigg), ~~ k=1,2
\end{eqnarray*}
where 
\begin{eqnarray*}
\Psi_k^{\text{Msplit-{\sc hr}}}(\hbtheta_n)
&=&
\frac{(-1)^k}{\calL}\sum_{\ell=1}^{\calL}
\lbrace
\tbmu_{d,\ell}^{\top} \tbSigma_{n,\ell}^{-1} 
(\tbmu_{a,\ell}-\bmu_{k,\ell})
-\frac{\bar{r}_{n,\ell}}{2} 
\rbrace , \\
\bar{r}_{n,\ell}
&=&\bigg ( \frac{1}{n'_1}-\frac{1}{n'_2} \bigg)
\frac{n'-2}
{n'-3- \vert \calS_{n,\ell}^{(1)} \vert } \times 
\vert \calS_{n,\ell}^{(1)} \vert ,\\
\Upsilon^{\text{Msplit-{\sc hr}}}(\hbtheta_n)
&=&
\frac{1}{\calL^2}\sum_{\ell=1}^{\calL} 
\lbrace
\tbmu_{d,\ell}^{\top} \tbSigma_{n,\ell}^{-1}
\bSigma_{\ell}
\tbSigma_{n,\ell}^{-1}\tbmu_{d,\ell} 
\rbrace .
\end{eqnarray*}
By Lemma \ref{lem:pan2016}, we obtain
\begin{eqnarray}
\label{denum_1}
\tbmu_{d,\ell}^{\top} \tbSigma_{n,\ell}^{-1}
\bSigma_{\ell}
\tbSigma_{n,\ell}^{-1}\tbmu_{d,\ell} =
\tbmu_{d,\ell}^{\top} \bSigma_{\ell}^{-1}\tbmu_{d,\ell} 
\bigg(1+O_p(m_{max}\sqrt{\log p/n_1})\bigg).
\end{eqnarray}
We consider the following decomposition
\begin{eqnarray*}
\tbmu_{d,\ell}^{\top} \bSigma_{\ell}^{-1}\tbmu_{d,\ell} 
&=&
(\tbmu_{d,\ell}-\bmu_{d,\ell})^{\top} 
\bSigma_{\ell}^{-1}(\tbmu_{d,\ell}-\bmu_{d,\ell})
+
2(\tbmu_{d,\ell}-\bmu_{d,\ell})^{\top}\bSigma_{\ell}^{-1}\bmu_{d,\ell}
\\
&+&\bmu_{d,\ell}^{\top}\bSigma_{\ell}^{-1}\bmu_{d,\ell}
=
\calA_1+2\calA_2+\calA_3
\end{eqnarray*} 
Now by Lemma \ref{lem:general_DACH} and Markov's inequality, 
also using the Condition (C2) in the Appendix \ref{app:lemma}, 
we have for a constant $C_1>0$,
\begin{eqnarray*}
\Pr\bigg( (\tbmu_{d,\ell}-\bmu_{d,\ell})^{\top} 
\bSigma_{\ell}^{-1}(\tbmu_{d,\ell}-\bmu_{d,\ell})  
> \eta \bigg)
\leq
\frac{C_1}{\eta}\frac{n}{n_1n_2}m_{max}.
\end{eqnarray*}
If $\eta=M\frac{m_{max}}{n_2}$, then for large $M>0$, 
$\calA_1=O_p(m_{max}/n_2)$. 
By Cauchy-Schwartz inequality, 
$\calA_2^2\leq (\tbmu_{d,\ell}-\bmu_{d,\ell})^{\top} 
\bSigma_{\ell}^{-1}(\tbmu_{d,\ell}-\bmu_{d,\ell}) 
\calA_3$. 
Hence $\calA_2=O_p(\sqrt{m_{max}/n_2})
\calA_3^{1/2}$. Therefore, by combining these results 
we have 
\begin{eqnarray}
\label{denum_2}
\tbmu_{d,\ell}^{\top} \bSigma_{\ell}^{-1}\tbmu_{d,\ell} 
=
O_p(m_{max}/n_2)
+O_p(\sqrt{m_{max}/n_2})\sqrt{\calA_3}
+\calA_3
\end{eqnarray}

Now for $\Psi_1^{\text{Msplit-{\sc hr}}}(\hbtheta_n)$, we have 
\begin{eqnarray}
\label{num_1}
\tbmu_{d,\ell}^{\top} \tbSigma_{n,\ell}^{-1} 
(\bmu_{1,\ell}-\tbmu_{a,\ell})
=
\tbmu_{d,\ell}^{\top} \bSigma_{\ell}^{-1} 
(\bmu_{1,\ell}-\tbmu_{a,\ell})
\bigg(1+O_p(m_{max}\sqrt{\log p/n_1})\bigg)
\end{eqnarray}
We decompose it as 
\begin{eqnarray*}
2\tbmu_{d,\ell}^{\top} \bSigma_{\ell}^{-1} 
(\bmu_{1,\ell}-\tbmu_{a,\ell})
&=&
(\tbmu_{1,\ell}-\bmu_{1,\ell})^{\top} \bSigma_{\ell}^{-1} 
(\tbmu_{1,\ell}-\bmu_{1,\ell})
\\&-&
(\tbmu_{2,\ell}-\bmu_{2,\ell})^{\top} \bSigma_{\ell}^{-1} 
(\tbmu_{2,\ell}-\bmu_{2,\ell})\\
&=&
2(\tbmu_{2,\ell}-\bmu_{2,\ell})^{\top} \bSigma_{\ell}^{-1} 
\bmu_{d,\ell}
-
\bmu_{d,\ell}^{\top}\bSigma_{\ell}^{-1} \bmu_{d,\ell}
\\
&=&
\calB_1-\calB_2-2\calB_3-\calA_3
\end{eqnarray*}
Similar to the proof of $\calA_1$, we have 
$\calB_1=O_p(m_{max}/n_1)$, and 
$\calB_2=O_p(m_{max}/n_2)$. Also similar 
to $\calA_2$, we have $\calB_3=O_p(\sqrt{m_{max}/n_2})\sqrt{\calA_3}$. 
Hence, 
\begin{equation}
\label{num_2}
\tbmu_{d,\ell}^{\top} \bSigma_{\ell}^{-1} 
(\bmu_{1,\ell}-\tbmu_{a,\ell})
=
O_p(\frac{m_{max}}{n_1})+O_p(\frac{m_{max}}{n_2})
+O_p(\sqrt{m_{max}/n_2})\calA_3^{1/2}
-\frac{1}{2}\calA_3 
\end{equation}
We recall that $\Delta_p^2=
\bmu_d^{\top}\bSigma^{-1}\bmu_d
=\bbeta^{\top}\bmu_d$ and 
$\calS_{n,\ell}^{(1)}=
\lbrace j:\vert \hbmu_{dj,\ell}^{(1)} \vert > \tau_n\rbrace$.  
For each $\ell=1,..,\calL$, and any $\eta>0$
\begin{eqnarray*}
&&\Pr \bigg(  
\vert \bmu_{d,\ell}^{\top}\bSigma_{\ell}^{-1}\bmu_{d,\ell}
-\Delta_p^2 \vert 
>\eta
\bigg)
=
\Pr \bigg(
\vert
\sum_{j\in \calS_{n,\ell}^{(1)}}\beta_{j}\mu_{dj}
-
\sum_{j'\in \calS}\beta_{j'}\mu_{dj'}
\vert >\eta
\bigg)\\
&=&
\Pr \bigg(
\bigg\vert ~~~
{\sum_{j\in \calS_{n,\ell}^{(1)}, j\not\in \calS}}  \beta_{j}\mu_{dj}
~ + ~
{\sum_{j\in \calS_{n,\ell}^{(1)}, j\in \calS}}  \beta_{j}\mu_{dj}
~ - ~
{\sum_{j'\in \calS, j'\in \calS_{n,\ell}^{(1)}}}  \beta_{j'}\mu_{dj'}
~ - ~
{\sum_{j'\in \calS, j'\not\in \calS_{n,\ell}^{(1)}}}  \beta_{j'}\mu_{dj'}
\bigg\vert
>\eta
\bigg)\\
&=&
\Pr \bigg(
\vert
\sum_{j\in \calS, j\not\in \calS_{n,\ell}^{(1)}}  \beta_{j}\mu_{dj}
\vert
>\eta
\bigg)
\leq 
\sum_{j=1}^p
\Pr \bigg(
j\in \calS \text{ and } j\not\in \calS_{n,\ell}^{(1)}
\bigg)
\end{eqnarray*}
By part (i) of Lemma \ref{lem:general_DACH}, 
the last term tends to zero, 
as $n_1,n_2\to\infty$. Therefore, 
$
\calA_3=\Delta_p^2+o_p(1). 
$
By combining this result 
together with \eqref{denum_1}-\eqref{num_2}, also with 
$\bar{r}_{n,t}
=O_p(m_{max}/n_2)$,  
we result  
\begin{eqnarray*}
\Pi_1^{\text{Msplit-{\sc hr}}}(\calD_n)
&=&
\Phi \bigg(
\frac{-\Delta_p}{2} 
\bigg\lbrace 1+O_p\bigg(\Delta_p^{-1}\sqrt{\frac{m_{max}}{n_2}}\bigg)
+O_p\bigg(m_{max}\sqrt{\frac{\log p}{n_1}}\bigg) \bigg\rbrace 
\bigg)\\
&=&
\Phi \bigg(
\frac{-\Delta_p}{2} 
( 1+O_p(\kappa'_n)) \bigg),
\end{eqnarray*}  
We can show the same result 
for $\Pi_2^{\text{Msplit-{\sc hr}}}(\calD_n)$.

\textbf{(b)} When $\Delta_p\to \infty$, 
the result follows from Lemma \ref{lem:lem1_shao} 
by condition $\Delta_p^2\kappa'_n=o(1)$.  
$\blacksquare$  \\

\noindent 
\textbf{Proof of Lemma \ref{lem:shao}}.
\textbf{(a)}
Recall the sequence 
$a_n=M_2(\log p/n)^{\alpha}$, 
with $0<\alpha<1/2$ and $M_2>0$. 
Let $c_1,c_2$ be some positive constants.  
Inspired by the proof of Lemma 2 of \cite{shao2011sparse}, 
we have 
\begin{eqnarray} 
&\Pr&\bigg( \bigcap _{\lbrace j: \vert \mu_{dj} \vert > ra_n\rbrace} 
\left\lbrace \vert \hat{\mu}_{dj} \vert > a_n \right\rbrace \bigg) 
\geq  
1- \sum _{j=1}^p 
\Pr\bigg(  \vert \hmu _{dj}- \mu_{dj} \vert > a_n(r-1) \bigg) \nonumber\\
& \geq  &
1-2 \sum _{j=1}^p 
\Phi \left( \frac{-a_n(r-1)}{\sigma_j\sqrt{n/n_1n_2}}  \right)
\nonumber \\
&\geq & 
1-p c_1 \exp 
\left\lbrace  -\left( \frac{\log p}{n}\right) ^{2\alpha}
.\frac{n_1n_2}{n}c_2  \right\rbrace  \label{a-i}.
\end{eqnarray}
Since $(\log p/n_2)(n_1/\log p)^{2\alpha}=o(1)$ and 
$n_2=o(n_1)$, as $n_1,n_2\to \infty$, 
\eqref{a-i} tends to $1$, and the result of part (a) holds. 

\noindent
\textbf{(b)}
Similar to part (a), for some positive constants $c_1,c_2$, we have 
\begin{eqnarray*}
\Pr\bigg( \bigcap _{\lbrace j: \vert \mu_{dj} \vert \leq a_n/r \rbrace} 
\left\lbrace \vert \hmu _{dj}\vert \leq a_n \right\rbrace 
\bigg) 
\geq 
1-p c_1 \exp \left\lbrace  
-\left( \frac{\log p}{n}\right) ^{2\alpha}
\frac{n_1n_2}{n}c_2  
\right\rbrace  .
\end{eqnarray*}
This together with $(\log p/n_2)(n_1/\log p)^{2\alpha}=o(1)$ 
and $n_2=o(n_1)$, prove that the right hand side of 
the above inequality tends to $1$, as $n_1,n_2 \to \infty$.  

\noindent 
\textbf{(c)} 
The result follows from parts  (a) and (b).
$\blacksquare$ \\

\noindent
\textbf{Proof of Theorem \ref{thrm:slda}}.
\textbf{(a)}
The misclassification rates of {\sc slda} in Class $k=1,2$, are given as
\begin{eqnarray*}
\Pi_k^{\text{\sc slda}}(\calD_n)
=
\Phi  \left( \frac{ (-1)^k\tbmu _d^{\top} \widetilde{\bSigma}_n^{-1}
	(\hat{\pmb{\mu}}_k-\pmb{\mu}_k)
	-\tbmu _d^{\top}\widetilde{\bSigma}_n^{-1}\hat{\pmb{\mu}}_d/2 }
{\sqrt{ \tbmu _d^{\top}\widetilde{\bSigma}_n^{-1} \bSigma \widetilde{\bSigma}_n^{-1}\tbmu _d } }
\right). 
\end{eqnarray*}

Recall $d_{n_1}=C_{h,p}(n_1^{-1}\log p)^{(1-h)/2}$, where 
$C_{h,p}=\max _{1\leq i\leq p}\sum_{j=1}^p \vert \sigma_{ij} \vert ^{h}$
for some $0\leq h <1$. It follows from Lemma \ref{lem:bickel_cov} 
in the Appendix A 
that  
\begin{eqnarray*}
\tbmu _d^{\top} \widetilde{\bSigma}_n^{-1} \bSigma \widetilde{\bSigma}_n^{-1} \tbmu _d
=
\tbmu _d^{\top} \widetilde{\bSigma}_n^{-1} \tbmu _d 
\lbrace 1+O_p(d_{n_1}) \rbrace. 
\end{eqnarray*}
Let $\Delta_p^2=\bmu_d^{\top}\bSigma^{-1}\bmu_d$, 
$\calJ_1=(\tbmu _d-\bmu_d)^{\top}\bSigma^{-1} (\tbmu _d-\bmu_d)$ and 
$\calJ_2=2\bmu_d^{\top}\bSigma^{-1}(\tbmu _d-\bmu_d)$. Now, 
\begin{eqnarray*}
\tbmu _d^{\top}\bSigma^{-1}\tbmu _d =
\calJ_1+\calJ_2+ \Delta_p^2
\label{denom}
\end{eqnarray*}
Following by the proof of Theorem 1 of \cite{shao2011sparse}, 
we have 
$$
\calJ_1
\leq
c_0 \left\lbrace 
\parallel \tbmu _{d1}-\pmb{\mu}_{d1} \parallel ^2 
+ \parallel \pmb{\mu}_{d0} \parallel ^2  
\right\rbrace,
$$ 
where  
$\tbmu _d^{\top}=(\tbmu _{d1}^{\top},\textbf{0}^{\top})$, 
$\pmb{\mu}_d^{\top}=(\pmb{\mu}_{d1}^{\top},\pmb{\mu}_{d0}^{\top})$, 
and $\tbmu _{d1}$ and $\pmb{\mu}_{d1}$ are two 
vectors of dimension $\hat{q}$, whose elements 
correspond to those features $x_j$s for which $| \hmu _{dj} | > a_n$. 
By condition \eqref{cond_lem2_shao}, we have 
$\parallel  \tbmu _{d1}-\bmu_{d1} \parallel ^2=O_p(q_n/n_2)$,
$\parallel \bmu_{d0} \parallel ^2=O_p\left( D_{g,p}.a_n^{2(1-g)}\right) $, 
and $\calJ_1=O_p(k_{n_2})$, where 
$k_{n_2}=\max \lbrace  \frac{q_n}{n_2}, \ D_{g,p}a_n^{2(1-g)} \rbrace $.
Consequently, by condition \eqref{cond_lem2_shao}, 
\begin{eqnarray*}
\calJ_2=(\tbmu _d-\pmb{\mu}_d)^{\top}\bSigma^{-1}\pmb{\mu}_d 
\leq 
\Delta _p 
\sqrt{\parallel \tbmu _{d1}-\pmb{\mu}_{d1} \parallel ^2
	+\parallel \pmb{\mu} _{d0} \parallel ^2 
} 
= 
\Delta_p ~O_p(\sqrt{k_{n_2}})	.
\end{eqnarray*}
Therefore in the denominator of 
$\Pi_k^{\text{\sc slda}}(\calD_n)$, we have 
\begin{eqnarray}
\tbmu _d^{\top}\widetilde{\bSigma}_n^{-1}\bSigma\widetilde{\bSigma}_n^{-1}\tbmu _d 
&=&
\bigg \{ O_p(k_{n_2})
+\Delta_p O_p(\sqrt{k_{n_2}})
+\Delta _p^2 \bigg \} ~O_p(d_{n_1}) 
\nonumber\\
&=&
\bigg \{
O_p\bigg(\sqrt{k_{n_2}/\Delta_p^2}\bigg)
+1
\bigg \} ~
\Delta _p^2 ~ O_p(d_{n_1}) . 
\label{denom_slda}
\end{eqnarray}
Now, the numerator of $\Pi_k^{\text{\sc slda}}(\calD_n)$ 
can be decomposed as
\begin{eqnarray}
&& (-1)^k\tbmu _d^{\top}\widetilde{\bSigma}_n^{-1}(\hbmu _k-\pmb{\mu}_k)
-\frac{1}{2}\hbmu _d^{\top}\widetilde{\bSigma}_n^{-1}\tbmu _d  
\nonumber\\
&=&
(-1)^k\tbmu ^{\top}_d\widetilde{\bSigma}_n^{-1}(\hbmu _k-\pmb{\mu}_k) 
- \frac{1}{2}(\hbmu _d-\pmb{\mu}_d)^{\top}\widetilde{\bSigma}_n^{-1}\tbmu _d 
-\frac{1}{2}(\pmb{\mu}_d-\tbmu _d)^{\top}\widetilde{\bSigma}_n^{-1}\tbmu _d \nonumber\\ 
&-&\frac{1}{2}\tbmu _d^{\top}\widetilde{\bSigma}_n^{-1}\tbmu _d  
\nonumber\\
&=&
\calJ_3+\calJ_4+\calJ_5- \frac{1}{2}\tbmu _d^{\top}\widetilde{\bSigma}_n^{-1}\tbmu _d  
\nonumber\\
&=&
\sqrt{\tbmu _d^{\top}\widetilde{\bSigma}_n^{-1}\tbmu }_d\left\lbrace O_p\bigg(\sqrt{q_n/n_k}\bigg)
+
O_p\bigg(\sqrt{q_nC_{h,p}/n_k}\bigg)\right\rbrace \sqrt{1+O_p(d_{n_1})}
\nonumber\\
&+& \sqrt{\tbmu _d^{\top}\widetilde{\bSigma}_n^{-1}\tbmu _d} 
O_p(\sqrt{q_n/n_2})\sqrt{1+O_p(d_{n_1})} \nonumber\\
&+&\left\lbrace O_p(k_{n_2})
+\Delta _pO_p(\sqrt{k_{n_2}}) \right\rbrace  
O_p(d_{n_1}) 
+\tbmu _d^{\top}\widetilde{\bSigma}_n^{-1}\tbmu _d .
\label{num_slda}
\end{eqnarray}

Again, by condition \eqref{cond_lem2_shao}  
we have 
\begin{eqnarray*}
\calJ_3&=&
\tbmu_d^{\top}\widetilde{\bSigma}_n^{-1}(\hbmu_k-\bmu_k)
\\&=&
\sqrt{\tbmu _d^{\top}\widetilde{\bSigma}_n^{-1}\tbmu }_d\left\lbrace O_p\bigg(\sqrt{q_n/n_k}\bigg)
+
O_p\bigg(\sqrt{q_nC_{h,p}/n_k}\bigg)\right\rbrace \sqrt{1+O_p(d_{n_1})}, 
\end{eqnarray*}
and
\begin{eqnarray*}
\calJ_4
=
(\hbmu _d-\bmu_d)^{\top}\widetilde{\bSigma}_n^{-1}\tbmu _d
=
\sqrt{  \tbmu _d^{\top} \widetilde{\bSigma}_n^{-1} \tbmu _d  }
O_p(\sqrt{q_n/n_2}).
\end{eqnarray*}
Also, similar to the expression of $\calJ_1$, we have 
\begin{eqnarray*}
\calJ_5
=
(\pmb{\mu}_d-\tbmu _d)^{\top}\widetilde{\bSigma}_n^{-1}\tbmu _d
=
\sqrt{\tbmu_d^{\top}\widetilde{\bSigma}_n^{-1}\tbmu_d}
O_p(\sqrt{k_{n_2}})\sqrt{1+O_p(d_{n_1})}.
\end{eqnarray*}
finally, by combining \eqref{denom_slda} and \eqref{num_slda} we arrive at 
\begin{eqnarray*}
&&\Pi_k^{\text{\sc slda}}(\calD_n)
=
\Phi \bigg( 
\frac{(-1)^k\tbmu _d^{\top}\widetilde{\bSigma}_n^{-1}(\hbmu _k-\pmb{\mu}_k) }
{\sqrt{\tbmu _d^{\top}\widetilde{\bSigma}_n^{-1}\bSigma\widetilde{\bSigma}_n^{-1}
		\tbmu _d}} 
-
\frac{1}{2} \frac{\hbmu _d^{\top}\widetilde{\bSigma}_n^{-1}\tbmu _d}
{\sqrt{\tbmu _d^{\top}\widetilde{\bSigma}_n^{-1}\bSigma\widetilde{\bSigma}_n^{-1}\tbmu _d}}  
\bigg) \\
&=& 
\Phi \bigg(
-\frac{1}{2}\Delta_p 
\left\lbrace 
O_p\bigg(\Delta_p^{-1}\sqrt{q_{n}C_{h,p}/n_k}\bigg) 
+
O_p\bigg(\sqrt{k_{n_2}/\Delta_p^2}\bigg)
+
1
+
O_p(d_{n_1}) 
\right\rbrace 
\bigg) \\
&=&  
\Phi \bigg(
-\frac{1}{2}\Delta_p 
\left\lbrace 1+O_p(b_{n_k}) \right\rbrace 
\bigg)~~,~~k= 1, 2
\end{eqnarray*}
as claimed, 
where
\begin{eqnarray*}
b_{n_k}
=
\max 
\left\lbrace  d_{n_1}, \
\frac{\sqrt{k_{n_2}}}{\Delta_p}, \
\frac{1}{\Delta_p}\sqrt{\frac{q_n}{n_k}C_{h,p}}  
\right\rbrace .
\end{eqnarray*}

\noindent
\textbf{(b)-i.}
If $\Delta_p$ is bounded, then $\Delta _p^2b_{n_k}\to 0$ 
is equivalent to $b_{n_2}\to 0$, which imply 
$\Pi_k^{\text{\sc slda}}(\calD_n)/\Pi^{\text{opt}}\overset{p}{\longrightarrow} 1$, for $k= 1, 2$.

\noindent
\textbf{(b)-ii.}
If $\Delta_p\to \infty$, by   
Lemma \ref{lem:lem1_shao} in the Appendix A, 
when $\Delta _p^2b_{n_2}\to 0$, 
and consequently $\Delta _p^2b_{n_1}\to 0$,  
we have 
$\Pi_k^{\text{\sc slda}}(\calD_n)/\Pi^{\text{opt}}\overset{p}{\longrightarrow} 1$, 
for $k=1,2$. 
$\blacksquare$ \\

\noindent
\textbf{Proof of Theorem \ref{thrm:ROAD}}.
The class-specific MCRs of the {\sc road} in \eqref{ROADF} are given by
\begin{eqnarray*}
\Pi_k^{\text{\sc road}}(\calD_n;c)
&=&
\Phi \left( \frac{(-1)^k\hat{\ww}_c^{\top}(\hat{\bmu}_a-\bmu_k)}
{(\hat{\ww}_c^{\top}\bSigma\hat{\ww}_c)^{1/2}} \right), 
~  k=1,2.
\end{eqnarray*} 
The oracle versions of the MCRs, 
evaluated at the true parameter values of $\bSigma$ and $\bmu_k$, are given by
\begin{eqnarray*}
\Pi_k^{\text{orc}}(c)
&=&
\Phi \bigg(
\frac{-\ww_c^{\top}\bmu_d}{2(\ww_c^{\top}\bSigma \ww_c)^{1/2}}
\bigg), ~ k=1,2.
\end{eqnarray*} 
By the tail probability inequality  
\begin{eqnarray*}
1-\Phi(\tau)\leq 
\frac{1}{\tau\sqrt{2\pi}}\exp \lbrace -\tau ^2/2 \rbrace,
~ ~~ \tau >0,
\end{eqnarray*} 
we have that, for $\eta_1>0$,
\begin{eqnarray*}
\Pr \bigg( \parallel \hbmu_k-\bmu_k \parallel_{\infty } >\eta_1 \bigg) 
\leq 
\sum_{j=1}^p \Pr \bigg(  \vert  \hmu_{jk}-\mu_{jk} \vert > \eta_1 \bigg)
\leq 
C_1p \exp \lbrace -C_2n_k \eta^2_1 \rbrace .
\end{eqnarray*} 
Thus, by choosing $\eta_1 = M_1 a_{n_k}$, for some $M_1>0$, we arrive at  
$\parallel \hbmu_k-\bmu_k \parallel_{\infty }=O_p(\sqrt{\log p/n_k})$. 
Also, by Lemma \ref{lem:A3bickel} in the Appendix A, for $\eta_2>0$, 
\begin{eqnarray*}
&&\Pr \bigg( \max_{j,l} \vert \hsigma_{j,l}-\sigma_{j,l} \vert >\eta_2 \bigg) \leq  \\
&\leq &
\sum_{j,l} \sum_{k=1}^2
\Pr \bigg( 
\vert  \sum_{i=1}^{n_k}(X_{ijk}X_{ilk}-\sigma_{jl}) \vert >(n-2) \eta_2 /4
\bigg)\\
&+&
\sum_{j,l} \sum_{k=1}^2
\Pr \bigg( 
\vert  n_k\hmu_{jk}\hmu_{lk}-\sigma_{jl}) \vert >(n-2) \eta_2 /4
\bigg)\\
&\leq &
p^2C_1 \exp \lbrace -C_2(n-2)^2\eta^2_2/n_k \rbrace
+
p^2C_3 \exp \lbrace -C_4(n-2)^2\eta_2^2 \rbrace.
\end{eqnarray*} 
Thus, by choosing $\eta_2 = M_2\sqrt{\log p /n_1}$, for some $M_2>0$, we arrive at
$\|\widehat \bSigma_n - \bSigma \|_{\infty} = O_p(\sqrt{\log p /n_1})$. 
Using the Lipschitz property of 
the cumulative distribution function of standard normal, 
$\Phi(.)$, we have 
\begin{eqnarray*}
&& ~\bigg\vert \Pi_2^{\text{\sc road}}(\calD_n;c)
-\Pi_2^{\text{orc}}(c) \bigg\vert 
\leq  
\bigg\vert  
\frac{-\hat{\ww}_c^{\top}(\bmu_2-\hbmu_a)}
{(\hat{\ww}_c^{\top}\bSigma\hat{\ww}_c)^{1/2}}
-
\frac{-\ww_c^{\top} \bmu_d}
{2(\ww_c^{\top}\bSigma\ww_c)^{1/2}} 
\bigg\vert \\
&= & 
\bigg\vert 
\frac{-\hat{\ww}_c^{\top}(\bmu_2-\hat{\bmu}_2+\hbmu_2-\hbmu_a)}
{(\hat{\ww}_c^{\top}\bSigma\hat{\ww}_c)^{1/2}}
-
\frac{-\ww_c^{\top} \pmb\mu_d}
{2(\ww_c^{\top}\bSigma\ww_c)^{1/2}} 
\bigg\vert 
\\
&\leq &
\bigg\vert 
\frac{\hat{\ww}_c^{\top}(\bmu_2-\hbmu_2)}
{(\hat{\ww}_c^{\top}\bSigma \hat{\ww}_c)^{1/2}}
\bigg\vert 
+
\bigg\vert  
\frac{\hat{\ww}_c^{\top}\hbmu_d}
{2(\hat{\ww}_c^{\top}\bSigma\hat{\ww}_c)^{1/2}}-
\frac{\ww_c^{\top} \bmu_d}
{2(\ww_c^{\top}\bSigma \ww_c)^{1/2}} 
\bigg\vert \\ \\
& = & E_1+E_2. 
\end{eqnarray*}
Now, 
\begin{eqnarray*}
E_1&=&\bigg\vert 
\frac{\hat{\ww}_c^{\top}(\bmu_2-\hbmu_2)}
{(\hat{\ww}_c^{\top}\bSigma \hat{\ww}_c)^{1/2}}
\bigg\vert
\leq 
\frac{\parallel \hat{\ww}_c \parallel _1}
{\parallel \hat{\ww}_c \parallel _2 .c_0}
\parallel \hbmu_2-\bmu_2 \parallel _{\infty}	 
\\
&\leq &
\sqrt{\parallel \hat{\ww}_c \parallel _0}
O_p(\sqrt{\log p /n_2})
=O_p(\sqrt{\hat{s}_c\log p /n_2})
\end{eqnarray*}
and 
\begin{eqnarray*}
E_2 &=&
\bigg\vert   
\frac{\hat{\ww}_c^{\top}\hbmu_d}
{2(\hat{\ww}_c^{\top}\bSigma\hat{\ww}_c)^{1/2}}-
\frac{\ww_c^{\top} \bmu_d}
{2(\ww_c^{\top} \bSigma \ww_c)^{1/2}} 
\bigg\vert  \\
&=& 
\bigg\vert
\frac{\hat{\ww}_c^{\top}\hbmu_d-\hat{\ww}_c^{\top}
	{\bmu}_d+\hat{\ww}_c^{\top}{\bmu}_d }
{2(\hat{\ww}_c^{\top}\bSigma\hat{\ww}_c)^{1/2}}-
\frac{\ww_c^{\top} \bmu_d}
{2(\ww_c^{\top}\bSigma\ww_c)^{1/2}} 
\bigg\vert   \\
&\leq & 
\bigg\vert 
\frac{\hat{\ww}_c^{\top}(\hbmu_d-{\bmu}_d)}
{2(\hat{\ww}_c^{\top}\bSigma\hat{\ww}_c)^{1/2}} 
\bigg\vert  
+ 
\bigg\vert  
\frac{\hat{\ww}_c^{\top}{\bmu}_d}
{2(\hat{\ww}_c^{\top}\bSigma\hat{\ww}_c)^{1/2}}
- 
\frac{{\ww}_c^{\top}{\bmu}_d}
{2(\ww_c^{\top}\bSigma\ww_c)^{1/2}}
\bigg\vert  \\
&\leq & 
\frac{\parallel \hat{\ww}_c \parallel_1}
{\parallel \hat{\ww}_c \parallel_2}
\frac{\parallel \hbmu_d-\bmu_d \parallel_{\infty}}
{\min _j \lambda_j}
+ 
E_3\\
&\leq & 
\sqrt{\parallel \hat{\ww}_c \parallel_0}
\frac{\parallel \hbmu_d-\bmu_d \parallel_{\infty}}{c_0}
+ 
E_3
=
O_p(\sqrt{\hat{s}_c\log p/n_2})+E_3.
\end{eqnarray*}

According to the same notations in \cite{fan2012road}, let 
$f_0(\ww)=\ww^{\top}\bmu_d/(\ww^{\top}\bSigma\ww)^{1/2}$, 
$f_1(\ww)=\ww^{\top}\hbmu_d/(\ww^{\top}\bSigma\ww)^{1/2}$, and 
$f_2(\ww)=\ww^{\top}\hbmu_d/(\ww^{\top}\widehat{\bSigma}\ww)^{1/2}$. 
By the proof of Theorem 1 of \cite{fan2012road}, we have 
\begin{eqnarray*}
E_3&=&\bigg\vert  
\frac{\hat{\ww}_c^{\top}{\bmu}_d}
{2(\hat{\ww}_c^{\top}\bSigma\hat{\ww}_c)^{1/2}}
- 
\frac{{\ww}_c^{\top}{\bmu}_d}
{2(\ww_c^{\top}\bSigma\ww_c)^{1/2}}
\bigg\vert =\frac{1}{2}\vert f_0(\hat{\ww}_c)-f_0(\ww_c)  \vert \\
&\leq &
\vert f_0(\hat{\ww}_c)-f_1(\hat{\ww}_c)  \vert 
+\vert f_1(\hat{\ww}_c)-f_2(\hat{\ww}_c)  \vert 
+\vert f_2(\hat{\ww}_c)-f_0(\ww_c)  \vert \\
&=&
O_p(\sqrt{\hat{s}_c\log p/n_2})+O_p(c^2\sqrt{\log p/n_1})+
O_p\bigg(\sqrt{\max\lbrace s_c , ~ s_c^{(1)}\rbrace \log p/n_2}\bigg).
\end{eqnarray*}
Therefore, we have 
$$E_1+E_2+E_3=O_p(c^2\sqrt{\log p/n_1})+
O_p\bigg(\sqrt{\max\lbrace s_c ,~ s_c^{(1),~ \hat{s}_c}\rbrace \log p/n_2}\bigg),$$ 
and finally 
\begin{equation*}
\bigg\vert \Pi_2^{\text{\sc road}}(\calD_n;c)-
\Pi_2^{\text{orc}}(c) \bigg\vert 
=
O_p(c^2\sqrt{\log p/n_1})+
O_p\bigg(\sqrt{\max\lbrace s_c ,~ s_c^{(1),~ \hat{s}_c}\rbrace \log p/n_2}\bigg)
\end{equation*}
Similarly, the same result holds for
$\vert \Pi_1^{\text{\sc road}}(\calD_n;c)-\Pi_1^{\text{orc}}(c) \vert$, and this completes 
the proof.  $\blacksquare$ \\
\section{Remaining Proofs}
\label{app:remain}
\noindent
In this Appendix, we provide the proofs of our claim in Remark \ref{remark:prop1}, and also 
the proofs of Propositions \ref{prop:DACH0} and 
\ref{prop:gap_fill_general}. 
\\

\noindent
\textbf{Proof of the Claim in Remark \ref{remark:prop1}}.
Recall $\calI_i$, $i=1,...,6$, defined in Theorem \ref{thrm:LDA}. 
When $p$ is fixed with respect to the sample size, and $n_2=o(n_1)$, then 
as $n_1,n_2\to  \infty$, we have, 
\[
Var(\calI_1)=\frac{2p}{n_1^2} \to 0~,~Var(\calI_2)=\frac{2p}{n_2^2} \to 0~,~
Var(\calI_3)=\frac{\Delta_p^2 }{n_2}\to 0~,~
\]
\[
Var(\calI_4)=\frac{\Delta_p^2 }{n_1} \to 0~,~
Var(\calI_5)=\frac{n^2p}{n_1^2n_2^2}\to 0~,~
Var(\calI_6)=\frac{\Delta_p^2 }{n_2}\to 0.
\]
On the other hand,   
$\mathbb{E}(\calI_1)=\frac{p}{n_1}$, $\mathbb{E}(\calI_2)=\frac{p}{n_2}$, 
$\mathbb{E}(\calI_3)=\mathbb{E}(\calI_4)=0$, $\mathbb{E}(\calI_5)=\frac{np}{n_1n_2}$, 
and $\mathbb{E}(\calI_6)=0$. Thus, by following the proof of Theorem \ref{thrm:LDA}, 
we have
\begin{eqnarray*}
\frac{\Psi _1^{\text{\sc lda}} (\hbtheta_n)}{\sqrt{\Upsilon ^{\text{\sc lda}}  (\hbtheta_n)}} 
&=& \frac{ \frac{p}{n_1}-\frac{p}{n_2}+o_p(1)-\Delta_p^2   }
{2\left\lbrace   \frac{np}{n_1n_2}+o_p(1)+\Delta_p^2  \right\rbrace ^{1/2} }
= -\frac{1}{2} \Delta_p +o_p(1), 
\\
\frac{\Psi _2^{\text{\sc lda}} (\hbtheta_n)}{\sqrt{\Upsilon^{\text{\sc lda}}  (\hbtheta_n) }} 
&=& \frac{ -\frac{p}{n_1}+\frac{p}{n_2}+o_p(1)-\Delta_p^2  }
{2 \left\lbrace   \frac{np}{n_1n_2}+o_p(1)+\Delta_p^2   \right\rbrace  ^{1/2}}
= -\frac{1}{2}\Delta_p +o_p(1).
\end{eqnarray*}
Therefore, $\Pi_k^{\text{\sc lda}}(\calD_n)/\Pi^{\text{opt}} \overset{p}{\longrightarrow} 1$, 
for $k=1,2$. $\blacksquare$ \\ 

\noindent
\textbf{Proof of Proposition \ref{prop:DACH0}}.
The MCRs of $\delta_0^{\text{Msplit-{\sc hr}}}$ are given by 
\begin{eqnarray*}
\Pi_{0,k}^{\text{Msplit-{\sc hr}}}(\calD_n)
=
\Phi \bigg(
\frac{\Psi_{0,k}^{\text{Msplit-{\sc hr}}}(\hbtheta_n)}
{\sqrt{\Upsilon_0^{\text{Msplit-{\sc hr}}}(\hbtheta_n)}}
\bigg), ~~ k= 1, 2,
\end{eqnarray*} 
where 
\[
\Psi_{0,k}^{\text{Msplit-{\sc hr}}}(\hbtheta_n)
=(-1)^{(k+1)}\frac{1}{\calL}\sum_{\ell=1}^{\calL}\sum_{j=1}^p
r_j(\bmu_k;\hbtheta_{n,\ell}^{(2)})h_j(\hbtheta_{n,\ell}^{(1)}).
\]
and 
\[
\Upsilon_0^{\text{Msplit-{\sc hr}}}(\hbtheta_n)
=\frac{1}{\calL^2}\sum_{\ell=1}^{\calL}\sum_{j=1}^p
\sigma_j^2 \bigg( 
\hmu_{dj,\ell}^{(2)}/\hsigma_{j,\ell}^{(2),2} 
\bigg)^2 h_j(\hbtheta_{n,\ell}^{(1)}).
\]
Now, due to the independence property of 
$\calD_{n,\ell}^{(1)}$ and $\calD_{n,\ell}^{(2)}$, 
for each replication $t$, we have,  
\begin{eqnarray*}
B_{0,n}^{\text{Msplit-{\sc hr}}} &=&
\mathbb{E} \lbrace
\Psi_{0,1}^{\text{Msplit-{\sc hr}}}(\hbtheta_n)
-
\Psi_{0,2}^{\text{Msplit-{\sc hr}}}(\hbtheta_n)
\rbrace \\
&=&
\frac{1}{\calL}\sum_{\ell=1}^{\calL}\sum_{j=1}^p
\mathbb{E}\bigg\lbrace r_j(\bmu_1;\hbtheta_{n,\ell}^{(2)})
+ r_j(\bmu_2;\hbtheta_{n,\ell}^{(2)})\bigg\rbrace
\mathbb{E} \bigg\lbrace  h_j(\hbtheta_{n,\ell}^{(1)}) \bigg\rbrace ,  
\end{eqnarray*} 
where 
$r_j(\bmu_k;\hbtheta_{n,\ell}^{(2)})=
\hmu_{dj,\ell}^{(2)}(\mu_{jk,\ell}-\hmu_{aj,\ell}^{(2)})
/\hsigma_{j,\ell}^{(2),2}$. 
Hence 
\begin{eqnarray*} 
\bar{r}_n =
\mathbb{E}
\bigg\lbrace r_j(\bmu_1;\hbtheta_{n,\ell}^{(2)})
+ r_j(\bmu_2;\hbtheta_{n,\ell}^{(2)})\bigg\rbrace
=
(\frac{1}{n'_1}-\frac{1}{n'_2})
\frac{\Gamma (f_{n'}-1 )}
{\Gamma (f_{n'})}
f_{n'},
\end{eqnarray*}
where $f_{n'}=n'/2-1$.
$\blacksquare$ \\

\noindent
\textbf{Proof of Proposition \ref{prop:gap_fill_general}}.
The MCRs of $\delta^{\text{Msplit-{\sc hr}}}$ in 
\eqref{DACH-general} are given by 
\begin{eqnarray*}
\Pi_{0,k}^{\text{Msplit-{\sc hr}}}(\calD_n)
=
\Phi \bigg(
\frac{\Psi_{0,k}^{\text{Msplit-{\sc hr}}}(\hbtheta_n)}
{\sqrt{\Upsilon_0^{\text{Msplit-{\sc hr}}}(\hbtheta_n)}}
\bigg),~k=1,2,
\end{eqnarray*} 
where 
\[
\Psi_{0,k}^{\text{Msplit-{\sc hr}}}(\hbtheta_n)
=\frac{(-1)^k}{\calL}\sum_{\ell=1}^{\calL} 
\tilde{\bmu}_{d,\ell}^{\top} \widetilde{\bSigma}_{\ell}^{-1} 
(\tilde{\bmu}_{a,\ell}-\bmu_k),
\]  
and 
\[ 
\Upsilon_0^{\text{Msplit-{\sc hr}}}(\hbtheta_n)
=\frac{1}{\calL^2}\sum_{\ell=1}^{\calL} 
\tilde{\bmu}_{d,\ell}^{\top} \widetilde{\bSigma}_{\ell}^{-1} 
\bSigma_{\ell} \widetilde{\bSigma}_{\ell}^{-1} \tilde{\bmu}_{d,\ell}.
\]
Hence, 
\begin{eqnarray*}
&&\mathbb{E} \lbrace 
\Psi_{0,1}^{\text{Msplit-{\sc hr}}}(\hbtheta_n)
-
\Psi_{0,2}^{\text{Msplit-{\sc hr}}}(\hbtheta_n)
\rbrace \\
&=&
\frac{1}{\calL}\sum_{\ell=1}^{\calL}
\mathbb{E}\bigg\lbrace 
\mathbb{E}\bigg\lbrace 
\tbmu_{d,\ell}^{\top}  \widetilde{\bSigma}_{\ell}^{-1}
(\bmu_{1,\ell}-\tbmu_{a,\ell}) 
- 
\tbmu_{d,\ell}^{\top}  \widetilde{\bSigma}_{\ell}^{-1}
(\tbmu_{a,\ell}-\bmu_{2,\ell})
\bigg\vert 
\calD_{n,\ell}^{(1)}
\bigg\rbrace 
\bigg\rbrace \\
&=&
\frac{1}{\calL}\sum_{\ell=1}^{\calL} 
\mathbb{E} \bigg\lbrace  
\mathbb{E} \bigg\lbrace  
\tbmu_{d,\ell}^{\top}  \widetilde{\bSigma}_{\ell}^{-1}
(\bmu_{1,\ell}+ \bmu_{2,\ell}- 2\tbmu_{a,\ell}) 
\bigg\rbrace 
\bigg\rbrace \\
&=&
\frac{1}{\calL}\sum_{\ell=1}^{\calL} 
\mathbb{E}\lbrace \bar{r}_{n,\ell}\rbrace .
\end{eqnarray*}
The second equation follows from the independence 
property of $\calD_{n,\ell}^{(1)}$ and $\calD_{n,\ell}^{(2)}$, 
for each $\ell$. Under 
normal assumption for the distribution of features, 
the matrix $\widetilde{\bSigma}_{\ell}^{-1}$ 
has the Inverse Wishart distribution 
with parameters $\bSigma_{\ell}^{-1}$ and $n'-2$, 
where $\bSigma_{\ell}$ is the covariance matrix corresponding to 
the features included in $\calS_{n,\ell}^{(1)}$, 
Thus, if $n'-3>\vert \calS_{n,\ell}^{(1)}  \vert$, then 
$\mathbb{E}\lbrace \widetilde{\bSigma}_{\ell}^{-1} \rbrace 
=\frac{n'-2}
{n'-2- \vert \calS_{n,\ell}^{(1)}  \vert  -1} \bSigma_{\ell}^{-1}$,  
and 
\begin{eqnarray*}
\bar{r}_{n,\ell}
&=&
\mathbb{E}\lbrace 
\tbmu_{d,\ell}^{\top}  \widetilde{\bSigma}_{\ell}^{-1}
(\bmu_{1,\ell}+\bmu_{2,\ell}-2\tbmu_{a,\ell}) 
\rbrace \\
&=&
tr \bigg\lbrace 
\bSigma_{\ell}^{-1}
\frac{n'-2}{n'-3- \vert \calS_{n,\ell}^{(1)} \vert  }
\bSigma_{\ell} (\frac{1}{n'_1}-\frac{1}{n'_2})
\bigg\rbrace \\
&=&
\frac{n'-2}{n'-3- \vert \calS_{n,\ell}^{(1)} \vert  }
\vert \calS_{n,\ell}^{(1)} \vert
(\frac{1}{n'_1}-\frac{1}{n'_2}).
\end{eqnarray*}
and the result follows. $\blacksquare$\\

\clearpage

\bibliographystyle{chicago}
\bibliography{ref}
\end{document}